%% file: girv.tex
\documentclass[useAMS,usenatbib]{mn2e}
\usepackage{graphicx,lscape,amssymb}

\newcommand{\Teff}{\mbox{$\mathrm{T}_{\mathrm{eff}}$}}
\newcommand{\Logg}{\mbox{$\log\,g$}}
\newcommand{\Msun}{\mbox{$\mathrm{M}_{\odot}$}}

\newcommand{\fuv}{\mbox{$\mathrm{m}_{\mathrm{FUV}}$}}
\newcommand{\nuv}{\mbox{$\mathrm{m}_{\mathrm{NUV}}$}}
\newcommand{\rcmc}{\mbox{$r_{\mathrm{CMC}}$}}
\newcommand{\rsdss}{\mbox{$r_{\mathrm{SDSS}}$}}
\newcommand{\ebv}{\mbox{$\mathrm{E}(\mathrm{B}-\mathrm{V})$}}
\newcommand{\frrK}{\mbox{$(\fuv-\rcmc)$} vs \mbox{$(\rcmc-K_s)$}}
\newcommand{\nrrK}{\mbox{$(\nuv-\rcmc)$} vs \mbox{$(\rcmc-K_s)$}}
\newcommand{\UKfrrK}{\mbox{$(\fuv-\rsdss)$} vs \mbox{$(\rsdss-K)$}}
\newcommand{\UKnrrK}{\mbox{$(\nuv-\rsdss)$} vs \mbox{$(\rsdss-K)$}}

\title{The Unseen Population of F to K-type Companions to Hot
Subdwarf Stars}

\author[J.Girven et al.]{
J. Girven$^{1\star}$, D. Steeghs$^1$, U. Heber$^2$, B.~T. G\"ansicke$^1$,
T.~R. Marsh$^1$, E. Breedt$^1$, \newauthor
~C.~M. Copperwheat$^{1}$, S. Pyrzas$^1$ and P. Longa Pe\~{n}a$^1$\\
$^{1}$ Department of Physics, University of Warwick, Coventry CV4 7AL,
UK \\
$^{2}$ Dr. Remeis-Sterwarte, Astronomisches Institut der Universit\"at
Erlangen-N\"urnberg, Sternwartstr. 7, 96049 Bamberg, Germany \\
$^{\star}$  j.m.girven@warwick.ac.uk
}

\begin{document}

\date{Started 2010}

\maketitle

\label{firstpage}

\begin{abstract}

\noindent We present a method to select hot subdwarf stars with A to
M-type companions using photometric selection criteria. We cover a wide
range in wavelength by combining GALEX ultraviolet data, optical
photometry from the SDSS and the Carlsberg Meridian telescope,
near-infrared data from 2MASS and UKIDSS. We construct two complimentary
samples, one by matching GALEX, CMC and 2MASS, as well as a smaller, but
deeper, sample using GALEX, SDSS and UKIDSS. In both cases, a large
number of composite subdwarf plus main--sequence star candidates were
found. We fit their spectral energy distributions with a composite model
in order to estimate the subdwarf and companion star effective
temperatures along with the distance to each system. The distribution of
subdwarf effective temperature was found to primarily lie in the
$20,000-30,000$\,K regime, but we also find cooler subdwarf candidates,
making up $\sim5-10$\,per\,cent. The most prevalent companion spectral
types were seen to be main--sequence stars between F0 and K0, while
subdwarfs with M-type companions appear much rarer. This is clear
observational confirmation that a very efficient first stable Roche-lobe
overflow channel appears to produce a large number of subdwarfs with F to
K-type companions. Our samples thus support the importance of binary
evolution for subdwarf formation.

\end{abstract}

\begin{keywords}
stars: fundamental parameters - subdwarfs - white dwarfs - ultraviolet:
stars - infrared: stars.
\end{keywords}

\section{Introduction}
\label{s-int}

Subluminous blue stars were first discovered by \citet{humason+zwicky47-1}
in a photometric survey of the North Galactic Pole region.
\citet{greenetal86-1} found many more hot subdwarfs in the
Palomar-Green (PG) survey, to the extent that they were the dominant
species among faint ($B \lesssim 16.1$) blue objects. In the PG survey
they outnumber white dwarfs (WD) and are prevalent enough to account for
the ultraviolet upturn in early-type galaxies \citep{brownetal97-1}. Hot
subdwarf stars are either core helium-burning stars at the end of the
horizontal branch or have evolved even beyond that stage
\citep{heberetal84-1, heber86-1}. They have a relatively well defined
mass around the canonical (theoretical) value of $0.46$\,\Msun\
\citep{safferetal94-1, hanetal03-1, politanoetal08-1} and radii of a few
tenths of a solar radius. Their very thin layers of hydrogen
($\mathrm{M}_{\mathrm{env}}<0.01\Msun$) are not able to support shell
burning after helium-core exhaustion. Thus instead of following the
asymptotic giant branch route, they evolve more or less directly into
white dwarfs. Observationally two classes are defined, those with
helium-poor spectra (sdBs) and those that are helium-rich (sdOs).

Formation scenarios of subdwarfs invoke either fine-tuned single star
evolution or rely on close-binary star interactions. In the \textit{late
hot-flasher scenario}, a low-mass star undergoes the He core-flash at the
tip of the red-giant branch. However, if sufficient mass is lost on
the red giant branch, the star can experience the He core-flash whilst
descending the white dwarf cooling track
\citep{castellani+castellani93-1}. Such a star would end up close to the
He main sequence (MS), at the very hot end of the extreme horizontal
branch \citep{dcruzetal96-1}.
Alternatively, the formation involves one or two phases of common-envelope
evolution and/or stable Roche-lobe overflow (RLOF) within a close
binary system \citep{mengeletal76-1}. Binary evolution could even take the
route of merging two helium white dwarfs followed by He ignition
\citep{webbink84-1,iben90-1,saio+jeffery00-1}. All formation scenarios
require substantial mass loss before the start of core He-burning, however
the specific physical mechanisms for this are still unclear. A detailed
review on this and the field as a whole is given by \citet{heber09-1}.

Since the first quantitative estimates of the contribution of different
binary channels to the population of subdwarf stars
\citep{tutukov+yungelson90-1}, it has been shown that a large fraction of
subdwarfs do reside in binaries. In the PG sample of subdwarfs, a
significant fraction show composite colours or spectra (at least
20\,per\,cent; \citealt{fergusonetal84-1}, $\sim54-66$\,per\,cent;
\citealt{allardetal94-2}). Radial velocity surveys
\citep[e.g.][]{maxtedetal01-1, morales-ruedaetal03-1} confirm the high
fraction of binaries with ratios as high as two-thirds. 
High-resolution optical spectra from the ESO Supernova Ia Progenitor
Survey \citep[SPY;][]{napiwotzkietal01-1}, led to binary star fractions of
30-40\,per\,cent \citep{napiwotzkietal04-1,liskeretal05-1}.
\citet{copperwheatetal11-1} estimate that the binary fraction in the sdB
population is somewhat higher at 46 - 56 per cent. This is only a
lower limit since the radial velocity variations that
\citet{copperwheatetal11-1} search for would be difficult to detect in
long period systems.

Other searches have used near-infrared photometry
\citep[e.g.][]{thejlletal95-1, ulla+thejll98-1, williamsetal01-1} or
photometric catalogues such as the Two Micron All Sky Survey
\citep[2MASS;][]{skrutskieetal06-1} to find subdwarfs with companions
\citep[e.g.][]{stark+wade03-1, greenetal06-1, vennesetal11-2}.
Ca\,\textsc{II} absorption can also be used to infer the
presence of a cooler companion star \citep{jeffery+pollacco98-1}.
The majority of companions found to date have either been M-type stars or
white dwarfs \citep{heber09-1}. However, some F, G and K-type companions
to subdwarfs have been seen in studies such as
\citet{cuadrado+jeffery01-1}, \citet{reed+stiening04-1},
\citet{liskeretal05-1}, \citet{wadeetal06-1}, \citet{stark+wade06-1},
\citet{wadeetal09-1}, \citet{monibidin+piotto10-1} and
\citet[MUCHFUSS;][]{geieretal11-1}. Depending on the study, and its
corresponding selection effects, the companions to subdwarfs have been
shown to be mostly main-sequence stars
\citep[e.g.][]{cuadrado+jeffery01-1} and giant or subgiant companions in
some cases (e.g. \citealt{allardetal94-2} and BD-$7^{\circ}5977$;
\citealt{heberetal02-1}).

Many of the previous surveys have been biased by selection effects and
inhomogeneous data sets. \citet{hanetal03-1} argued that a large number of
sdB stars may be missing from current samples. Early-type main--sequence
stars of spectral type A and earlier would outshine a subdwarf at optical
wavelengths. F to K-type companions on the other hand, have generally been
avoided because the spectral analysis of the composite spectrum becomes
difficult. Systems with earlier type companions are actually predicted, in 
some cases, to be far more common than the M-type companions that have
primarily been found so far. In the \citet{hanetal03-1} study, subdwarfs 
with early type companions are produced in the very efficient first stable 
RLOF channel and are expected to be in systems with subdwarfs as cool as
$15,000$\,K. \citet{clausenetal12-1}, however, do not find the same
multitude of F-type companions. Identifying this predicted population, and
determining their relative contribution to the subdwarf population  would
offer important constraints on the prior binary evolution that led to
their formation. In addition, the distribution of orbital periods and
subdwarf temperatures of such a sample will provide direct constraints on
key parameters that underpin subdwarf population synthesis models
\citep{clausenetal12-1}.

In this study, we take advantage of recent large-area ultraviolet,
optical and infrared photometric surveys to search for new composite
systems comprised of subdwarfs plus main--sequence star companions of
mid-M-type and earlier. Cuts in colour-colour space are employed to
separate these objects from possible contaminants. We also develop a
fitting technique to simultaneously determine the subdwarf and companion
effective temperatures from the photometric magnitudes. This permits the
recovery of composite systems with much earlier type companions than seen
in previous studies. Furthermore, we are sensitive to a wide range of
separations and binary periods in that we only limit ourselves to
spatially unresolved systems. Finally, we discuss the distribution of
objects in effective temperature and distance to the system.

\section{Synthetic models}
\label{s-mod}

To aid our search for subdwarfs with companions, we produced a grid of
synthetic sdB and main--sequence star spectra, which allowed us to
produce synthetic colours of the composite systems.

The sdB spectra were calculated using the model atmosphere code described
by \citet{heberetal00-1}, covering $\Teff=11,000-40,000$\,K in steps of
$1,000$\,K. The corresponding surface gravities were chosen to ensure that
our temperature sequence tracks the (extreme) horizontal-branch stars
\citep{dormanetal93-1}. This translates into $\Logg=4.0$ for
$\Teff=11,000-13,000$\,K objects, $\Logg=4.5$ for
$\Teff=14,000-16,000$\,K, $\Logg=5.0$ covering $\Teff=17,000-20,000$\,K,
$\Logg=5.5$ for $\Teff=21,000-28,000$\,K and $\Logg=6.0$ for
$\Teff=29,000-40,000$\,K. Surface gravity does not significantly affect
spectral slope, but does affect the width of line profiles, which is
a negligible feature when fitting photometry as we do here. It also
corresponds to a significant change in the size of the subdwarf and
therefore the relative brightness of the subdwarf and the companion.

A range of solar metalicity main--sequence star templates of effective
temperatures from $4,250$\,K to $25,000$\,K in 48 steps were taken from
the \citet{castelli+kurucz03-1} ATLAS9 model atmosphere library. For
models below $4,250$\,K, \citet{pickles98-1} stellar spectral library
models are substituted because of the problems with
\citet{castelli+kurucz03-1} model colours in this region
\citep{bertoneetal04-1}. A \citet{pickles98-1} M0V star is used as a
proxy for a $4,000$\,K model. Similarly, M1V, M2V, M3V and M5V replace
$3,750$\,K, $3,500$\,K, $3,250$\,K and $3,000$\,K models, respectively.
We restrict the models to unevolved main--sequence stars because, as we
will see in Section\,\ref{ss-cuts}, sub-giant and giant companions do not
contribute significantly to our sample. The impact of this will be
further discussed in Sections\,\ref{s-res} and \ref{s-dis}.
Both the sdB and main--sequence star spectra cover the wavelength range
$1,150-25,000$\,\AA. To normalise the \citet{castelli+kurucz03-1}
main--sequence star models to a flux at $10$\,pc, we rescale the models
to match the luminosities from the (zero age main sequence) isochrones of
\citet{girardietal00-1}.

The two grids of spectra were folded through all relevant filter
transmission curves to calculate absolute magnitudes. The two components
could therefore be added at a common distance. To separate composite
subdwarf plus companion systems from single subdwarfs and single
main--sequence stars in colour-colour space, a large wavelength range
must be sampled. The combination of a very blue and a red colour allows
for a significant contribution from both the subdwarf and companion
components to be seen in a colour-colour diagram. We therefore chose to
cross-match an ultraviolet survey with a series of optical and
near-infrared surveys.

\section{Cross-matching}

\subsection{Sample I: GALEX, CMC and 2MASS}
\label{s-cm1}

The Carlsberg Meridian Telescope (CMT) has a 2k by 2k CCD camera with a
Sloan $r$ filter operating in a drift scan mode. The CMT maps the sky from
La Palma (Spain) covering the declination range $-30^{\circ}$ to
$+50^{\circ}$ with a magnitude limit of $\rcmc=17$. The
Carlsberg Meridian Catalogue, Number 14 \citep[Version 1.0:
CMC][]{cmc14} is an astrometric and photometric catalogue of 95.9
million stars covering $9<\rcmc<17$. We cross-matched the CMC catalogue
with 2MASS using a $2\arcsec$ matching radius. Because the surveys used
here avoid the Galactic plane, the contamination from matching to other
stars within $2\arcsec$ will be relatively small \citep{girvenetal11-1}.
With these combined catalogues, we were able to calculate an $(\rcmc-J)$
colour as a diagnostic for spectral type, as well as $(J-K)$, indicative
of  strong companion star contributions in composite systems (see
Figure\,\ref{f-rrK}).

Based upon the $(\rcmc-J)$ colour of the composite models described in
Section\,\ref{s-mod}, the CMC sample was cut to include only stars bluer
than a G0V star ($5750$\,K on the \citealt{castelli+kurucz03-1} grid),
i.e. $(\rcmc-J)<0.9$. The cut includes all possible combinations of
subdwarf plus companion, but removes a significant fraction of
contaminants.
This does not limit our selection of subdwarfs with companions as
discussed in Section\,\ref{s-uve}. The sample was also limited to
$\rcmc<16.0$, primarily to match the magnitude limit of 2MASS ($K_s
\simeq 14.3$). This resulted in $\sim1.9$ million objects.

All objects within the $(\rcmc-J)$ colour cut were cross-matched with the
Galaxy Evolution Explorer (GALEX) all-sky ultraviolet survey
\citep{martinetal05-1} Data\,Release\,6. This provides magnitudes in two
bandpasses, \fuv\ and \nuv, centered around 1500 and
2300\AA, respectively. The matching was performed using the predefined
cross-matching tables in GALEX CasJobs \citep{budavarietal09-1} searching
for all sources within $2\arcsec$. The resulting catalogue of neighbours
contains approximately $560,000$ matched objects and hereafter will be
referred to as the ``\textit{C2M}'' sample. The mean of any multiple
GALEX observations was taken where available and both bands were
corrected for non-linearity according to \citet{morrisseyetal07-1}.

Finally, the objects from the match between CMC, 2MASS and GALEX were further
cross-matched with the Sloan Digital Sky Survey (SDSS) Data Release 7
\citep[DR7;][]{abazajianetal09-1}. This sample will hereafter be referred
to as the ``\textit{C2MS}'' sample which is smaller and
photometrically deeper. The SDSS CasJobs predefined cross-matching tables
\citep{li+thakar08-1} were utilised. Objects were limited to have good
quality photometric magnitudes (see Table\,\ref{t-sel}). This resulted in
a sample of $\sim105,000$ objects for which good SDSS $u$, $g$, $r$, $i$
and $z$ magnitudes were available along with GALEX, CMC and 2MASS
photometry.
For $\sim1.5$\,per\,cent of objects within this sample, SDSS optical
spectra are available.

\begin{table*}
\caption{\label{t-sel} Colour selection for finding subdwarfs with
companions, for both the \textit{C2M} and \textit{C2MS} (with or
without SDSS magnitudes) and \textit{SU} samples. Constraints with a
``Sample'' flag were only applied to that sample, whereas constraints
with no flag were applied to both samples. ``bad\_flags'' is defined as
saturated or bright or edge or nodeblend and ``nChild'' is the number of
children objects detected by SDSS.}
\begin{tabular}{lcll}
\hline
Colour & Constraint &  & Sample \\

\hline
$\rcmc$     & $\le$ & $16.0$ & \textit{C2M} \\
$(\rcmc-J)$ & $<$   & $0.9$  & \\

$\rcmc$ Uncertainty & $\le$ & $0.10$ & \\
$\ebv$              & $\le$ & $0.15$ & \\
\hline

$(\fuv-\rcmc)$     & $\le$ & $3.8 * (\rcmc-K_s) - 0.3$  & \\
$(\fuv-\rcmc)$     & $\le$ & $-2.7 * (\rcmc-K_s) + 4.7$ & \\
$(\fuv-\rcmc)$     & $\ge$ & $-3.0$ & \\
FUV artifact flag  & $\le$ & $1$    & \\
$\fuv$ Uncertainty & $\le$ & $0.05$ & \textit{C2M} \\
                   & $\le$ & $0.10$ & \textit{SU} \\

 & AND & \\

$(\nuv-\rcmc)$     & $\le$ & $1.3 * (\rcmc-K_s) + 0.54$  & \\
$(\nuv-\rcmc)$     & $\le$ & $-1.45 * (\rcmc-K_s) + 3.3$ & \\
$(\nuv-\rcmc)$     & $\le$ & $3.5 * (\rcmc-K_s) + 0.12$  & \\
$(\nuv-\rcmc)$     & $\ge$ & $-2.0$ & \\
NUV artifact flag  & $\le$ & $1$    & \\
$(\rcmc-K_s)$       & $\le$ & $1.75$ & \textit{C2M} \\
$\nuv$ Uncertainty & $\le$ & $0.05$ & \textit{C2M} \\
$(\rsdss-K)$       & $\le$ & $1.5$  & \textit{SU} \\
$\nuv$ Uncertainty & $\le$ & $0.10$ & \textit{SU} \\
\hline

SDSS specific: & & & \\
flags \& bad\_flags & $=$ & $0$ & \\
nChild              & $=$ & $0$ & \\
\hline

\end{tabular}
\end{table*}

\subsection{Sample II: GALEX, SDSS and UKIDSS}
\label{s-cm2}

The GALEX, CMC and 2MASS cross-matched sample discussed above benefits
from covering a large area (limited by the GALEX footprint), but is
relatively shallow with a limiting magnitude of $r=16.0$ and
$K_{s}=14.3$. This restricts our ability to construct volume-limited
samples.

A second, complimentary sample was selected from GALEX, SDSS and UKIDSS.
One of the five UKIDSS sub-surveys, the Large Area Survey (LAS), aims to
be the infrared counterpart to the SDSS. UKIDSS LAS will eventually
provide imaging over $4028\,\deg^2$ in four broad band colours, $Y$, $J$,
$H$, and $K$, with limiting (Vega) magnitudes of 20.2, 19.6, 18.8 and
18.2, respectively. This adds a significant increase in depth over 2MASS.
Here, we made use of UKIDSS Data\,Release\,9 (see \citealt{dyeetal06-1}),
which covers $\sim60$\,per\,cent of the total, planned, LAS area. SDSS
and UKIDSS were cross-matched to find the closest
match within $2\arcsec$ using the UKIDSS-SDSS pre-match tables. This
sample was then matched to GALEX within $2\arcsec$, using the CasJobs
neighbours search, returning approximately $120,000$ objects. Again,
multiple GALEX neighbours were combined into a single measurement and
fluxes were corrected for non-linearity \citep{morrisseyetal07-1}. This
sample will hereafter be referred to as the ``\textit{SU}'' sample. It is
limited in area by the current size of UKIDSS, but extends several
magnitudes deeper than 2MASS in $K$. Because the UKIDSS LAS area is
entirely encompassed by the SDSS footprint, we can make use of the
higher precision, deeper SDSS photometry, rather than CMC.
The number of objects at each stage of the analysis is given in
Table\,\ref{t-nums}.

\begin{table*}
 \caption{\label{t-nums} Summary of numbers at each stage of the
processing. The left hand columns shows which surveys were included at
that stage in the processing. The ``In cuts'' columns are those objects
satisfying the criteria from Figure\,\ref{f-rrK} and Table\,\ref{t-sel}
and the ``$\Sigma$'' column displays the total number of objects in this
catagory.}
 \begin{tabular}{lllllcccc}
  \hline \hline
  Sample & \multicolumn{4}{l}{Surveys} & Total & \multicolumn{3}{c}{In
      cuts} \\
  Name & Ultraviolet & \multicolumn{2}{c}{Optical} & Infrared & (approx)
      & $\Sigma$ & SDSS Spectra & SIMBAD \\ \hline
  & & & CMC & 2MASS & $1,900,000$ & - & - & -  \\
  \textit{C2M} & GALEX & & CMC & 2MASS & $560,000$ & 449 & - & 58 \\
  \textit{C2MS} & GALEX & SDSS & CMC & 2MASS & $105,000$ & 93 & 25 & 24
      \\
      \hline
  & & SDSS & & UKIDSS & $220,000$ & - & - & - \\
  \textit{SU} & GALEX & SDSS & & UKIDSS & $120,000$ & 134 & 72 & 47 \\
  \hline
 \end{tabular}
\end{table*}

\section{Selecting ultraviolet excess objects}
\label{s-uve}

\subsection{Colour-colour diagrams}
\label{ss-cc}

\begin{figure*}
 \begin{minipage}{2\columnwidth}
  \includegraphics[width=0.5\columnwidth]{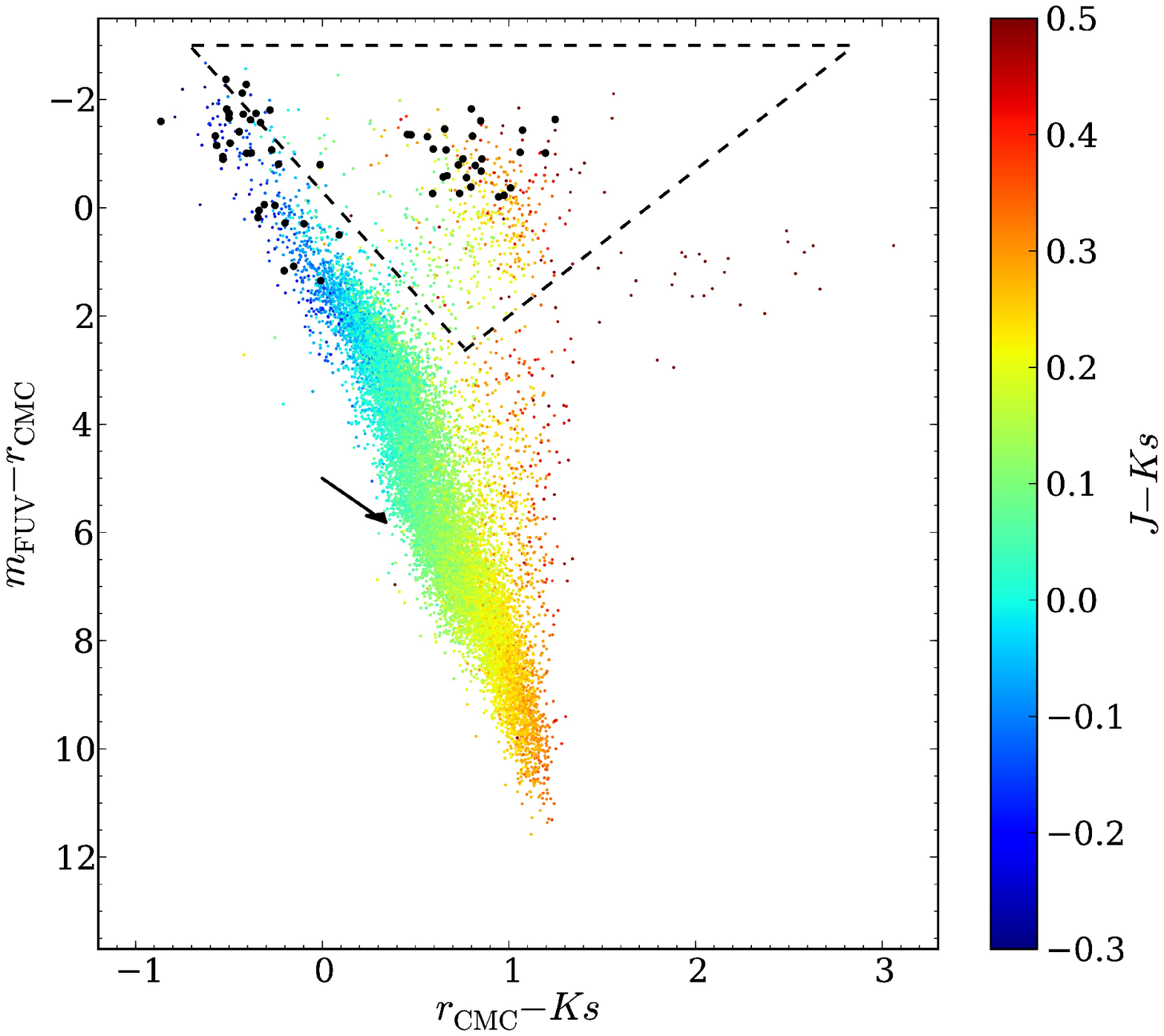}
  \includegraphics[width=0.5\columnwidth]{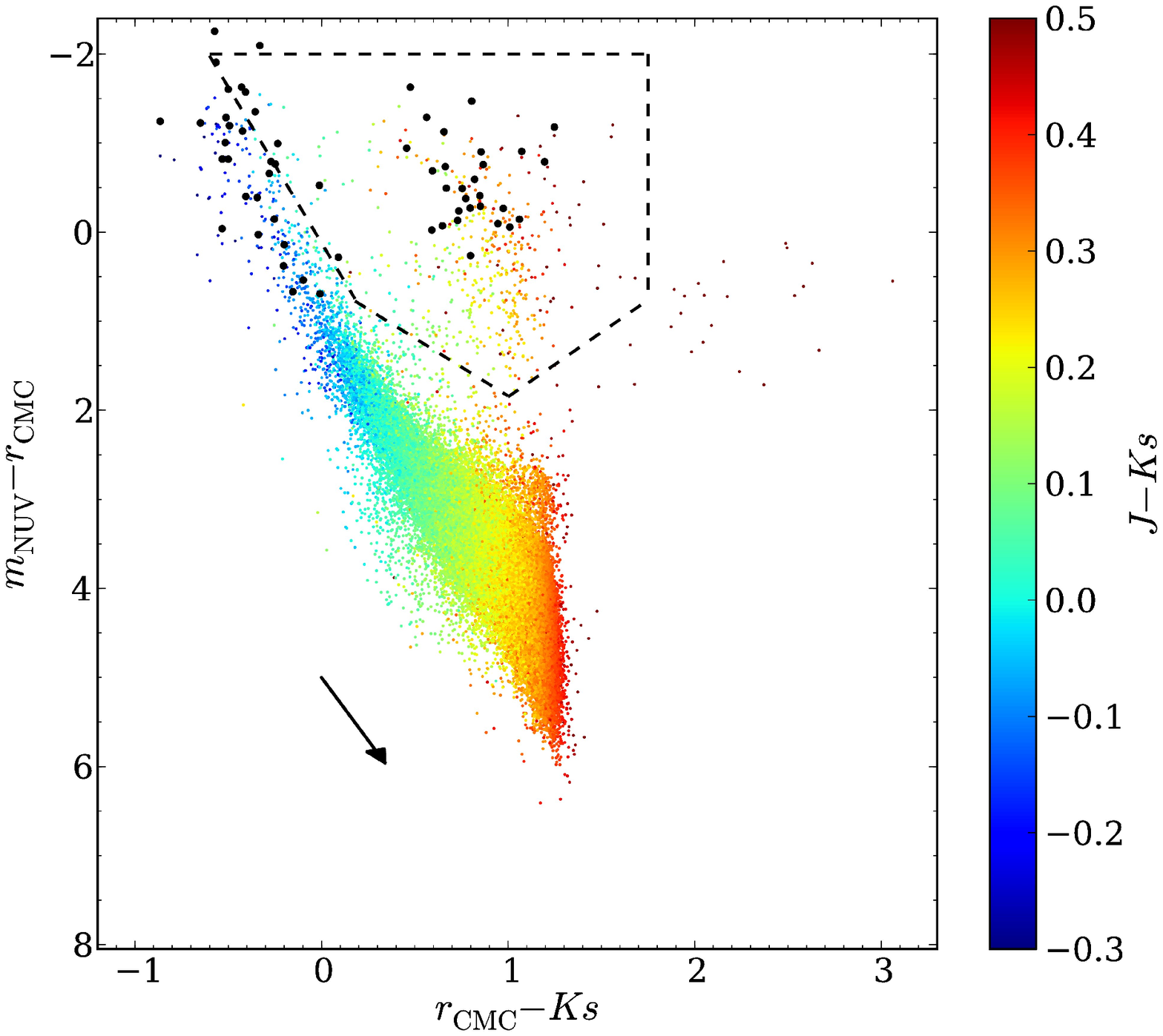}
  \caption{\label{f-rrK} Colour-colour diagrams of \frrK\ (left;
$\sim15,000$ objects) and \nrrK\ (right; $\sim105,000$ objects) for
objects with matches in GALEX, CMC and 2MASS catalogues (the
\textit{C2M} sample) and satisfying the magnitude cuts in
Table\,\ref{t-sel}. The majority of objects are main--sequence stars and
therefore do not have \fuv\ magnitudes. This causes the factor of 10
difference in numbers of objects in the two diagrams.
\ebv, taken from the GALEX catalogue \citet{schlegeletal98-1} map value,
is also limited to $<0.15$ and for clarity, $\nuv \leq 16.5$. The plotted
colour for each point corresponds to the $(J-K_s)$ value, where colours
above 0.5 or below -0.3 are shown as 0.5 and -0.3, respectively. For
single stars, this range corresponds to spectral types O5 to K0. The
sdB/sdO star candidates from \citet{kilkennyetal88-1} with matches in the
\textit{C2M} catalogue are shown as black circles (totalling 84 objects).
A reddening vector corresponding to $\ebv=0.15$ is shown as a black arrow
centered on $(0,5)$ in both diagrams. The dashed black lines show the
colour-colour selections as discussed in Section\,\ref{ss-cuts} and
Table\,\ref{t-sel}.}
 \end{minipage}
\end{figure*}

Figure\,\ref{f-rrK} shows colour-colour diagrams for the objects with
detections in GALEX, CMC and 2MASS. We compare \frrK\ and \nrrK, where
the $(J-K_s)$ colour of each object is colour-encoded in the plot. For
single stars, the $(J-K_s)$ range corresponds to spectral types O5 to K0.
The colour indices are tailored to highlight in colour-colour space the
position of composite blue plus red objects. The $(\rcmc-K_s)$
colour of an object is a relatively good indication of stellar spectral
type and $(\fuv-\rcmc)$ will indicate objects with an excess in the
ultraviolet in contrast to single main--sequence stars. The truncation at
$(\rcmc-K_s)\sim1.5$ is caused by our imposed cut of $(\rcmc-J)<0.9$.

\begin{figure*}
 \begin{minipage}{2\columnwidth}
  \includegraphics[width=0.5\columnwidth]{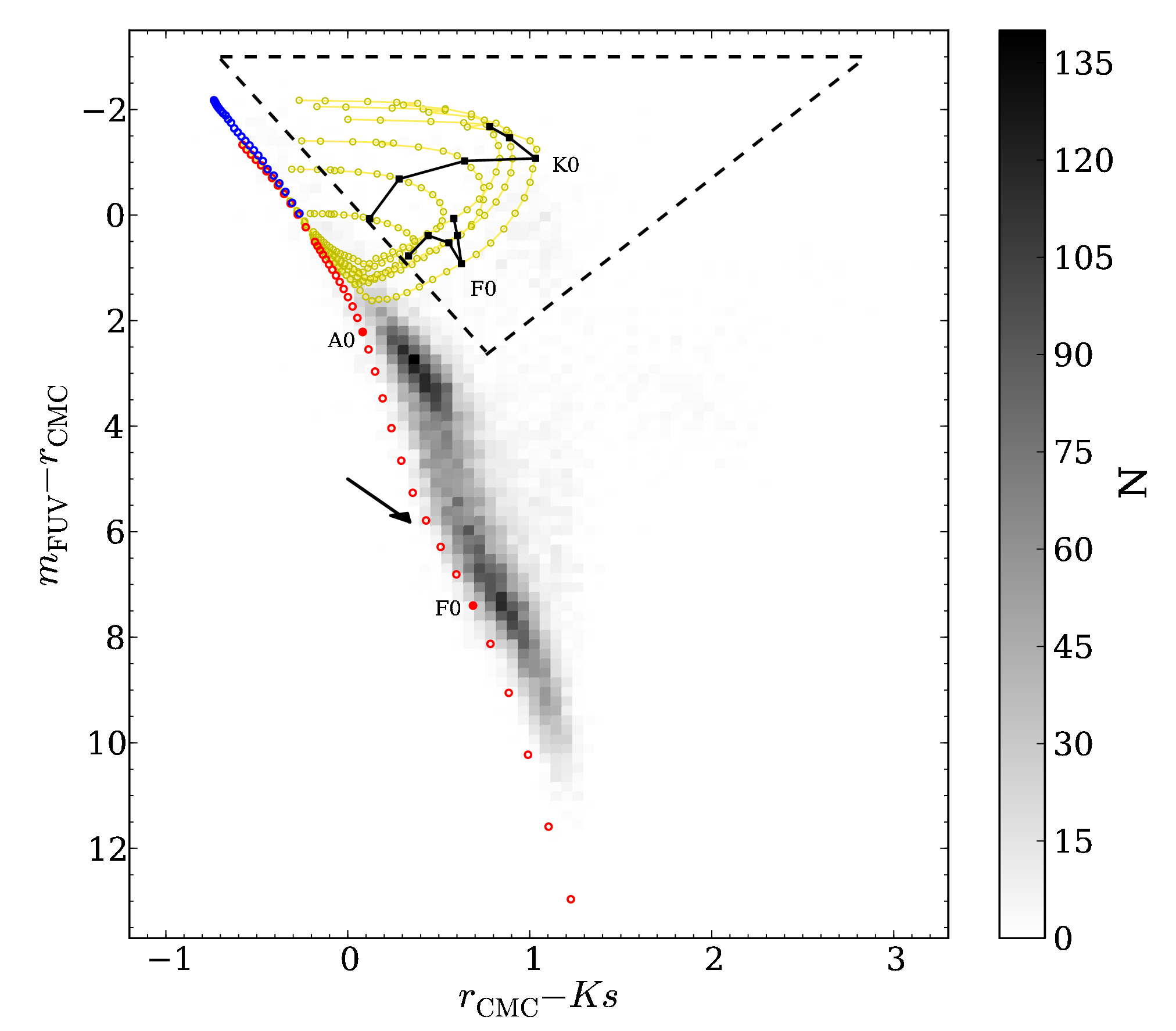}
  \includegraphics[width=0.5\columnwidth]{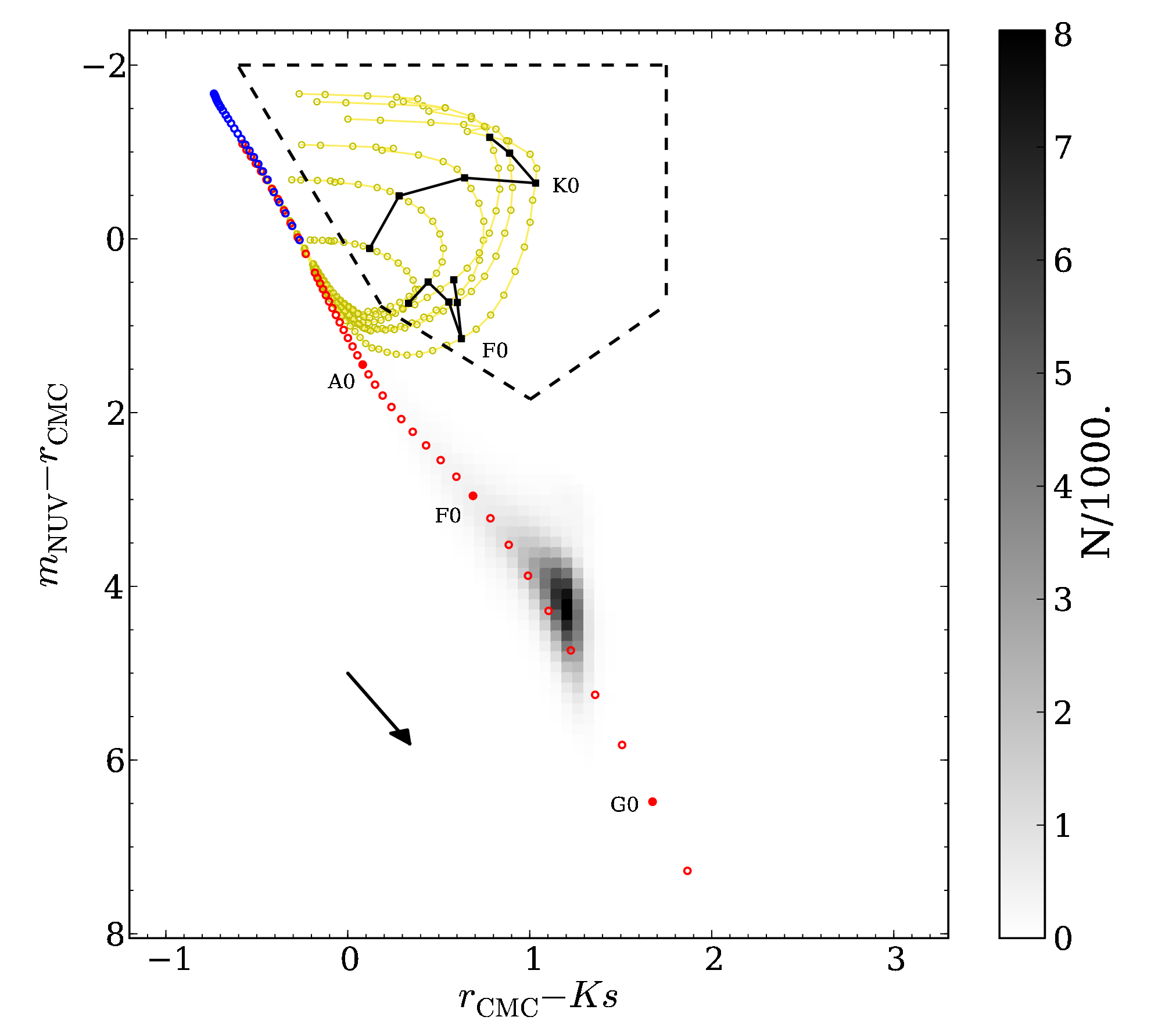}
  \caption{\label{f-hist} 2D density plots of the \frrK\ and \nrrK\
colour-colour diagrams in Figure\,\ref{f-rrK}, however, \nuv\ is no
longer limited to $16.5$. The grey scale is shown on the respective
colour bars, labelled by the number of objects (N) on the left and N/1000
on the right. The reddening vector again corresponds to $\ebv=0.15$ and
is shown as a black arrow centered on $(0,5)$. The dashed black lines
show the colour-colour selections as discussed in Table\,\ref{t-sel}.
Additionally, subdwarf ($15,000\leq\Teff\leq40,000$\,K) and
main--sequence star model colours (Section\,\ref{s-mod}) are plotted as
blue and red open circles, respectively. The yellow lines show a single
subdwarf ($15,000 \leq \Teff \leq 40,000$\,K in steps of $5,000$\,K)
paired with the sequence of main--sequence stars described in
Section\,\ref{s-mod}. As a reference, subdwarfs (of the above
temperatures) with approximately a K0 or F0-type companion are joined
with a black line.}
 \end{minipage}
\end{figure*}

In Figure\,\ref{f-hist}, the same sources are plotted but now encoding
the density of sources on a grey scale to better represent relative
numbers. The main sequence is found along the bottom edge of the main
group of objects in the \frrK\ plane, and more centrally through the main
group in the \nrrK\ plane (Figure\,\ref{f-rrK}). Simulated colours
derived from our main--sequence star model (Section\,\ref{s-mod}) confirm
that this is the expected position of the main sequence in our chosen
colours. Similarly, composite sdB plus companion star models are also
shown in Figure\,\ref{f-hist}, highlighting the region of colour-colour
space where we expect to find such systems.

The large scatter in $(\fuv-\rcmc)$ or $(\nuv-\rcmc)$ for a given
$(\rcmc-K_s)$, especially at the red end, can be explained due to a few
factors. First of all, even though we formally require the \rcmc\
uncertainty to be less than $0.1$, there appears to be additional
systematic scatter in the \rcmc\ magnitudes. Investigating the vertically
extended regions in our colour-colour diagrams (Figure\,\ref{f-rrK}) when
using the much more reliable \rsdss\ instead of \rcmc, we find that the
spread is significantly reduced. However, the larger sky coverage of the
CMC is far more important for our study especially as the subdwarf plus
companion systems fall in a relatively clean part of the diagram. Another
reason for the observed spread is the fact that the GALEX magnitudes have
been shown to suffer from non-linearities for bright stars, amongst other
problems \citep[e.g.][]{morrisseyetal07-1, wadeetal09-1}. Although we
corrected for non-linearity using the method described in
\citet{morrisseyetal07-1}, the equations are empirical and there may be a
significant scatter in individual measurements.

In addition, despite limiting $\ebv\leq0.15$, much of the spread around
the main sequence can be accounted for by considering the effects of
interstellar reddening. The \ebv\ magnitude for each object is taken from
the GALEX catalogue, which is itself calculated from the Galactic
reddening maps of \citet{schlegeletal98-1}. The interstellar reddening is
illustrated by the reddening vectors in Figure\,\ref{f-rrK}. These are
calculated by folding the mean extinction curve of
\citet{fitzpatrick+massa07-1} through the relevant filter transmission
curves. In the \frrK\ plane, reddening of blue objects moves them above
the main sequence in $(\fuv-\rcmc)$, a region populated by a number of
objects. However, reddening in the \nrrK\ plane approximately moves
objects along the main sequence. The components of the reddening vectors
are approximately the same in both $(\fuv-\rcmc)$ and $(\nuv-\rcmc)$
because of the $2200$\AA\ bump in the reddening function
\citep{papoular+papoular09-1} coincides with the central wavelength of
\nuv.
However, the intrinsic and significant variations in the reddening law
along different lines of sight affect the ultraviolet magnitudes more so
than the optical values. \citet{fitzpatrick+massa07-1} show that even
when considering the standard stars that are used to calculate the
adopted reddening function, a significant spread around the mean
extinction curve is observed. This leads to large departures from the
mean law, affecting the ultraviolet region in particular. These
variations in the extinction curve, along with the variation of the true
reddening to the subdwarf compared with that calculated in the
\citet{schlegeletal98-1} maps, are thus likely responsible for the
stellar sources populating a vertically extended region in the \nrrK\
plane. In any case, the outliers form only a small fraction of the total
source population and the reddening vector does not move main--sequence
stars into the colour selections we discuss below.

\subsection{Isolating subdwarfs in binaries}
\label{ss-cuts}

In order to classify our sources and check for known objects within our
sample, we resolved all positions using
SIMBAD\footnote{http://simbad.u-strasbg.fr/simbad/}, and also consulted
any available SDSS optical spectra. In the upper-left corner of the
\frrK\ colour-colour diagram, one would expect to find white dwarfs and
single-star subdwarfs, which is corroborated by classifications in the
SIMBAD database. 
Unfortunately, none of our sources with colours consistent with single
subdwarfs have SDSS spectra that could conclusively confirm their
classification (due to them saturating in SDSS). The objects towards the
right of the diagram, with $(\rcmc-K)\sim2.0$, prove to be galaxies.
These are removed by use of the point source flag in SDSS.

\citet{kilkennyetal88-1} created a catalogue of subdwarf stars and
candidates from previous studies, including work on the PG survey. This
includes subdwarfs both with and without companions. We matched this
catalogue to the \textit{C2M} catalogue, resulting in 1704 objects. The
subset for which appropriate quality limits are satisfied are plotted in
Figure\,\ref{f-rrK} (84 sources). We see that this sample splits into two
distinct groups. A significant fraction falls in the region where single
subdwarfs and white dwarfs are expected to lie. However, a good number of
these ($\sim35$\,per\,cent) lie at a much redder $(\rcmc-K_s)$ colour,
where, from the synthetic magnitudes calculated in Section\,\ref{s-mod},
we expect subdwarfs with main--sequence star companions. The objects in
this redder region (inside the black dashed lines in Figures \ref{f-rrK}
and \ref{f-hist}), would appear to be main--sequence F or G-type stars
from their $(\rcmc-K_s)$ colour, but have an ultraviolet excess in
$(\fuv-\rcmc)$ and/or $(\nuv-\rcmc)$ colour. This confirms that a
significant fraction of the \citet{kilkennyetal88-1} sample show
photometric evidence for being composite, but also that we have detected
a large number of new sources within that same region of colour space.

For the new \textit{C2M} objects in this region, where SDSS spectra are
available, they can be seen to be mostly subdwarfs along with one white
dwarf and two cataclysmic variable stars (CV: see Table\,\ref{t-box}).
SIMBAD, however, only returns four known subdwarfs in this region of
colour-colour space. This may be expected as previous work has
intentionally focused on single-lined sdB systems that are therefore
dominated by the subdwarf. The number of objects grouped under a few
broad classifications are summarised in Table\,\ref{t-box}. Note that
close to $90$\,per\,cent of the \textit{C2M} sources (without SDSS)
within this region are unknown.

In order to isolate composite subdwarfs while avoiding obvious
contaminants, we devised cuts in colour-colour space (Table\,\ref{t-sel})
guided by our simulated composite subdwarf colours and the SIMBAD and
SDSS spectroscopic classifications discussed above.
The right hand side of the cuts was chosen to avoid contamination from
galaxies and quasars, and similarly on the lower side the main sequence
was avoided. At the left hand edge, the cuts were chosen to avoid
early-type stars and single subdwarfs. We require objects to be in both
the \frrK\ and \nrrK\ cuts because objects residing in just an individual
box are likely to arise from spurious \textit{GALEX} fluxes.
Contamination of this region due to interstellar reddening is small
because very few objects will be moved from the main sequence, along the
reddening vector, into the box, as shown in Figure\,\ref{f-rrK}.
Similarly, the scatter from a poor \rcmc\ magnitude does not lead to a
large contamination, because the subdwarfs with companions region is
sufficiently far from the main sequence.

\begin{table*}
 \caption{\label{t-box} Table of classifications for the 449, 93 and 134
objects in the \textit{C2MS}, \textit{C2MS} and \textit{SU}
samples, respectively, and inside the colour-colour selection boxes from
Figure\,\ref{f-rrK} and Table\,\ref{t-sel}. The SDSS spectra column is
from visual inspection of the optical spectra. The number of galaxies
seen in the SDSS spectra is virtually zero because the flags used to
select the SDSS objects remove any extended objects.}
 \begin{tabular}{lccccc}
  \hline \hline
                 & \textit{C2M} &
                   \multicolumn{2}{c}{\textit{C2MS}} &
                   \multicolumn{2}{c}{\textit{SU}} \\
  Classification & SIMBAD & SIMBAD & SDSS spectra & SIMBAD & SDSS spectra
                   \\ \hline
  SD        & 7   & 4  & 22 & 7  & 62  \\
  Composite & 9   & 1  & 0  & 4  & 0   \\
  CV/Nova   & 21  & 8  & 2  & 10 & 4   \\
  Galaxy    & 2   & 0  & 0  & 0  & 2   \\
  Quasar    & 0   & 0  & 0  & 0  & 0   \\
  WD        & 19  & 11 & 1  & 26 & 4   \\
  \hline
  Total with classification    & 58  & 24 & 25 & 47 & 72 \\
  Total without classification & 391 & 69 & 68 & 87 & 62  \\
  \hline
 \end{tabular}
\end{table*}

We repeated a similar selection using the \textit{SU} sample, again using
\UKfrrK\ and \UKnrrK\ colour-colour diagrams (not shown). All magnitudes
were limited to have uncertainties less than 0.1 mag and $\ebv\leq0.15$.
An increase in the number of quasars was seen, which encroached on the
cuts used for \nrrK. The upper limit on $(\rsdss-K)$ was therefore
reduced, as shown in Table\,\ref{t-sel}, however the contamination was
not completely removed. The cuts on \UKfrrK\ remained unchanged, where we
ignore the small differences between UKIDSS $K$ magnitude versus 2MASS
$K_s$ magnitudes\footnote{Assuming a $J-K_s$ colour of $\sim0.3$ and
using the transformations of \citet{carpenter01-1}, the difference
between the $K_s$ and $K$ magnitude is $\sim0.003$ and therefore
negligible.}.
After these adjustments, 134 objects reside within the cuts, 72 of which
have SDSS spectra. This is significantly more than the \textit{C2MS}
sample because many of the \textit{C2MS} objects are saturated in SDSS.
As for the 2MASS sample, we provide broad classifications for the
\textit{SU} sample in Table\,\ref{t-box}.

With our selection cuts in place, we can use the tracks of our synthetic
subdwarf-companion pairs to consider the completeness of our
composite subdwarf sample. We find that our region covers only a limited
range in companion type for a given subdwarf temperature, as systems that
are either dominated by the companion or the subdwarf fall outside our
region. This choice is required to reduce contamination from single
stars.
Based on our simulated colours, we find that subdwarfs with temperatures
up to $30,000$\,K would fall in the \frrK\ colour cut for even the
coolest main--sequence companion in our grid ($3,000$\,K:
$\sim$M5). $35,000$\,K and $40,000$\,K subdwarfs, however, would require
$\gtrsim3,750$\,K ($\lesssim$M0) and $\gtrsim5,000$\,K ($\lesssim$\,K0)
companions, respectively, to make them stand out from the main sequence
populations. In the case of early-type companions, subdwarfs plus O-type
and B-type stars are also lost as they merge back into the blue end of
the main sequence. A $15,000$\,K, $20,000$\,K, $30,000$\,K and
$40,000$\,K subdwarf would be identified if it had an $\lesssim7,500$\,K
($\gtrsim$F0), $\lesssim8,250$\,K ($\gtrsim$A5), $\lesssim8,000$\,K
($\gtrsim$A5) or $\lesssim8,500$\,K ($\gtrsim$A5) companion,
respectively.

\begin{figure*}
 \begin{minipage}{2\columnwidth}
  \begin{center}
   \includegraphics[width=0.8\columnwidth]{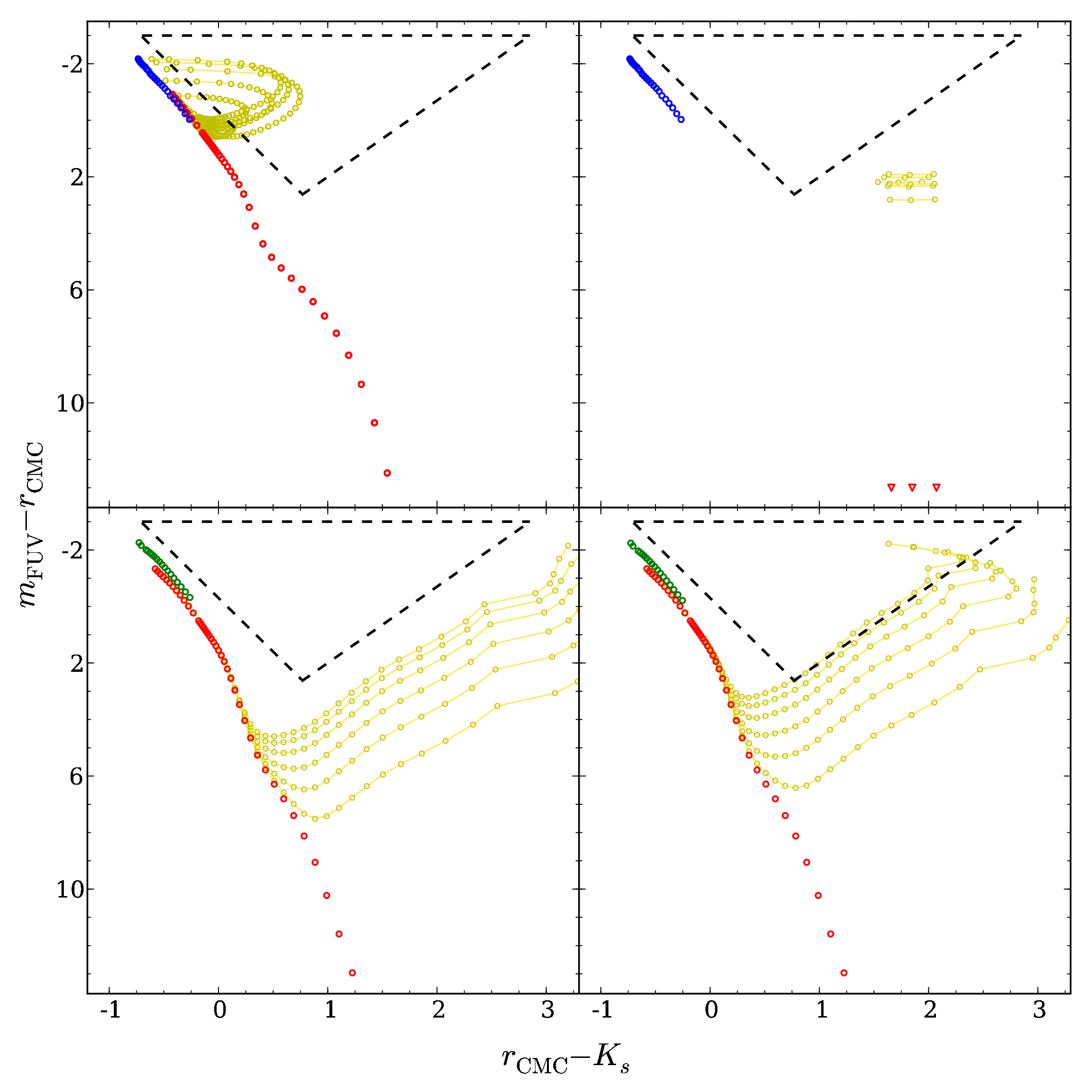}
   \caption{\label{f-model} Potential contaminants of the subdwarf plus
main--sequence star sample, following the same format as
Figure\,\ref{f-hist}. See Section\,\ref{ss-cuts} for a discussion.
Top-left: Subdwarfs (blue open circles: described in
Section\,\ref{s-mod}) with metal-poor ($\log ([\mathrm{M/H}] /
[\mathrm{M/H}]_{\mathrm{solar}}) = 1.5$: red open circles) main--sequence
star companions. Composite objects are shown in yellow. Top-right:
Subdwarfs with giant star companions (ranging in spectral type from
approximately G7 to K3). Single G and K-type stars do not fall in the
range of the Figure, and therefore we mark their $(\rcmc-K_s)$ position
by downward pointing red triangles. Bottom-left: $\log(g)=8$ DA white
dwarfs (green open circles) with main--sequence companion stars
(described in Section\,\ref{s-mod}). Bottom-right: $\log(g)=7$, and
therefore larger radii, DA white dwarfs with main--sequence companion
stars. The DA white dwarf model grid was kindly provided by D. Koester
\citep[for a description, see][]{koester10-1}, and ranges from $15,000
\leq \Teff \leq 40,000$\,K in steps of $5,000$\,K.}
  \end{center}
 \end{minipage}
\end{figure*}

For the colour-colour tracks, the companions are restricted to be
main--sequence stars. However, we may also expect to find a population of
subdwarfs with sub-giant or giant companions similar to HD\,185510
\citep{fekel+simon85-1}, HD\,128220 \citep{howarth+heber90-1} and
BD-$7^{\circ}5977$ \citep{vitonetal91-1, heberetal02-1}. In fact, the
binary population synthesis of \citet{hanetal03-1} predicted that the
majority of K-type companions to subdwarfs should be evolved companions.
We calculated the \frrK\ location of G7 to K3-type giant stars
(Figure\,\ref{f-model}: upper-right panel) by taking the solar
metalicity, zero age horizontal branch stars from the
\citet{castelli+kurucz03-1} model atmosphere library and again rescaling
the fluxes to a corresponding zero age horizontal branch luminosity from
the isochrones of \citet{girardietal00-1}. All combinations of subdwarf
plus giant star systems fall outside of the colour cuts described in
Table\,\ref{t-box}.
Systems with either overluminous subdwarfs, or companions in an
intermediate state between the main-sequence and the
horizontal branch may, however, fall within the colour cuts. We do not
expect these to be a significant population in our sample.
Since we do not expect specific formation mechanisms to become more or
less prevalent as a function of distance, we can still use our sample to
study the spatial distribution of subdwarfs even if the sub-sample of
subdwarfs with evolved companions is selected against.

We may also expect that some detached white dwarf plus main--sequence
type companion systems are found to be contaminants of the sample, since
these are composite systems with a hot component and a cooler companion.
However, we simulated the colours of such systems and, with the
exception of very low gravity white dwarfs, they do not fall in the
colour-colour region selected in Table\,\ref{t-box} (see
Figure\,\ref{f-model}: lower two panels). In this colour space, the small
radius of the white dwarf means that the flux is dominated by all but the
latest of main sequence companions and so they lie closer to the main
sequence in both diagrams. They are thus unlikely to constitute a
significant contaminant.

Looking out of the Galactic plane to distances of over $1$\,kpc, we may
expect to see a sizeable fraction of thick disk and halo stars.
Therefore, the companions to the subdwarfs in our samples may be
metal-poor. In Figure\,\ref{f-model} (upper-left panel), we show that
subdwarfs with metal-poor ($\log ([\mathrm{M/H}] /
[\mathrm{M/H}]_{\mathrm{solar}}) = 1.5$: ATLAS9:
\citealt{castelli+kurucz03-1}) companions indeed still fall in
our colour selection. We discuss the associated possible biases on our
fitting technique in Section\,\ref{s-fit}.

A summary of our sample sizes at various stages of the analysis can be
found in Table\,\ref{t-nums}. The full list of 449 objects inside our
\textit{C2M} sample can be found in Table\,\ref{t-full_2m}.

\section{Spectroscopic Observations}
\label{s-obs}

We discuss here some spectroscopic follow-up obtained to verify that
the \textit{C2M} sample objects likely contain a subdwarf component
before turning to the modelling of their spectral energy distributions
(SED) in Section \ref{s-fit}.

\subsection{WHT}

Nine objects falling within the colour-colour cuts described in
Table\,\ref{t-sel} were observed in July and December 2010, using the
4.2m William Herschel Telescope (WHT) at the Roque de los Muchachos
Observatory, La Palma, Spain. We used the ISIS dual-beam spectrograph
mounted at the Cassegrain focus of the telescope, with a R600 grating on
both the blue and the red arms, and a 1\arcsec\, slit. The blue arm of
the spectrograph is equipped with a $2048\times4096$ pixel EEV12 CCD,
which we binned by factors of 3 (spatial direction) and 2 (spectral
direction). The $2048\times4096$ pixel REDPLUS CCD on the red arm was
binned similarly. This setup delivers a wavelength coverage of $3772 -
5136$\AA\, on the blue arm, with an average dispersion of 0.88\AA\, per
binned pixel, and $5983 - 7417$\AA\, on the red arm, with an average
dispersion of 0.98\AA\, per binned pixel. We determined the resolution to
be 1.2\AA, from measurements of the full width at half maximum of
night-sky lines. The setup during the December observations was
identical, except that the CCDs were binned $2\times2$.

The spectra were debiased and flatfielded using the {\sc
starlink}\footnote{Maintained and developed by the Joint Astronomy Centre
and available from http://starlink.jach.hawaii.edu/starlink} packages
{\sc kappa} and {\sc figaro} and then optimally extracted using the {\sc
pamela} code \citep{marsh89-1}. We derive the wavelength calibration from
Copper-Neon and Copper-Argon arc lamp exposures taken during the night,
selecting the arc lamp exposure nearest in time to each science spectrum.

Finally, the raw spectra were converted to flux units and the telluric
absorption lines removed. For the July run, the flux calibration was done
using a model spectrum of a ``flux standard'' DA white dwarf, observed on
the same night.
The December run suffered from poor weather and no flux standard was
observed. We calibrated these two spectra using an earlier observation of
SP1446+259, taken with the same instrumental setup. The shape of the
spectrum is therefore reliable, but the absolute flux level is not. Our
analysis does not depend on the absolute flux of the targets, so our
conclusions are unaffected.

\begin{figure*}
 \begin{minipage}{2\columnwidth}
  \includegraphics[angle=270, width=\columnwidth]{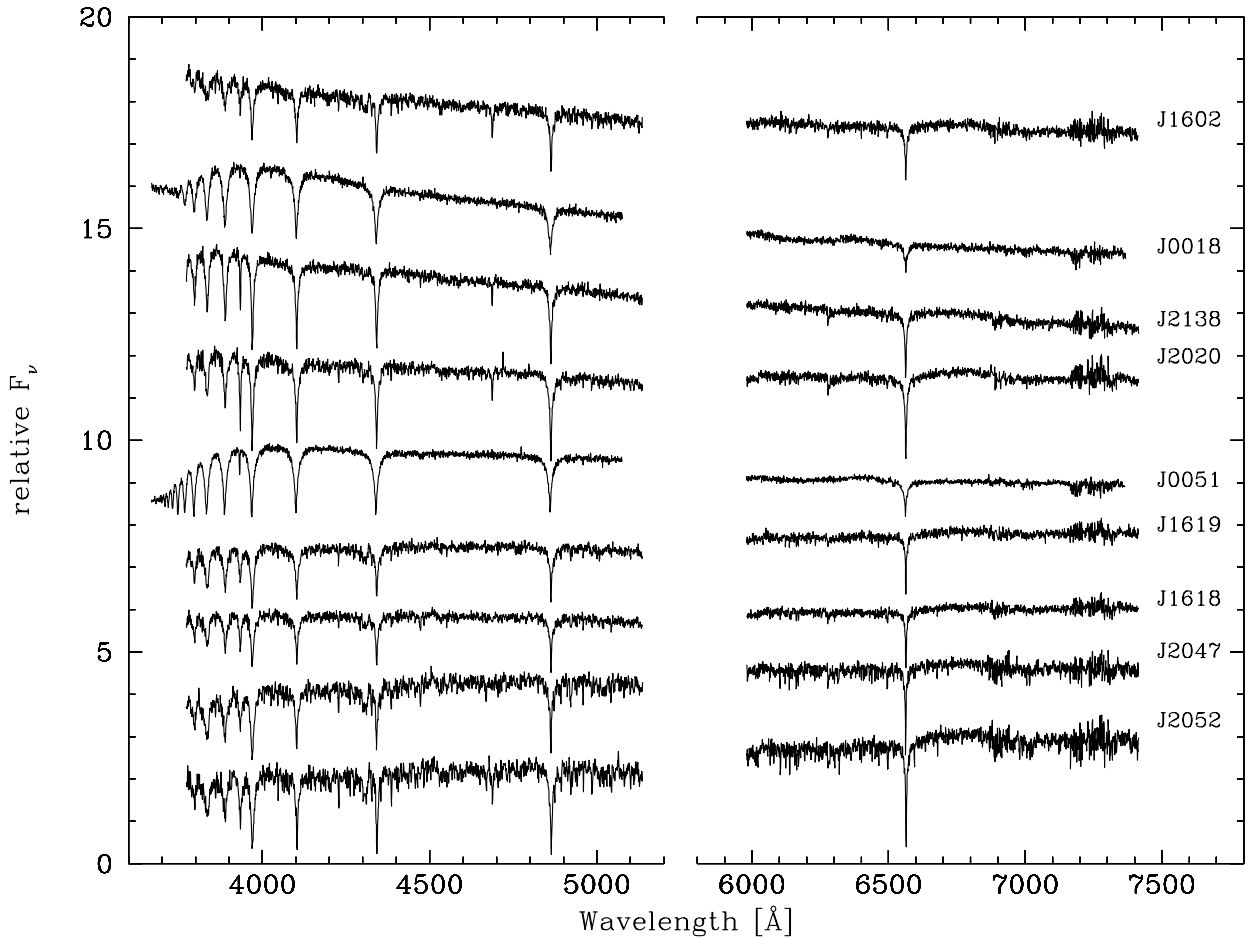}
  \caption{\label{f-sdb} WHT optical spectra of nine candidate
subdwarf plus companion stars chosen from the colour-colour selection
seen in Figure\,\ref{f-rrK}. Eight of the nine targets are subdwarfs with
hints of absorption lines from the companion star (see also
Section\,\ref{s-uve}). Spectra are ordered approximately by effective
temperature of the subdwarf, and shifted in flux by appropriate amounts.
0051$-$0955 is probably an A-type star rather than a subdwarf.}
 \end{minipage}
\end{figure*}

We plot the resultant spectra in Figure\,\ref{f-sdb} and find that all
but one of the nine objects chosen from the colour-colour selection are
sdB stars with companions (Table\,\ref{t-fup}).

\begin{table*}
 \caption{\label{t-fup} Follow-up spectroscopic observations and
 classifications.}
 \begin{tabular}{llllll}
  \hline \hline
  Name       & R.A.      & Dec       & \rcmc   & Classification &
    Telescope \\
             &           &           & [mag]   &                & \\
    \hline
  0018$+$0101 & $00^{\rm h}18^{\rm m}43.51^{\rm s}$ &
$+01^{\circ}01'23\farcs6$ & 15.1  & sdB        & WHT \\
  0051$-$0955 & $00^{\rm h}51^{\rm m}20.33^{\rm s}$ &
$-09^{\circ}55'23\farcs2$ & 14.4  & A-type star& WHT \\
  1602$+$0725 & $16^{\rm h}02^{\rm m}09.07^{\rm s}$ &
$+07^{\circ}25'10\farcs9$ & 14.7  & sdB        & WHT \\
  1618$+$2141 & $16^{\rm h}18^{\rm m}06.46^{\rm s}$ &
$+21^{\circ}41'25\farcs4$ & 14.9  & sdB        & WHT \\
  1619$+$1453 & $16^{\rm h}19^{\rm m}49.30^{\rm s}$ &
$+14^{\circ}53'09\farcs9$ & 14.7  & sdB        & WHT \\
  2020$+$0704 & $20^{\rm h}20^{\rm m}27.21^{\rm s}$ &
$+07^{\circ}04'13\farcs5$ & 14.3  & sdB        & WHT \\
  2047$-$0542 & $20^{\rm h}47^{\rm m}42.37^{\rm s}$ &
$-05^{\circ}42'31\farcs0$ & 14.9  & sdB        & WHT \\
  2052$-$0457 & $20^{\rm h}52^{\rm m}26.23^{\rm s}$ &
$-04^{\circ}57'45\farcs3$ & 14.5  & sdB        & WHT \\
  2138$+$0442 & $21^{\rm h}38^{\rm m}00.82^{\rm s}$ &
$+04^{\circ}42'11\farcs6$ & 14.8  & sdB        & WHT \\
  2331$-$2515 & $23^{\rm h}31^{\rm m}03.65^{\rm s}$ &
$-25^{\circ}15'47\farcs9$ & 14.5  & sdB        & MagE \\
  2342$-$2750 & $23^{\rm h}42^{\rm m}41.41^{\rm s}$ &
$-27^{\circ}50'01\farcs7$ & 15.1  & sdB        & MagE \\
  \hline
 \end{tabular}
\end{table*}

\subsection{MagE}

In addition to the WHT spectra, two candidates were observed on 7-8
June 2010, using the MagE (Magellan Echellette) spectrograph mounted on
the Magellan-Clay telescope at Las Campanas Observatory, Chile. We used
the $1\arcsec$ slit with the 175 lines/mm grating to cover
$\sim3100-11200$\,\AA\ at a resolution of $R=4100$. The data were
unbinned and we used the slow readout mode.

The spectra were reduced with the Carnegie pipeline written by
D.\,Kelson. This Python-driven pipeline performs typical calibrations:
flat-fielding, sky background subtraction followed by optimal extraction
and wavelength calibration. The wavelength calibrations were derived from
Thorium-Argon lamp exposures taken during the night, which provided ample
suitable lines over the entire wavelength range. The pipeline selects the
closest lamp exposures in time to each science spectrum. Raw spectra were
then flux calibrated using a spectrum of the flux standard Feige 110,
observed at the end of each night.
We find that both objects observed have spectra consistent with
being sdB stars with some evidence for a companion.

This initial exploration of eleven of our candidates thus offers strong
evidence that we are primarily selecting composite subdwarf systems with
our colour cuts, with a low contamination rate. We discuss contamination
of our samples further in Sections\,\ref{ss-overlap} and \ref{ss-dist}.

\section{Fitting composite systems}
\label{s-fit}

To quantify the likely composition of our subdwarf candidates, we
pursued SED fitting exploiting the broad wavelength range of the
photometric data that is available. The subdwarf star dominates the
ultraviolet flux while the main--sequence companion clearly dominates in
the infrared. This permits the decomposition of the SED into two
components at a common distance. In this section we demonstrate that good
constraints on both the subdwarf and companion star effective temperature
can be derived from such fits.
The observed magnitudes were fitted with the grid of subdwarf plus
main--sequence star magnitudes discussed in Section\,\ref{s-mod}, with
the additional option of having a subdwarf with no companion (shown as
MS\,$\Teff=0$\,K in Table\,\ref{t-ind1} onwards). This was performed by
minimising a weighted $\chi^2$ whilst varying the distance, subdwarf and
companion effective temperatures. Uncertainties were taken from the one
sigma contours in the $\chi^2$ surface. 
This fitting was restricted to the sub-samples where SDSS photometry is
available, since we require multi-band optical photometry in order to
decompose the SED.

Reddening from interstellar dust can potentially have a significant
effect on the shape of the subdwarf SED, especially at short wavelengths.
It would therefore primarily affect the inferred subdwarf effective
temperature. The slope will be flattened and thus a systematically lower
effective temperature would be found. Without prior knowledge of the
reddening to the system, this is not easily corrected for. To estimate an
upper limit for this effect, we calculate the reddening at the position
of the subdwarfs from the \citet{schlegeletal98-1} maps and use these
values to first deredden the magnitudes. Refitting these values gives a
second set of system parameters that will, in general, be overcorrected
for reddening in comparison to the fits without any reddening. The true
parameters will lie somewhere in between these two limits.

As shown in Figure\,\ref{f-model}, subdwarfs with metal-poor companions
fall in the colour cuts defined in Table\,\ref{t-box}. They are not
a contaminant, but fitting the metal-poor systems with solar metalicity
models will lead to biased system parameters. Less absorption in the
ultraviolet from metal lines means the companions will contribute a
fairly significant amount of flux at short wavelengths. To test the
effect of this, we fitted the \textit{C2MS} and \textit{SU} samples with
a grid of subdwarfs plus metal-poor ($\log ([\mathrm{M/H}] /
[\mathrm{M/H}]_{\mathrm{solar}}) = 1.5$) companions from the
\citet{castelli+kurucz03-1} ATLAS9 model atmosphere library. This has the
effect of reducing all subdwarf effective temperatures by a few thousand
Kelvin and shifting the distribution of companion types later by a few
hundred Kelvin. If anything, this accentuates the conclusions we draw in
Section\,\ref{s-dis}.

A final potential bias to our fitting method is that approximately
$10$\,per\,cent of subdwarfs are evolved and therefore will have lower
surface gravities and bloated radii compared with their unevolved
equivalent \citep{heber09-1}. Fitting a system with an evolved
subdwarf using our subdwarf plus main--sequence star model grid
(described in Section\,\ref{s-mod}), we would find that the companion
star is cooler and the subdwarf is hotter than the true temperature.
However, this situation will most likely result in a high minimum
$\chi^2$ and therefore be flagged as a bad fit.

\section{Fit Results and Individual Objects}
\label{s-res}

\begin{figure*}
 \begin{minipage}{2\columnwidth}
  \includegraphics[width=\columnwidth]{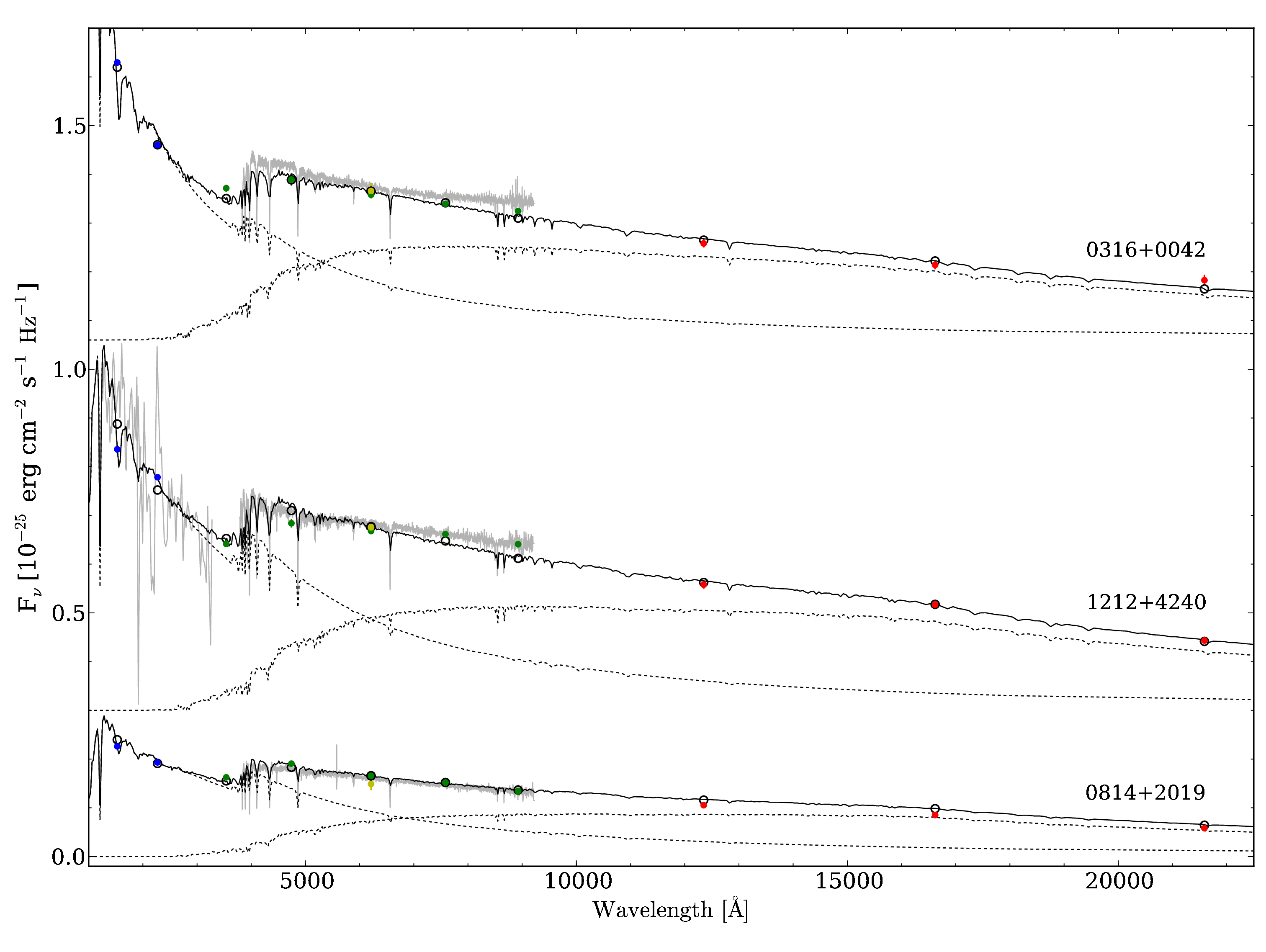}
  \caption{\label{f-fit} The SEDs of, and fits to, 0316$+$0042
(PG\,0313$+$005), 0814$+$2019 and 1212$+$4240
(PG\,1210$+$429). The optical SDSS spectra are shown in grey. The GALEX,
SDSS, CMC and 2MASS magnitudes are plotted in blue, green, yellow and
red, respectively with corresponding errorbars. The fit to
0316$+$0042 comprises a $\Teff=28,000$\,K sdB model and a
$6,250$\,K star (black dashed lines). Similarly, a $\Teff=21,000$\,K sdB
model and a $5,500$\,K star is used for 0814$+$2019 and a
$\Teff=23,000$\,K sdB model and a $5,750$\,K star for 1212$+$4240.
The composite spectra and magnitudes are the solid black line and open
black circles, respectively. The absolute flux level of the SDSS spectrum
does not match the model well in all cases. This is most likely related
to calibration issues of the SDSS spectroscopy, as it disagrees with the
SDSS photometry. For 1212$+$4240, an archive IUE ultraviolet
spectrum plotted in grey. 0316$+$0042 and 1212$+$4240 are
offset in flux by $0.30$ and $1.05$ units, respectively, for clarity.}
 \end{minipage}
\end{figure*}

\begin{figure*}
 \begin{minipage}{2\columnwidth}
  \includegraphics[width=\columnwidth]{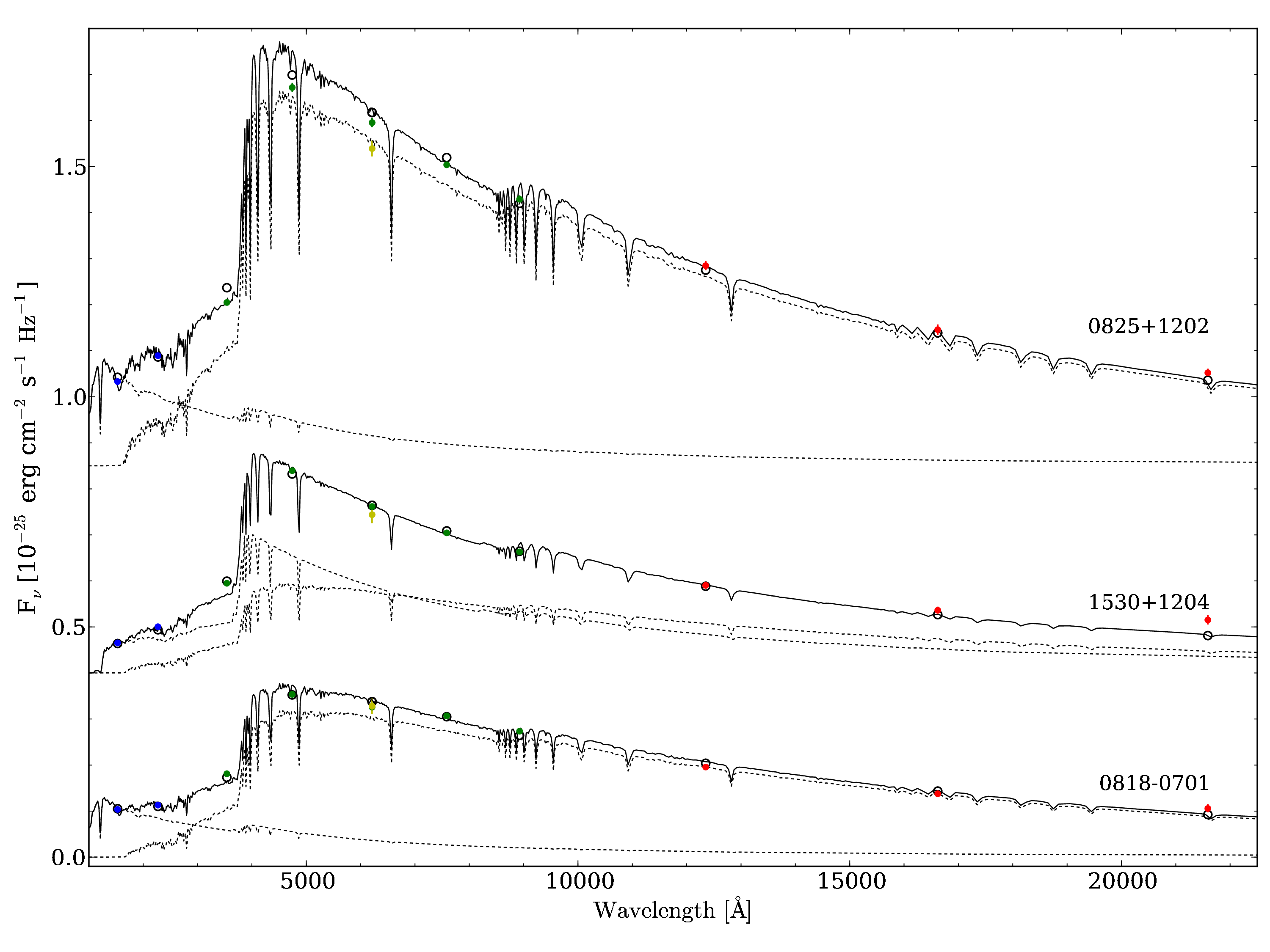}
  \caption{\label{f-fit2}
The SEDs of, and fits to, 0818$-$0701, 0825$+$1202 and 1530$+$1204,
following the same format as Figure\,\ref{f-fit}. The fit to 0818$-$0701
comprises a $\Teff=22,000$\,K sdB model and a $7,750$\,K star (black
dashed lines). Similarly, a $\Teff=22,000$\,K sdB model and a $8,250$\,K
star is used for 0825$+$1202 and a $\Teff=11,000$\,K sdB model and a
$8,000$\,K star for 1530$+$1204. 1530$+$1204 and 0825$+$1202 are offset
in flux by $0.40$ and $0.85$ units, respectively, for clarity.
}
 \end{minipage}
\end{figure*}

All the fit parameters for the 93 objects from the \textit{C2MS}
sample (Table\,\ref{t-nums}) are given in Table\,\ref{t-obj1}.
Similarly, the 134 \textit{SU} objects are shown in Table\,\ref{t-obj2}.
We adopt a somewhat unusual notation for the upper and lower
uncertainties, denoted by the ``\{'' symbol, because the subdwarf and
companion effective temperature uncertainties are strongly correlated.
``\{'' indicates the upper and lower $1\sigma$ uncertainties added to the
best fit value. The upper values all correspond to the same fit solution
and similarly for the lower values.
As an example, consider a hypothetical system where SD\,$\Teff =
15,000\{^{25,000}_{10,000}$\,K, and MS\,$\Teff =
2,000\{^{1,000}_{3,000}$\,K. This corresponds to three solutions: the
best fit (a $15,000$\,K subdwarf with a $2,000$\,K companion), a
$1\sigma$ uncertainty in the direction of increased subdwarf temperature
(a $25,000$\,K subdwarf with a $1,000$\,K companion), and a $1\sigma$
uncertainty in the direction of decreased subdwarf temperature (a
$10,000$\,K subdwarf with a $3,000$\,K companion). One cannot mix and
match these combinations. For example, a $10,000$\,K subdwarf with a
$1,000$\,K companion, or a $25,000$\,K subdwarf with a $3,000$\,K
companion, are \textit{not} valid solutions.
A minimum uncertainty is set at one grid point and therefore is also
limited by the extent of the grid: a minimum and maximum subdwarf
temperature of $11,000$ and $40,000$\,K, respectively.
We examine systematic uncertainties in Section\,\ref{ss-overlap},
leading to estimates of a few thousand Kelvin for a more realistic error.
We show example SEDs and fits to a few objects in Figures\,\ref{f-fit}
and \ref{f-fit2}. Objects in Figures\,\ref{f-fit} and \ref{f-fit2} are
found to have approximately G0 and A7-type companions, respectively.

We compared our results to published effective temperatures and/or known
companions for the \textit{C2MS} and \textit{SU} samples, shown in
Table\,\ref{t-ind1} and \ref{t-ind2}, respectively.
The best fit is not always satisfactory, indicated by a high $\chi^2$.
We include the ``Q'' (Quality) column to show where this is the case.
``Q'' values correspond to; 1:Good fit, 2:Average fit, 3:Poor fit,
4:WD/WD+MS/CV and 5:Quasar/Galaxy. Values of three and above are excluded
from the histograms shown in Figure\,\ref{f-hist1}, \ref{f-histd} and
\ref{f-hist3}. The classifications in this catagory between values of 1,
2 and 3 are purely qualitative.
SIMBAD has an entry for many more objects, but without any specific
details. All objects which were previously known (in one or more
of:
\citealt{fergusonetal84-1}, 
\citealt{kilkennyetal88-1}, 
\citealt{allardetal94-2}, 
\citealt{safferetal94-1}, 
\citealt{thejlletal95-1}, 
\citealt{ulla+thejll98-1}, 
\citealt{jeffery+pollacco98-1}, 
\citealt{cuadrado+jeffery01-1}, 
\citealt{maxtedetal01-1}, 
\citealt{williamsetal01-1}, 
\citealt{cuadrado+jeffery02-1}, 
\citealt{maxtedetal02-1}, 
\citealt{edelmannetal03-1}, 
\citealt{morales-ruedaetal03-1}, 
\citealt{stark+wade03-1}, 
\citealt{napiwotzkietal04-1}, 
\citealt{reed+stiening04-1}, 
\citealt{liskeretal05-1}, 
\citealt{ostensen06-1}, 
\citealt{wadeetal06-1}, 
\citealt{stark+wade06-1}, 
\citealt{stroeeretal07-1}, 
\citealt{wadeetal09-1}, 
\citealt{geieretal11-2} and
\citealt{vennesetal11-2}
) to be composite
subdwarf plus companion systems are highlighted in Tables\,\ref{t-obj1}
and \ref{t-obj2}.

\subsection{Potential systematic temperature differences}
\label{ss-tdiff}

When comparing the system parameters calculated herein and those from
the literature, there are a number of possible causes for discrepancies:
Firstly, one must consider the fact that often in the literature fitting
is performed on the absorption line profiles of the subdwarf with a
single star model \citep[e.g.][]{safferetal94-1}, whereas our
study suggests that these systems all have a significant contribution
from the companion. The single subdwarf fit would then result in biased
system parameters.

Secondly, if the subdwarf's companion is a sub-giant or giant type
star, our method would underestimate the subdwarf's effective temperature
because we only use main--sequence star models for the companion. While
this may affect isolated cases, we do not expect a significant population
of sub-giant and giant companion stars to be present in our sample given
the colour selection cuts we employed (see Section\,\ref{ss-cuts}).

Finally, the suppression of the subdwarf's ultraviolet flux due to line
blanketing could cause a biased effective temperature.
Subluminous B stars show peculiar abundance patterns. Some metals (mostly
the lighter ones) are found to be strongly depleted, while heavier
elements can be strongly enriched \citep{otoole+heber06-1,
blanchetteetal08-1}. The abundance patterns are caused by atomic
diffusion, which depends on various parameters (see
\citealt{michaudetal11-1}, for the state-of-the-art of modelling),
however, metalicity may not be an important one.
Because the abundance pattern differs from star to star, the ultraviolet
line blocking for any individual subdwarf will deviate from that
predicted from the solar metalicity models adopted here. Therefore, we
cannot quantify the systematic uncertainty in the temperature
determination of the subdwarf stars. \citet{otoole+heber06-1} regard
solar metalicity models as appropriate for sdB stars cooler than about
$30000$\,K, but prefer models of scaled supersolar abundances for hotter
stars as a proxy for enhanced ultraviolet line blocking.
Because the effective temperatures of our program stars are mostly
below $30000$\,K, we stay with solar metalicity model spectra.

\subsection{0018$+$0101}
\label{ss-0018}

\citet{liskeretal05-1} calculated an effective temperature for
0018$+$0101 (HE\,0016$+$0044) of $28,264\pm800$\,K. This compares
relatively well with our \textit{SU} sample estimate of
$25,000-23,000$\,K, however, a significantly higher temperature is
measured when using the \textit{C2MS} sample ($39,000-40,000$\,K). Either
a $23,000$ or a $40,000$\,K subdwarf provide an adequate fit to the SED,
and small changes in the $\chi^2$ surface lead to the alternate
solution. The flat $\chi^2$ surface comes about from a very blue
$(\fuv-\nuv)$ colour ($-0.64$) that is difficult to reconcile with the
rest of the SED.

\subsection{1300$+$0057 and 1538$+$0934}
\label{ss-1300}

The published effective temperatures for 1300$+$0057
\citep[$39359$\,K: HE\,1258$+$0113:][]{stroeeretal07-1} and
1538$+$0934 \citep[$35114$\,K:
HS\,1536$+$0944:][]{liskeretal05-1}, both in the \textit{SU} sample, are
only upper limits on the effective temperatures. \citet{liskeretal05-1}
note the presence of a cool ($\sim$K0-type) companion in the spectrum of
1538$+$0934 and therefore specifically state that the estimated
temperature is an upper limit.
\citet{stroeeretal07-1} also note the presence of a cool companion based
on the $B-J$ colour for 1300$+$0057 and therefore one may assume
the temperature is also an overestimate.
In both cases, the best fit model ($30,000$ and $23,000$\,K for
1300$+$0057 and 1538$+$0934, respectively) corresponds to a
bluer $(\fuv-\nuv)$ colour than the \textit{GALEX} fluxes. Therefore
using the higher published effective temperature model would not agree
with the data.

\subsection{1517$+$0310 and 1518$+$0410}
\label{ss-1517}

In the case of 1517$+$0310 and 1518$+$0410 (PG\,1514$+$034
and PG\,1515$+$044, respectively: \textit{SU} sample), the companion
effective temperatures measured ($6,000\{^{6,250}_{5,750}$\,K and
$5,500\{^{5,750}_{5,250}$\,K, respectively) are significantly different
from that in the catalogue of \citet{ostensen06-1} (K2 and K4.5;
corresponding to effective temperatures of $\sim4,800$ and $4,300$\,K,
respectively). The whole SED of 1517$+$0310 is not particularly
well fit by the calculated best model. The system has a very blue
$(\fuv-\nuv)$ colour and therefore the best fit model is forced to be a
hot subdwarf, which leads to a correspondingly increased companion
effective temperature.

\subsection{1709$+$4054}
\label{ss-1709}

1709$+$4054 (PG\,1708$+$409: \textit{C2MS} sample), was
classified by \citet{safferetal94-1} to be a subdwarf
with an effective temperature of $28,500$\,K. We determined
$25,000-28,000$\,K if we apply no reddening and $27,000-29,000$\,K when
applying the full \citet{schlegeletal98-1} reddening. However,
\citet{safferetal94-1} fit the line profiles of
this composite system with a single star subdwarf model, and therefore
comparing the two sets of temperatures is not comparing like for like.

\subsection{2138$+$0442}
\label{ss-2138}

For the case of 2138$+$0442 (PG\,2135$+$045;
\textit{C2MS} sample), we find a slightly lower effective temperature
($24,000-26,000$\,K) compared with the published value of
\citet[$\sim28,000$\,K:][]{cuadrado+jeffery02-1}. Including the full
\citet{schlegeletal98-1} reddening ($26,000-28,000$\,K), however, the
temperatures agree. \citet{cuadrado+jeffery02-1} treat 2138$+$0442
as a composite system fitting both objects in the blue region of the
spectrum, thus the above mentioned problem of fitting a single star model
(Section\,\ref{ss-tdiff}) does not apply.

\subsection{2244$+$0106}
\label{ss-2244}

2244$+$0106 (PB\,5146) was found to be a post-EHB star with a high
velocity in \citet[Hyper-MUCHFUSS;][]{tillichetal11-1}. They estimate a
$\Teff = 33580\pm680$\,K, $\Logg = 4.75\pm0.20$ and a distance of
$18.29\pm2.45$\,kpc, compared with our $22,000-26,000$\,K at
$6.1-7.9$\,kpc. However, the companion star is not accounted for in
\citet{tillichetal11-1} and therefore the subdwarfs effective temperature
is probably overestimated. This is also consistent with the unusually
low surface gravity.

\subsection{Overlap}
\label{ss-overlap}

\begin{figure*}
 \begin{minipage}{2\columnwidth}
  \includegraphics[width=\columnwidth]{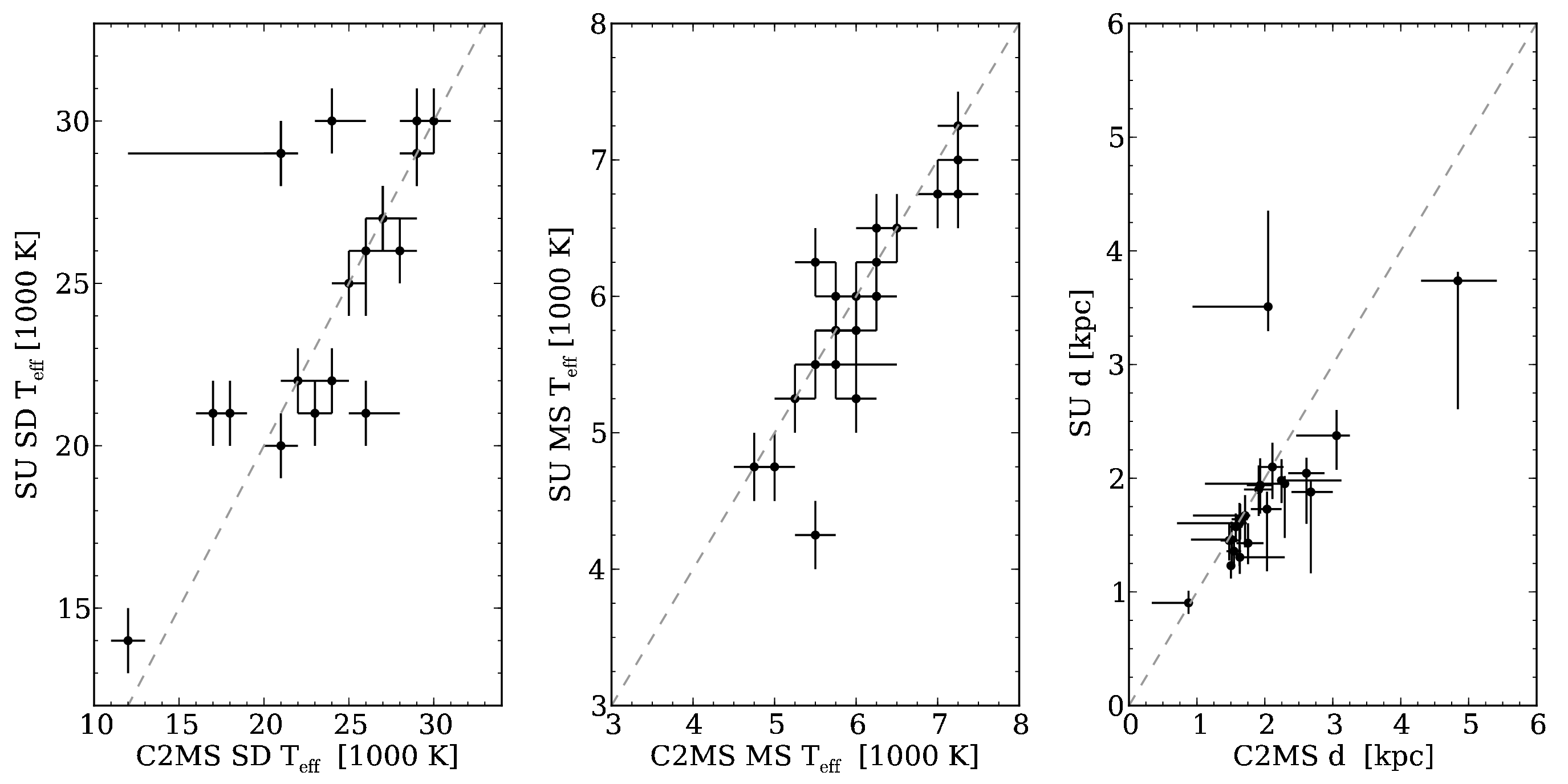}
  \caption{
\label{f-overlap} A comparison of fits using the \textit{C2MS} sample
versus that using the \textit{SU} sample, where there is overlap. Objects
with a ``Q''$\geq3$ in Table\,\ref{t-overlap} are excluded. The dashed
line shows a one to one relation between the parameters.
}
 \end{minipage}
\end{figure*}

Where the \textit{C2MS} and \textit{SU} samples overlap, a comparison of
the fits is given in Table\,\ref{t-overlap} and shown in
Figure\,\ref{f-overlap}. The two sets of fits appear consistent within
the uncertainties. We analysed the distribution of the difference between
all the \textit{C2MS} and \textit{SU} parameters (distance, subdwarf and
companion temperature) and find that the distributions are all
approximately Gaussian, centered about zero. We do not find any evidence
to suggest that the two samples effective temperatures are systematically
offset. The errors on the subdwarf effective temperature from the
$\chi^2$ fit may be slightly underestimated, and a more realistic error
is a few thousand Kelvin.
The one difference is that the UKIDSS data should better constrain the
companion star effective temperature due to the greater depth and higher
photometric accuracy of the near-infrared data.

Overall, in individual cases, we must bear in mind that we may
occasionally select the wrong solution (in cases where the $\chi^2$
surface is relatively flat), nor can we identify the exact
amount of reddening that should be corrected for. However, this study is
aimed at providing a statistical analysis of the sample rather than
correct parameters for all individual systems. The errors in the
measured parameters should be randomly distributed and therefore not
effect the distributions. It is thus not a significant issue for the
analysis presented here, but these uncertainties should be considered
when consulting the fitted parameters of individual systems.

We saw earlier that the key contaminants in our colour box are composite
systems containing white dwarfs (Table\,\ref{t-box}). Indeed, from the
\textit{C2MS} sample 0018$+$0101, 0141$+$0614,
0923$+$0652, 2117$-$0015 and 2117$-$0006 are
candidates for being DA white dwarfs with infrared excesses based on
their photometry \citep{girvenetal11-1}. However, such a classification
can only be confirmed through follow-up spectroscopy. SDSS spectroscopy
is available for 2117$-$0006, and \citet{girvenetal11-1} classify
it as a ``Narrow Line Hot Star'' (NLHS), which they believe to be a
group primarily made up of subdwarfs. 0018$+$0101, discussed in
Section\,\ref{ss-0018}, is also catalogued as a NLHS by
\citet{girvenetal11-1}, corroborating the subdwarf label.

\begin{table*}
\begin{minipage}{2\columnwidth}
\begin{center}
\caption{\label{t-ind1} Individual objects of interest from the
\textit{C2MS} sample. Only subdwarfs with measured system parameters are
displayed, but all known CVs and white dwarfs are shown. The ``\{''
notation is described in Section\,\ref{s-res} and does not simply
represent uncertainties. The fit parameters shown are not corrected for
interstellar reddening. The comments quoted for the possible white dwarfs
matched in \citet{girvenetal11-1} are the classifications according to
the SDSS spectra found therein. NLHS corresponds to Narrow Line Hot Star
(probable subdwarf).}
\begin{tabular}{llcccccl}
\hline \hline
 &  &  & This Paper &  &  & Literature &  \\
Name & Identifier & sdB \Teff\ & MS \Teff\ & d & sdB \Teff\ & MS Type &
Ref / Comments \\
 &  & (1000K) & (1000K) & (kpc) & (K) &  &  \\
\hline
\textbf{Subdwarfs} & & & & & & & \\
0018$+$0101 & HE\,0016$+$0044 & $40\{^{40}_{39}$ &
$5.50\{^{5.75}_{5.25}$ & $1.5\{^{1.5}_{1.4}$ & 28264 &  &
\citet{liskeretal05-1} \\
1212$+$4240 & PG\,1210$+$429 & $23\{^{24}_{22}$ &
$5.75\{^{6.00}_{5.50}$ & $1.5\{^{1.7}_{1.4}$ &  & K2.5 &
\citet{ostensen06-1} \\
1517$+$0310 & PG\,1514$+$034 & $40\{^{40}_{39}$ &
$6.00\{^{6.25}_{5.75}$ & $1.1\{^{1.1}_{1.0}$ &  & K2 &
\citet{ostensen06-1} \\
1709$+$4054 & PG\,1708$+$409 & $26\{^{28}_{25}$ &
$5.50\{^{5.75}_{5.25}$ & $1.7\{^{1.9}_{1.6}$ & 28500 &  &
\citet{safferetal94-1, safferetal98-1}
\\
2138$+$0442 & PG\,2135$+$045 & $25\{^{26}_{24}$ &
$5.00\{^{5.25}_{4.75}$ & $1.2\{^{1.3}_{1.1}$ & $\sim28000$ & $\sim$K2 &
\citet{cuadrado+jeffery02-1} \\
\hline
\textbf{CV} & & & & & & & \\
0141$+$0614 & HS\,0139$+$0559 & $12\{^{13}_{11}$ &
$7.25\{^{7.50}_{7.00}$ & $4.8\{^{5.4}_{4.3}$ &  &  &
\citet{heberetal91-1}, \\
 &  &  &  &  &  &  & \citet{aungwerojwitetal05-1} \\
0812$+$1911 &  & $14\{^{15}_{13}$ & $7.25\{^{7.50}_{7.00}$ &
$4.8\{^{6.9}_{4.8}$ &  &  & \citet{szkodyetal06-1} \\
1015$-$0308 & SW Sex & $18\{^{19}_{17}$ & $7.00\{^{7.25}_{6.75}$
& $2.6\{^{2.9}_{2.3}$ &  &  & e.g. \citet{greenetal82-1}, \\
 &  &  &  &  &  &  & \citet{penningetal84-1} \\
2143$+$1244 &  & $30\{^{31}_{29}$ & $6.25\{^{6.50}_{6.00}$ &
$2.9\{^{3.3}_{2.6}$ &  &  & \citet{szkodyetal05-1} \\
\hline
\textbf{Possible WD} & & & & & & & \citet{girvenetal11-1} \\
0018$+$0101 & HS\,0016$+$0044 & $40\{^{40}_{39}$ &
$5.50\{^{5.75}_{5.25}$ & $1.5\{^{1.5}_{1.4}$ &  &  & NLHS \\
0141$+$0614 & HS\,0139$+$0559 & $12\{^{13}_{11}$ &
$7.25\{^{7.50}_{7.00}$ & $4.8\{^{5.4}_{4.3}$ &  &  & \\
0923$+$0652 &  & $29\{^{30}_{28}$ & $6.75\{^{7.00}_{6.50}$ &
$2.1\{^{2.9}_{2.0}$ &  &  & \\
2117$-$0015 &  & $13\{^{14}_{12}$ & $6.75\{^{7.00}_{6.50}$ &
$3.5\{^{3.5}_{2.3}$ &  &  & \\
2117$-$0006 &  & $21\{^{22}_{20}$ & $6.25\{^{6.50}_{6.00}$ &
$2.0\{^{2.9}_{1.9}$ &  &  & NLHS \\
\hline
\end{tabular}
\end{center}
\end{minipage}
\end{table*}

\begin{table*}
\begin{minipage}{2\columnwidth}
\begin{center}
\caption{\label{t-ind2} Individual objects of interest from the
\textit{SU} sample. The ``\{'' notation is described in
Section\,\ref{s-res} and does not simply represent uncertainties. The
comments quoted for the possible white dwarfs matched in
\citet{girvenetal11-1} are the classifications according to the SDSS
spectra found therein. DA and NLHS correspond to DA white dwarf and
Narrow Line Hot Star (probable subdwarf), respectively.}
\begin{tabular}{llcccccl}
\hline \hline
 &  &  & This Paper &  &  & Literature &  \\
Name & Identifier & sdB \Teff\ & MS \Teff\ & d & sdB \Teff\ & MS Type &
Ref / Comments \\
 &  & (1000K) & (1000K) & (kpc) & (K) &  &  \\
\hline
\textbf{Subdwarfs} & & & & & & & \\ \vspace{0.1cm}
0018$+$0101 & HE\,0016$+$0044 & $24\{^{25}_{23}$ &
$4.25\{^{4.50}_{4.00}$ & $1.2\{^{1.3}_{1.2}$ & 28264 &  &
\citet{liskeretal05-1} \\
1300$+$0057 & HE\,1258$+$0113 & $30\{^{31}_{29}$ &
$3.50\{^{3.75}_{3.25}$ & $1.7\{^{1.8}_{1.6}$ & 39359$^a$ &  &
\citet{stroeeretal07-1} \\
1517$+$0310 & PG\,1514$+$034 & $40\{^{40}_{39}$ &
$6.00\{^{6.25}_{5.75}$ & $1.1\{^{1.2}_{1.0}$ &  & K2 &
\citet{ostensen06-1} \\
1518$+$0410 & PG\,1515$+$044 & $26\{^{27}_{25}$ &
$5.50\{^{5.75}_{5.25}$ & $1.8\{^{2.0}_{1.7}$ &  & K4.5 &
\citet{ostensen06-1} \\
1538$+$0934 & HS\,1536$+$0944 & $23\{^{24}_{22}$ &
$5.00\{^{5.25}_{4.75}$ & $1.8\{^{2.0}_{1.7}$ & 35114$^a$ & K0 &
\citet{liskeretal05-1} \\
\hline
\textbf{CV} & & & & & & & \\
0141$+$0614 & HS\,0139$+$0559 & $14\{^{15}_{13}$ &
$7.25\{^{7.50}_{7.00}$ & $3.7\{^{4.9}_{3.7}$ &  &  &
\citet{heberetal91-1} \\
0813$+$2813 &  & $20\{^{21}_{19}$ & $6.25\{^{6.50}_{6.00}$ &
$6.4\{^{6.7}_{4.6}$ &  &  & \citet{szkodyetal05-1} \\
0920$+$3356 & BK\,Lyn & $20\{^{21}_{18}$ & $6.75\{^{7.00}_{6.25}$
& $2.3\{^{2.3}_{1.7}$ &  &  & \citet{dobrzycka+howell92-1}, \\
 &  &  &  &  &  &  & \citet{ringwald93-1} \\
1015$-$0308 & SW\,Sex & $21\{^{22}_{20}$ & $6.75\{^{7.00}_{6.50}$
& $2.0\{^{2.5}_{1.9}$ &  &  & e.g. \citet{ballouz+sion09-1}, \\
 &  &  &  &  &  &  & \citet{ritter+kolb09-1} \\
2333$+$1522 &  & $17\{^{18}_{16}$ & $6.75\{^{7.00}_{6.50}$ &
$12.6\{^{17.8}_{12.3}$ &  &  & \citet{szkodyetal05-1} \\
\hline
\textbf{WDMS} & & & & & & & \citet{rebassa-mansergasetal11-1} \\
0032$+$0739 &  & $21\{^{22}_{20}$ & $5.25\{^{5.50}_{5.00}$ &
$4.4\{^{6.9}_{4.4}$ &  &  & \\
0300$-$0023 & WD\,0257$-$005 & $38\{^{39}_{34}$ &
$5.00\{^{5.25}_{4.75}$ & $3.2\{^{3.4}_{2.8}$ &  &  &  \\
0920$+$1057 &  & $34\{^{35}_{33}$ & $4.75\{^{5.00}_{4.50}$ &
$3.3\{^{3.5}_{3.1}$ &  &  &  \\
1016$+$0443 &  & $29\{^{30}_{28}$ & $4.50\{^{4.75}_{4.25}$ &
$4.8\{^{7.9}_{4.8}$ &  &  &  \\
1352$+$0910 &  & $29\{^{30}_{28}$ & $4.00\{^{4.25}_{3.75}$ &
$4.0\{^{6.5}_{4.0}$ &  &  &  \\
\hline
\textbf{Possible WD} & & & & & & & \citet{girvenetal11-1} \\
0018$+$0101 & HS\,0016$+$0044 & $24\{^{25}_{23}$ &
$4.25\{^{4.50}_{4.00}$ & $1.2\{^{1.3}_{1.2}$ &  &  & NLHS \\
0032$+$0739 &  & $21\{^{22}_{20}$ & $5.25\{^{5.50}_{5.00}$ &
$4.4\{^{6.9}_{4.4}$ &  &  & DA \\
0141$+$0614 & HS\,0139$+$0559 & $14\{^{15}_{13}$ &
$7.25\{^{7.50}_{7.00}$ & $3.7\{^{4.9}_{3.7}$ &  &  &  \\
0814$+$2811 &  & $22\{^{23}_{21}$ & $6.00\{^{6.25}_{5.75}$ &
$3.7\{^{4.0}_{3.3}$ &  &  & NLHS \\
0854$+$0853 & PN\,A66\,31 & $40\{^{40}_{39}$ &
$3.00\{^{3.25}_{3.00}$ & $1.2\{^{1.2}_{1.2}$ &  &  &  \\
0920$+$3356 & BK\,Lyn & $20\{^{21}_{18}$ & $6.75\{^{7.00}_{6.25}$
& $2.3\{^{2.3}_{1.7}$ &  &  &  \\
0925$-$0140 &  & $17\{^{18}_{16}$ & $5.50\{^{5.75}_{5.25}$ &
$9.4\{^{14.9}_{9.4}$ &  &  &  \\
0951$+$0347 &  & $23\{^{24}_{22}$ & $4.00\{^{4.25}_{3.75}$ &
$1.9\{^{2.0}_{1.7}$ &  &  & NLHS \\
0959$+$0330 & PG\,0957$+$037 & $31\{^{32}_{30}$ &
$3.00\{^{3.25}_{3.00}$ & $1.1\{^{1.2}_{1.1}$ &  &  &  \\
1006$+$0032 & PG\,1004$+$008 & $26\{^{27}_{25}$ &
$5.00\{^{5.25}_{4.75}$ & $3.3\{^{3.6}_{3.1}$ &  &  &  \\
1100$+$0346 &  & $34\{^{36}_{33}$ & $3.75\{^{4.25}_{3.50}$ &
$2.7\{^{2.9}_{2.6}$ &  &  & NLHS \\
1116$+$0755 &  & $28\{^{29}_{27}$ & $5.00\{^{5.25}_{4.75}$ &
$2.3\{^{2.3}_{1.4}$ &  &  &  \\
1135$+$0731 &  & $29\{^{30}_{28}$ & $6.25\{^{6.50}_{6.00}$ &
$6.4\{^{8.2}_{5.7}$ &  &  & NLHS \\
1215$+$1351 &  & $21\{^{22}_{20}$ &
$4.50\{^{4.75}_{4.25}$ & $3.1\{^{5.1}_{3.1}$ &  &  & NLHS \\
1228$+$1040 & WD\,1226$+$110 & $21\{^{22}_{20}$ &
$3.00\{^{3.25}_{3.00}$ & $2.2\{^{3.8}_{2.2}$ &  &  & DA:
\citet{gaensickeetal06-3} \\
1237$-$0151 &  & $23\{^{25}_{22}$ & $4.75\{^{5.00}_{4.50}$ &
$3.9\{^{4.3}_{3.6}$ &  &  &  \\
1300$+$0057 & HE\,1258$+$0113 & $30\{^{31}_{29}$ &
$3.50\{^{3.75}_{3.25}$ & $1.7\{^{1.8}_{1.6}$ &  &  & NLHS \\
1315$+$0245 &  & $33\{^{34}_{32}$ & $3.25\{^{3.50}_{3.00}$ &
$0.9\{^{1.0}_{0.8}$ &  &  &  \\
1323$+$2615 &  & $20\{^{21}_{19}$ & $5.00\{^{5.25}_{4.75}$ &
$5.8\{^{5.8}_{3.6}$ &  &  &  \\
1352$+$0910 &  & $29\{^{30}_{28}$ & $4.00\{^{4.25}_{3.75}$ &
$4.0\{^{6.5}_{4.0}$ &  &  & DA \\
1422$+$0920 &  & $26\{^{27}_{25}$ & $4.75\{^{5.00}_{4.50}$ &
$3.5\{^{3.7}_{3.2}$ &  &  & NLHS \\
1442$+$0910 &  & $26\{^{27}_{24}$ & $5.00\{^{5.25}_{4.75}$ &
$6.6\{^{7.1}_{5.9}$ &  &  &  \\
1443$+$0931 &  & $28\{^{29}_{27}$ & $4.50\{^{4.75}_{4.25}$ &
$4.5\{^{4.5}_{2.7}$ &  &  & NLHS \\
1500$+$0642 &  & $27\{^{28}_{26}$ & $4.25\{^{4.50}_{3.75}$ &
$3.9\{^{4.2}_{3.7}$ &  &  & NLHS \\
1507$+$0724 &  & $27\{^{28}_{26}$ & $4.50\{^{4.75}_{4.25}$ &
$4.1\{^{4.4}_{3.8}$ &  &  &  \\
1510$+$0409 &  & $26\{^{27}_{25}$ & $4.00\{^{4.25}_{3.75}$ &
$3.4\{^{3.6}_{3.2}$ &  &  & NLHS \\
1525$+$0958 &  & $29\{^{30}_{28}$ & $3.25\{^{4.25}_{3.00}$ &
$2.8\{^{4.8}_{2.8}$ &  &  & NLHS \\
1538$+$0644 & HS\,1536$+$0944 & $14\{^{15}_{13}$ & $7.25\{^{7.50}_{7.00}$
& $6.5\{^{8.6}_{6.4}$ &  &  &  \\
1543$+$0012 & WD\,1541$+$003 & $21\{^{22}_{20}$ &
$4.75\{^{5.00}_{4.50}$ & $2.9\{^{4.5}_{2.8}$ &  &  & NLHS \\
1554$+$0616 &  & $29\{^{30}_{28}$ & $4.75\{^{5.00}_{4.50}$ &
$3.8\{^{5.8}_{3.6}$ &  &  &  \\
1619$+$2407 &  & $24\{^{25}_{23}$ & $6.50\{^{6.75}_{6.25}$ &
$4.3\{^{4.8}_{3.8}$ &  &  & NLHS \\
2049$-$0001 &  & $18\{^{19}_{17}$ & $5.25\{^{5.50}_{5.00}$ &
$6.0\{^{6.4}_{5.6}$ &  &  &  \\
2117$-$0006 &  & $30\{^{31}_{29}$ & $6.50\{^{6.75}_{6.25}$ &
$2.1\{^{2.3}_{1.8}$ &  &  & NLHS \\
2147$-$0112 & FBS\,2145$-$014 & $25\{^{26}_{24}$ &
$3.25\{^{3.50}_{3.00}$ & $1.6\{^{1.7}_{1.5}$ &  &  &  \\
\hline
\end{tabular}
\\
$^a$ Noted presence of a cool companion, therefore temperature is an
upper limit, see Section\,\ref{ss-1300}.
\end{center}
\end{minipage}
\end{table*}

\begin{table*}
 \begin{minipage}{2\columnwidth}
  \begin{center}
   \caption{\label{t-overlap} Comparison of fits using the \textit{C2MS}
sample against that using the \textit{SU} sample where there is overlap.
The ``\{'' notation is described in Section\,\ref{s-res} and does not
simply represent uncertainties.
The ``Q'' (Quality) column values correspond to; 1:Good fit, 2:Average
fit, 3:Poor fit, 4:WD/WD+MS/CV and 5:Quasar/Galaxy.}
   \begin{tabular}{llcccccccc}
   \hline \hline
   & &\multicolumn{4}{c}{\textit{C2MS}}&\multicolumn{4}{c}{\textit{SU}}\\
   & & sdB \Teff & MS \Teff & d & Q & sdB \Teff & MS \Teff & d &
   Q \\
   Name & Identifier & (1000K) & (1000K) & (kpc) & & (1000K) & (1000K) &
   (kpc) & \\ \hline
0018+0101 & HE\,0016$+$0044 & $40\{^{40}_{39}$ &
$5.50\{^{5.75}_{5.25}$ & $1.5\{^{1.5}_{1.4}$ & 2 & $24\{^{25}_{23}$ &
$4.25\{^{4.50}_{4.00}$ & $1.2\{^{1.3}_{1.2}$ & 1 \\
0054+1508 &  & $21\{^{22}_{20}$ & $7.25\{^{7.50}_{7.00}$ &
$3.2\{^{4.8}_{3.2}$ & 2 & $29\{^{30}_{28}$ & $7.25\{^{7.50}_{7.00}$ &
$2.8\{^{3.3}_{2.7}$ & 3 \\
0141+0614 & HS\,0139$+$0559 & $12\{^{13}_{11}$ &
$7.25\{^{7.50}_{7.00}$ & $4.8\{^{5.4}_{4.3}$ & 1 & $14\{^{15}_{13}$ &
$7.25\{^{7.50}_{7.00}$ & $3.7\{^{4.9}_{3.7}$ & 2 \\
0316+0042 & PG\,0313$+$005 & $28\{^{29}_{27}$ &
$6.25\{^{6.50}_{6.00}$ & $2.2\{^{2.2}_{1.4}$ & 1 & $26\{^{27}_{25}$ &
$6.00\{^{6.25}_{5.75}$ & $2.0\{^{2.2}_{1.8}$ & 1 \\
0737+2642 &  & $25\{^{26}_{24}$ & $5.50\{^{5.75}_{5.25}$ &
$1.6\{^{1.8}_{1.5}$ & 1 & $25\{^{26}_{24}$ & $5.50\{^{5.75}_{5.25}$ &
$1.6\{^{1.8}_{1.5}$ & 1 \\
0755+2128 &  & $17\{^{18}_{16}$ & $7.25\{^{7.50}_{7.00}$ &
$2.3\{^{3.5}_{2.3}$ & 1 & $21\{^{22}_{20}$ & $7.00\{^{7.25}_{6.75}$ &
$2.0\{^{2.4}_{1.9}$ & 1 \\
0814+2019 &  & $21\{^{22}_{20}$ & $5.50\{^{5.75}_{5.25}$ &
$2.0\{^{3.2}_{2.0}$ & 1 & $20\{^{21}_{19}$ & $6.25\{^{6.50}_{6.00}$ &
$3.5\{^{3.7}_{2.7}$ & 1 \\
0829+2246 &  & $26\{^{27}_{24}$ & $6.00\{^{6.25}_{5.75}$ &
$2.7\{^{3.0}_{2.4}$ & 1 & $21\{^{22}_{20}$ & $5.25\{^{5.50}_{5.00}$ &
$1.9\{^{2.6}_{1.8}$ & 1 \\
0833-0006 &  & $29\{^{30}_{28}$ & $7.25\{^{7.50}_{7.00}$ &
$3.1\{^{3.6}_{2.9}$ & 2 & $29\{^{30}_{28}$ & $6.75\{^{7.00}_{6.50}$ &
$2.4\{^{2.7}_{2.1}$ & 2 \\
0929+0603 &  & $21\{^{30}_{20}$ & $5.75\{^{6.00}_{5.50}$ &
$1.6\{^{2.5}_{1.5}$ & 2 & $29\{^{30}_{28}$ & $6.00\{^{6.25}_{5.75}$ &
$1.6\{^{2.0}_{1.4}$ & 1 \\
0937+0813 & PG\,0935$+$084 & $23\{^{24}_{22}$ &
$6.00\{^{6.25}_{5.75}$ & $2.0\{^{2.3}_{1.8}$ & 1 & $21\{^{22}_{20}$ &
$5.75\{^{6.00}_{5.50}$ & $1.7\{^{2.3}_{1.6}$ & 1 \\
0941+0657 & PG\,0939$+$072 & $21\{^{22}_{20}$ &
$6.25\{^{6.50}_{6.00}$ & $1.7\{^{2.5}_{1.6}$ & 1 & $29\{^{30}_{28}$ &
$6.50\{^{6.75}_{6.25}$ & $1.7\{^{2.0}_{1.5}$ & 2 \\
1015-0308 & SW\,Sex & $18\{^{19}_{17}$ & $7.00\{^{7.25}_{6.75}$ &
$2.6\{^{2.9}_{2.3}$ & 1 & $21\{^{22}_{20}$ & $6.75\{^{7.00}_{6.50}$ &
$2.0\{^{2.5}_{1.9}$ & 2 \\
1018+0953 &  & $28\{^{29}_{27}$ & $5.75\{^{6.00}_{5.00}$ &
$1.6\{^{1.6}_{1.0}$ & 1 & $35\{^{36}_{34}$ & $5.50\{^{5.75}_{5.25}$ &
$1.3\{^{1.5}_{1.2}$ & 1 \\
1113+0413 & PG\,1110$+$045 & $29\{^{30}_{28}$ &
$4.75\{^{5.00}_{4.50}$ & $0.9\{^{1.4}_{0.9}$ & 1 & $30\{^{31}_{29}$ &
$4.75\{^{5.00}_{4.50}$ & $0.9\{^{1.0}_{0.8}$ & 1 \\
1203+0909 & PG\,1200$+$094 & $27\{^{28}_{25}$ &
$5.75\{^{6.00}_{5.50}$ & $1.5\{^{1.6}_{1.3}$ & 1 & $27\{^{28}_{26}$ &
$5.75\{^{6.00}_{5.50}$ & $1.5\{^{1.6}_{1.3}$ & 1 \\
1233+0834 &  & $30\{^{31}_{29}$ & $6.00\{^{6.25}_{5.75}$ &
$1.9\{^{2.1}_{1.7}$ & 2 & $30\{^{31}_{29}$ & $6.00\{^{6.25}_{5.75}$ &
$1.9\{^{2.2}_{1.7}$ & 1 \\
1325+1212 & PG\,1323$+$125 & $26\{^{27}_{25}$ &
$5.75\{^{6.00}_{5.50}$ & $2.1\{^{2.3}_{1.9}$ & 1 & $26\{^{28}_{25}$ &
$5.75\{^{6.00}_{5.50}$ & $2.1\{^{2.4}_{1.9}$ & 1 \\
1326+0357 & PG\,1323$+$042 & $24\{^{25}_{23}$ &
$5.00\{^{5.25}_{4.75}$ & $1.5\{^{1.7}_{1.4}$ & 2 & $22\{^{23}_{21}$ &
$4.75\{^{5.00}_{4.50}$ & $1.4\{^{1.5}_{1.2}$ & 1 \\
1402+3215 &  & $22\{^{23}_{21}$ & $6.25\{^{6.50}_{6.00}$ &
$1.9\{^{2.1}_{1.7}$ & 1 & $22\{^{23}_{21}$ & $6.25\{^{6.50}_{6.00}$ &
$1.9\{^{2.1}_{1.7}$ & 1 \\
1421+0753 & KN\,Boo & $27\{^{28}_{26}$ & $5.25\{^{5.50}_{5.00}$ &
$1.6\{^{1.7}_{1.5}$ & 1 & $27\{^{28}_{26}$ & $5.25\{^{5.50}_{5.00}$ &
$1.6\{^{1.7}_{1.5}$ & 1 \\
1502-0245 & PG\,1459$-$026 & $24\{^{25}_{22}$ &
$6.25\{^{6.50}_{6.00}$ & $1.8\{^{1.9}_{1.5}$ & 1 & $30\{^{31}_{29}$ &
$6.00\{^{6.25}_{5.75}$ & $1.4\{^{1.6}_{1.3}$ & 1 \\
1542+0056 &  & $29\{^{30}_{28}$ & $6.50\{^{6.75}_{6.25}$ &
$1.5\{^{2.1}_{1.4}$ & 1 & $29\{^{30}_{28}$ & $6.50\{^{6.75}_{6.25}$ &
$1.5\{^{1.7}_{1.3}$ & 1 \\
    \hline
   \end{tabular}
  \end{center}
 \end{minipage}
\end{table*}

\subsection{Distributions of fits -- \textit{C2MS} sample}
\label{ss-dist}

The distribution of subdwarf and companion effective temperatures for the
\textit{C2MS} sample is shown in Figure\,\ref{f-hist1} and the
distribution of distances for the \textit{C2MS} and \textit{SU} samples
in Figure\,\ref{f-histd}. Here we compare the parameters with and
without reddening corrections. Objects that are known to be contaminants,
such as white dwarfs and CVs, have been removed from all three
(distance, subdwarf and companion temperature) histograms. For galaxies,
these should be flagged by SDSS and are therefore removed by the flags in
Table\,\ref{t-sel}. Using Table\,\ref{t-box} to estimate the
remaining fraction of contaminants, we know that $12$\,per\,cent
($3/25$) of the objects with SDSS spectra are contaminants. Therefore,
approximately eleven ($12$\,per\,cent of $93$) of the whole \textit{C2MS}
sample will be contaminants. The two CVs and one white dwarf with SDSS
spectra (Table\,\ref{t-box}) can be removed from the histograms. Thus,
the contamination of the \textit{C2MS} sample (now with and without SDSS
spectra) used for calculating distributions will be $9$\,per\,cent
($8/90$). Since any such contaminants will be distributed right across
our fit parameters, we believe they do not distort our statistical
analysis to a significant degree.

\subsubsection{Subdwarf temperature distribution}

Taking the system parameters calculated without correcting for reddening,
we find that the subdwarf effective temperatures (Figure\,\ref{f-hist1})
are spread from $20,000-30,000$\,K and peak in the $20,000-24,000$\,K range.
We do see a pronounced drop in numbers below $20,000$\,K. 
Reddening is not the issue here; applying the full
\citet{schlegeletal98-1} reddening correction to the objects before
fitting does not lead to a significant shift in the distribution, though
it is slightly smoothed.

Based on our theoretical tracks for composite
systems, we know that we have a reduced completeness below
$\sim25,000$\,K (see Section\,\ref{ss-cuts}). For example, cool subdwarfs
of $\sim15,000$\,K with an M-type companion will be missed by the colour
selection. 
This could lead to a bias towards hotter subdwarfs, which we
select over a wider range of companion types. In addition, the redder
$(\rcmc-K_s)$ colours (Figure\,\ref{f-hist}) means that at the $K_s$
magnitude limit ($14.3$), systems will be detected down to fainter \rcmc\
magnitudes. This is however offset by the increasing intrinsic brightness
of cooler subdwarfs (because of decreasing \Logg\ and increasing radius).
We previously discussed a bias towards cooler subdwarfs if our
assumption of a main--sequence type companion is incorrect
(Section\,\ref{ss-tdiff}). However, we believe this to be a relatively
small fraction given our sample selection (Section\,\ref{ss-cuts}).

To quantify these possible biases, the limitations on distance introduced
by various magnitude cuts can be seen in Table\,\ref{t-dlim}. These are
derived by taking the absolute magnitudes of the composite system and
calculating the distance the object would have to be moved to in order to
have an apparent magnitude at the relevant limit. The primary effects in
this case are caused by the saturation limit of SDSS ($\rsdss>14.1$),
corresponding to a minimum distance, and the faint $K_s$ magnitude limit
of 2MASS ($K_s<14.3$), setting a maximum distance. These significantly
depend on companion spectral type (see below) and, to a lesser extent,
on subdwarf effective temperature. It can be seen that the imposed \rcmc\
magnitude limit does not have an effect because the $K_s$ limit is always
more restrictive. In essence, in the \textit{C2MS} sample, the 2MASS
depth limits the volume over which we are reasonably complete.

\subsubsection{Companion type distribution}
\label{sss-dist_comp}

As discussed in Section\,\ref{ss-cuts}, the way in which we select
subdwarfs with companions introduces a bias in companion type.
We expect our selection to be complete for subdwarfs with $20,000
\leq \Teff \leq 35,000$\,K and companions in the range A5 to M5-type.
Similarly, including the more extreme subdwarf temperatures ($15,000
\leq \Teff \leq 40,000$\,K), we are complete for F0 to K0-type
companions. The companion type range is smaller in the latter case
because, for example, a $40,000$\,K subdwarf with a M5-type
companion does not fall in our colour selection, whereas a $35,000$\,K
subdwarf with a M5-type companion does.

\begin{figure*}
 \begin{minipage}{2\columnwidth}
  \includegraphics[width=\columnwidth]{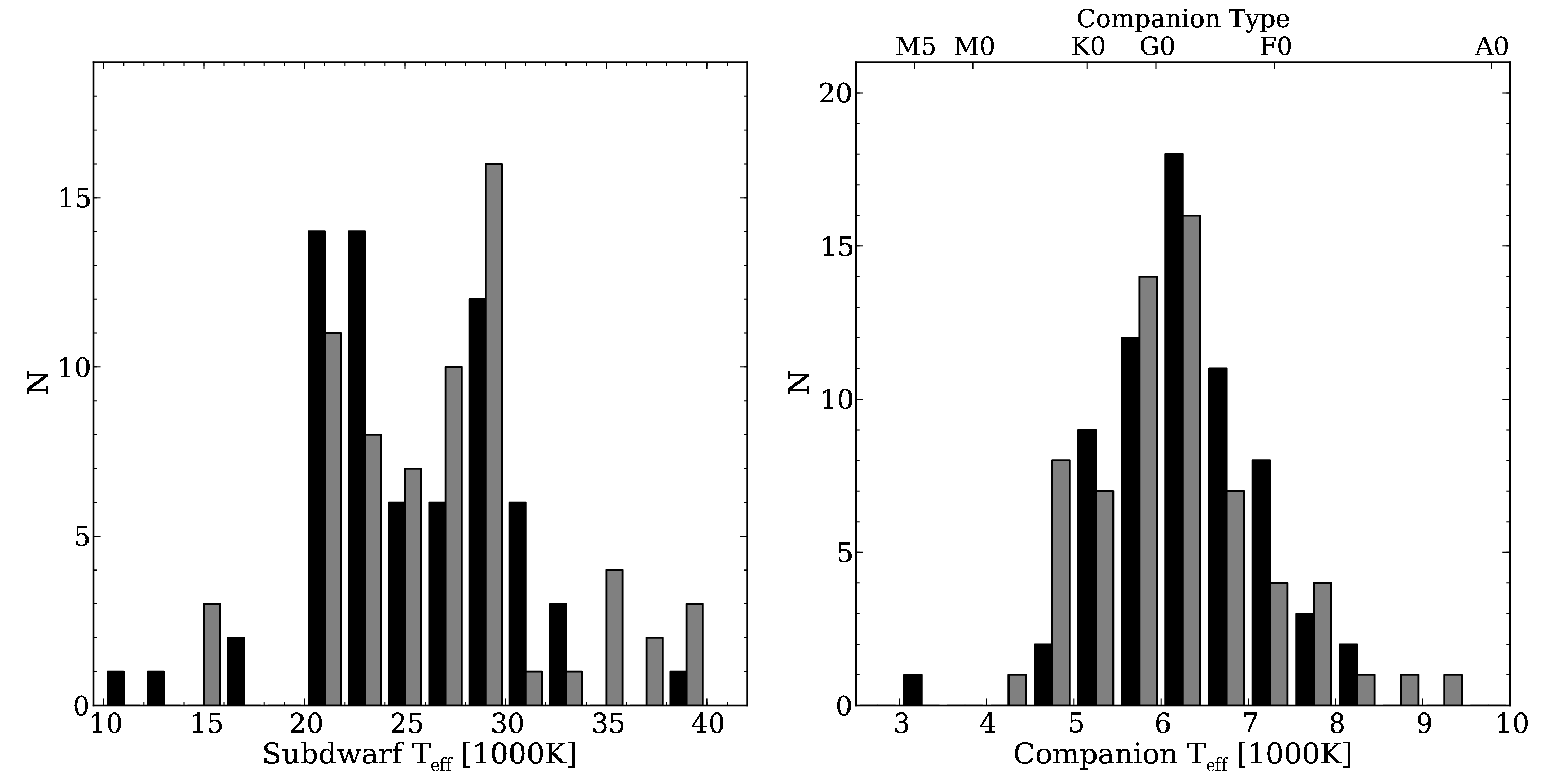}
  \caption{\label{f-hist1} Distributions of the subdwarf (left) and
companion (right) effective temperatures calculated from the fitting
method described in Section\,\ref{s-fit} when applied to the
\textit{C2MS} sample. The grey and black histograms show the system
parameters when calculated with and without the (maximum) reddening
correction, respectively. Adjoining pairs of histogram show the number of
objects in the same bin. A total of 66 objects are included in the
histograms, where 27 objects that are known to be contaminants (from
their SIMBAD classification or their SDSS spectra), or the
subdwarf--companion model provides a bad fit (``Q''$\geq3$ in
Table\,\ref{t-obj1}), have been removed. The subdwarf effective
temperature histogram is grouped in bins of $2,000$\,K and the companion
star histogram uses bins of $500$\,K.}
 \end{minipage}
\end{figure*}

The distribution of companion effective temperature in
Figure\,\ref{f-hist1} ramps up from early spectral types towards
$\sim$\,G0, as might be expected from the initial mass function (IMF). On
the other hand, the subsequent turn over and drop towards mid-K-type may
be a product of our selection biases.
A $15,000$\,K subdwarf with a M0-type companion saturates in SDSS at
$d\leq1.7$\,kpc and is too faint for 2MASS at $d\geq1.7$\,kpc
(Table\,\ref{t-dlim}). Therefore we are not sensitive to all subdwarfs
with M0-type companions. The best way to reduce such biases and test our
completion is by probing to fainter $K_s$-band magnitudes. This was the
key motivation behind our second sample, using  \textit{SU} which extends
several magnitudes deeper and reaches $K\sim17.8$, though at the
expense of limited sky coverage.

\begin{table*}
 \begin{minipage}{2\columnwidth}
  \begin{center}
   \caption{\label{t-dlim} Limitations on the distance of subdwarf plus
main--sequence star candidates caused by the relative magnitude cuts.
This is calculated for $15,000$, $20,000$, $30,000$ and $40,000$\,K
subdwarfs and companions with effective temperatures of $3,250$\,K (M0),
$5,000$\,K (K0), $7,250$\,K (F0). The important limits considered are;
the saturation of \rsdss\ at 14.1 (therefore a minimum distance), the cut
made on \rcmc\ at 16.0 (therefore a maximum distance), the $K_s$-band
magnitude limit of 2MASS at 14.3 and the $K_s$-band magnitude limit of
UKIDSS at 17.8.}
  \begin{tabular}{llcccccc}
   \hline \hline
sdB \Teff & MS \Teff & Abs & \multicolumn{2}{c}{d (kpc)} & Abs &
\multicolumn{2}{c}{d (kpc)} \\
(K)       & (K)      & $r$ & \rsdss=14.1 & \rcmc=16.0    & $K$ &
$Ks$=14.3 & $K$=17.8        \\ \hline
$15,000$ & $7,250$ & 2.2 & 2.4 & 5.8 & 1.9 & 3.1 & 15.6 \\
         & $5,000$ & 2.9 & 1.7 & 4.1 & 2.8 & 2.0 & 9.9 \\
         & $3,250$ & 3.0 & 1.7 & 4.0 & 3.2 & 1.7 & 8.5 \\
$20,000$ & $7,250$ & 2.5 & 2.1 & 5.0 & 2.0 & 2.8 & 14.3 \\
         & $5,000$ & 3.6 & 1.3 & 3.0 & 3.4 & 1.5 & 7.7 \\
         & $3,250$ & 3.7 & 1.2 & 2.8 & 4.0 & 1.1 & 5.8 \\
$30,000$ & $7,250$ & 2.8 & 1.8 & 4.4 & 2.2 & 2.7 & 13.5 \\
         & $5,000$ & 5.0 & 0.7 & 1.6 & 3.9 & 1.2 & 6.0 \\
         & $3,250$ & 5.5 & 0.5 & 1.2 & 5.3 & 0.6 & 3.1 \\
$40,000$ & $7,250$ & 2.7 & 1.9 & 4.5 & 2.2 & 2.7 & 13.5 \\
         & $5,000$ & 4.6 & 0.8 & 1.9 & 3.9 & 1.2 & 6.2 \\
         & $3,250$ & 5.0 & 0.7 & 1.6 & 5.1 & 0.7 & 3.5 \\
   \hline
   \end{tabular}
  \end{center}
 \end{minipage}
\end{table*}

\subsubsection{Distance distribution}
\label{sss-dist_d}

\begin{figure*}
 \begin{minipage}{2\columnwidth}
  \includegraphics[width=0.49\columnwidth]{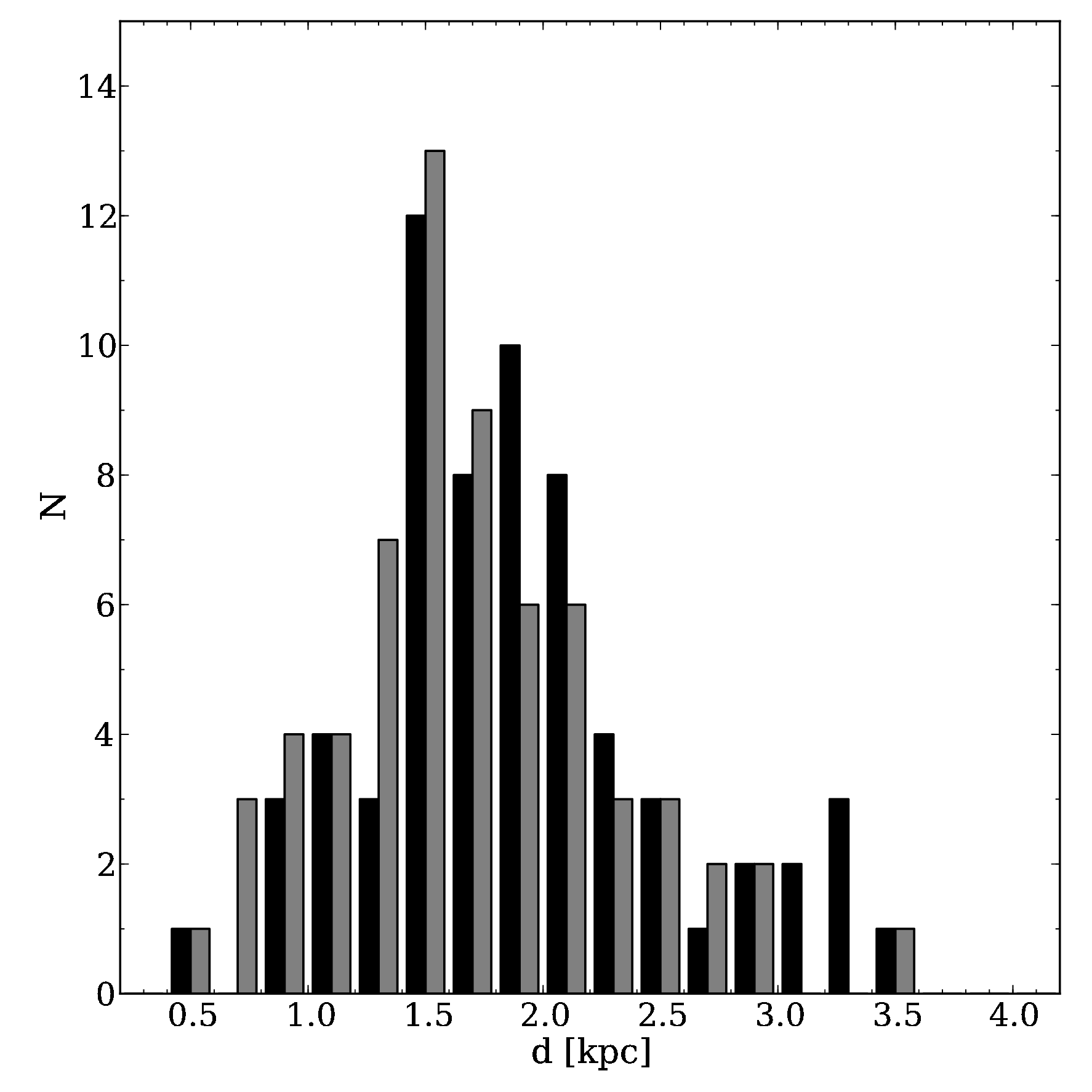}
  \includegraphics[width=0.49\columnwidth]{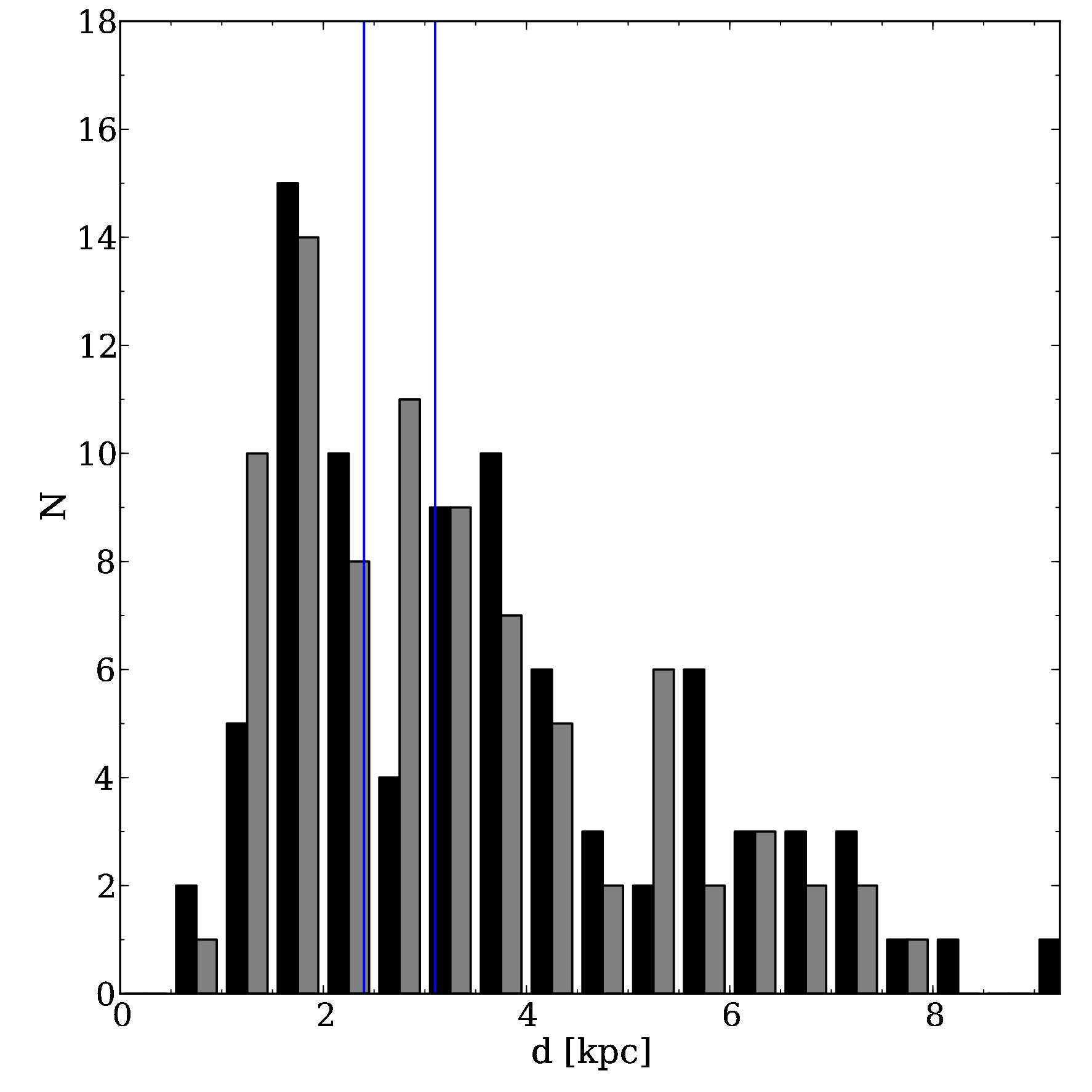}
  \caption{\label{f-histd} Distribution of the distance to the
subdwarf-companion star systems as calculated from the fitting method
described in Section\,\ref{s-fit} when applied to the \textit{C2MS}
(left) and \textit{SU} samples (right). The grey and black histograms
show the system parameters when calculated with and without the reddening
correction, respectively. Adjoining pairs of histogram show the number of
objects in the same bin. Objects that are known to be contaminants have
been removed. A total of 66 objects are included in the \textit{C2MS}
histogram (left), where 27 objects that are known to be contaminants
(from their SIMBAD classification or their SDSS spectra), or the
subdwarf--companion model provides a bad fit (``Q''$\geq3$ in
Table\,\ref{t-obj1}~or~\ref{t-obj2}), have been removed. In the
\textit{SU} histogram, 84 objects are included, where 50 have been
removed. The distances are given in kpc and the bin sizes are $0.2$ and
$0.5$\,kpc for the \textit{C2MS} and \textit{SU} samples, respectively.
The vertical lines in the right hand plot show the region where the
volume limit sample is defined ($2.4$ to $3.1$\,kpc).}
 \end{minipage}
\end{figure*}

The calculated distance distribution seen in Figure\,\ref{f-histd} shows
a rapid increase towards $\sim2$kpc, followed by an extended tail.
As we discussed previously, the limitations on distance due to our
magnitude cuts and limits are important and are a complex function of
subdwarf effective temperature and companion type (Table\,\ref{t-dlim}).
There are no clean regions where all temperatures and companion types are
sampled evenly to give a complete, volume-limited sample.
If one assumes that all subdwarfs (independent of temperature and
companion type) are drawn from same parent distance distribution, and we
select each subdwarf--companion system with equal probability, the
distribution shown in Figure\,\ref{f-histd} would represent the true
distance distribution. Therefore we would be relatively confident that
the peak appears at $1.5-2.0$kpc. However, Table\,\ref{t-dlim} does show
that for some combinations of subdwarf and companion temperature we are
no longer complete at this peak distance. Here again the deeper
\textit{SU} sample can provide us with a more complete sample.

\subsection{Distribution of fits -- \textit{SU} sample}
\label{ss-dist2}

\begin{figure*}
 \begin{minipage}{2\columnwidth}
  \includegraphics[width=\columnwidth]{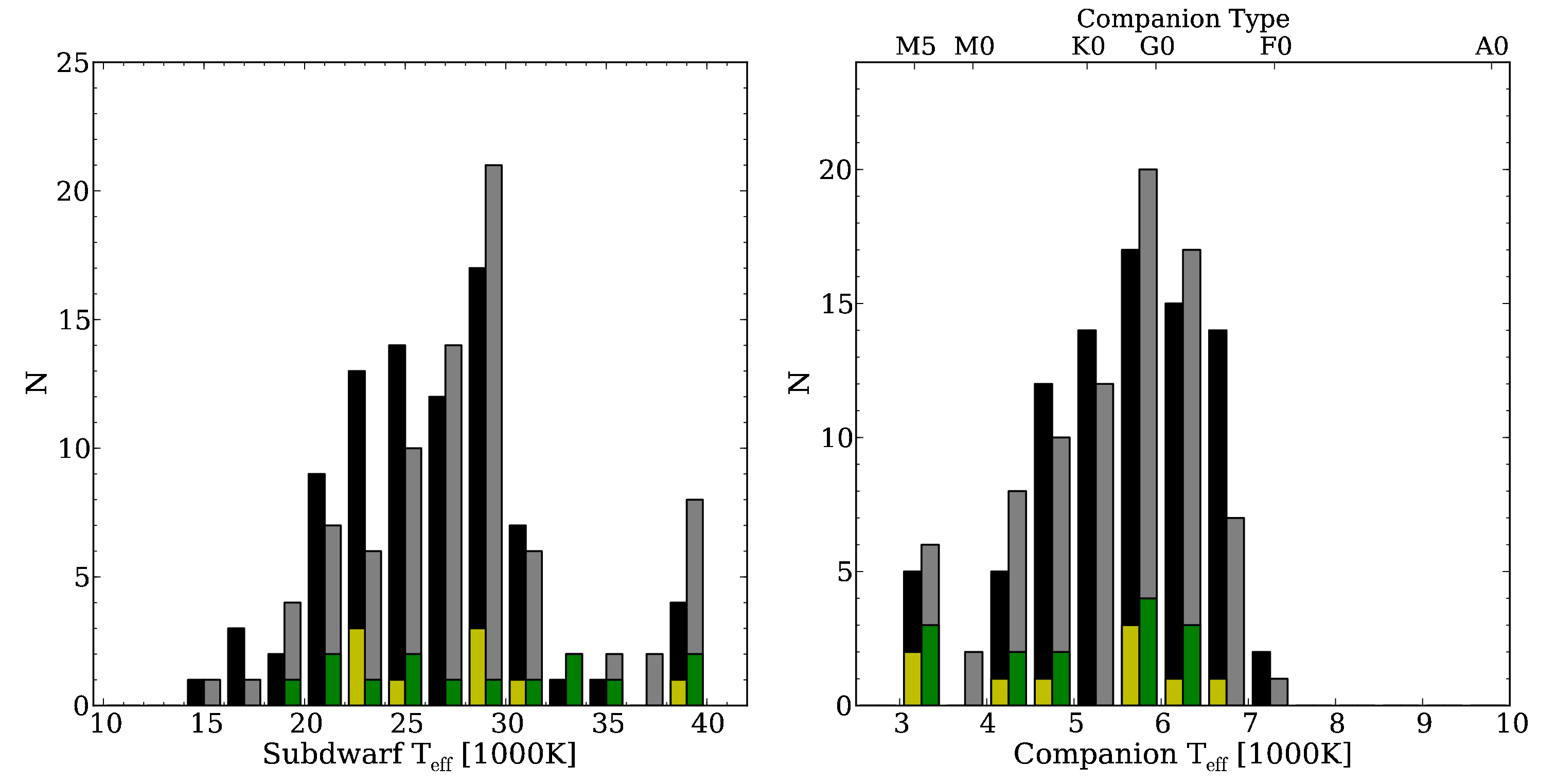}
  \caption{\label{f-hist3} Distributions of the subdwarf (left) and
companion (right) effective temperatures calculated from the fitting
method described in Section\,\ref{s-fit} when applied to the \textit{SU}
sample. The grey and black histograms show the system parameters when
calculated with and without the reddening correction, respectively.
Adjoining pairs of histogram bars show the number of objects in the same
bin. A total of 84 objects are included in the histograms, where
50 objects that are known to be contaminants (from their SIMBAD
classification or their SDSS spectra), or the subdwarf--companion model
provides a bad fit (``Q''$\geq3$ in Table\,\ref{t-obj1}), have been
removed.
Similarly, the green and yellow histograms shows distributions (with and
without reddening corrections) when the distances to the objects are
limited to be between $2.4$ and $3.1$kpc, such that the histogram is a
volume limited sample. 11 and 16 objects are included in the histograms,
respectively. The subdwarf effective temperature histogram is grouped in
bins of $2,000$\,K and the companion star histogram uses bins of
$500$\,K.}
 \end{minipage}
\end{figure*}

The corresponding distributions of the subdwarf and companion effective
temperature for the \textit{SU} sample are shown in Figure\,\ref{f-hist3}
with the distribution of distances in Figure\,\ref{f-histd}. The subdwarf
effective temperature distribution is broadly consistent with that of the
\textit{C2MS} sample, with most subdwarf temperatures between
$20,000-30,000$\,K. It is also similar to that shown for uncontaminated
sdBs by \citet[][Figure\,1]{greenetal06-1}.
To establish a volume--limited sample, we again refer to
Table\,\ref{t-dlim} where we contrast the impact of the 2MASS versus
UKIDSS $K$-band limits. The distances sampled are significantly larger,
though as before dependent on subdwarf and companion temperature.
Overall, the \textit{SU} sample should be less biased against finding
lower temperature subdwarfs compared to the \textit{C2MS} sample (see the
second example given in Section\,\ref{sss-dist_comp}; a $15,000$\,K
subdwarf with a M0-type companion would now be detected to $8.5$\,kpc).
This does not appear to have increased the numbers of low temperature
subdwarfs found and thus it appears that their absence is not due to our
sample biases, but represents an intrinsic deficit of cool subdwarfs
within the subdwarf population.
Accounting for reddening (as seen in the grey histogram) does not have a
large effect, although it shifts the calculated subdwarf effective
temperatures systematically higher by $1,000-2,000$\,K.

Comparing the distribution of companion effective temperatures to the
\textit{C2MS} sample, the \textit{SU} sample has a larger number of
objects with early M-type companions. Hence, the \textit{SU} sample
overcomes the main limitation found within the \textit{C2MS} sample, the
shallow $K_s$-band data. The increased depth of UKIDSS allows us to probe
significantly more systems with M-type companions, however we still see a
deficit compared with K-type companions and earlier. This also seems
obvious from the lack of systems populating the subdwarf plus M-type
companion region of the colour-colour diagram in Figure\,\ref{f-rrK}.
Selecting subdwarfs with companions later than $\sim$M5-type is still
limited by the colour selection method as discussed previously
(Section\,\ref{ss-cuts}). Probing deeper in the $K$-band does not help
for companion types later than $\sim$M5.
Accounting for reddening has a complementary effect to that on effective
temperature. As the subdwarfs become hotter, the required companion also
shifts to higher temperatures.

We searched for a correlation between subdwarf and companion effective
temperatures, but none was found at a level above the parameter
uncertainties. Better statistics, from larger samples, are needed to
investigate the subtleties of population.

Overall, when considering confirmed subdwarf systems, we believe that the
fitting method is producing temperatures accurate to within a few
thousand Kelvin and companion temperatures to within several hundred
Kelvin (a few spectral types). There is some disagreement between
individual fit results when compared with the literature. However, our
principal goal is not to achieve superior parameters for individual
systems. Indeed, more data are required to accurately establish
parameters for individual systems.
Our method does appear to be efficient in finding composite subdwarf
binaries, while our SED fitting is accurate enough to allow us to
consider the broad statistical parameter distributions within our
samples. There will be some influence from contaminants. However, the
numbers of contaminants are a relatively small fraction
(Table\,\ref{t-box}) and wherever possible they have been removed from
the distributions.

\subsection{A volume-limited sample}

The advantage of using the \textit{SU} sample is that significantly
larger distances are probed. Referring to Table\,\ref{t-dlim}, the sample
is complete for F0 to M0-type companions over distances of $2.4$ to
$3.1$kpc. Although we can therefore construct a volume limited sample in
this region, only 14 objects with good quality fits to the SED fall
within this region (Figure\,\ref{f-hist3}). It is impossible to draw
meaningful conclusions about parameter distributions for such a small
sample.

Following the assumptions described in Section\,\ref{sss-dist_d}, and
thus assuming the distribution seen in Figure\,\ref{f-hist3} is
representative of the true distance distribution, the peak at $2\pm1$kpc
in our distance distribution may then be associated with representing the
spatial distribution of the bulk of the subdwarf binaries.
Then, assuming the distribution follows a simple disk population of the
form $r^2\mathrm{exp}(-r/H)$, where $r$ is the distance from the center
of the disk and $H$ is the scale height, the turning point in a distance
histogram should represent $2H$. Therefore, the scale height of the
subdwarf population in the \textit{SU} sample is $1\pm0.5$kpc.

\section{Discussion}
\label{s-dis}

Existing samples of subdwarfs have shown that a substantial fraction of
them reside in binaries. \citet{hanetal03-1} used population synthesis
models to calculate that the intrinsic binary fraction should be
$76-89$\,per\,cent.
Our samples explicitly target composite systems and thus should be
dominated by subdwarfs with bound binary companions.
\citet{heber09-1} states that the vast majority of subdwarfs have a
temperature between $20,000-40,000$\,K. The temperature distribution
found here appears approximately consistent with this range, however we
do find a sub-sample of cooler subdwarfs with temperatures below
$20,000$\,K. In the \textit{SU} sample, where sample biases against
cooler subdwarfs are smallest, they make up $\sim5-10$\,per\,cent.
This is true whether or not we account for the full
\citet{schlegeletal98-1} reddening value, and thus cannot be an artifact
due to reddening. Fitting of the ultraviolet part of the SED is
especially important for calculating reliable subdwarf effective
temperature, because this is the region where the subdwarf dominates.

Utilising the SED from the ultraviolet down to the infrared, we have a
large range over which both the subdwarf and the companion can dominate a
region of the spectrum. We show that both samples here are sensitive to
companions of spectral type A5 to M5 for $20,000$ to $35,000$\,K subdwarf
effective temperatures and F0 to K0-type if $15,000$ and $40,000$\,K
subdwarfs are included. These ranges can be seen visually in
Figure\,\ref{f-hist}. Many subdwarfs are found to indeed have companions
in this regime.
In the \textit{C2MS} sample (Figure\,\ref{f-hist1}), the distribution of
companion type is seen to be a broad peak from F-type companions to
$\sim$\,K0-type. A significant turnover is then seen towards late K and
M-type companions. this can be explained, for the \textit{C2MS} sample,
because we are only sensitive to these systems over a very small distance
range. However, the \textit{SU} sample extends several magnitudes deeper
in the $K$-band and therefore removes this bias, but still shows a clear
deficit of early M-type companions. This is contrary to the relative
abundance of late type companions found in many previous surveys.
If the M-type companions were only paired with cool subdwarfs, they would
not have been selected by the colour cuts, but this is not consistent
with the results of the radial velocity studies. It therefore appears
that subdwarfs with F, G and K-type main--sequence companions are
intrinsically much more common than those with lower mass M-type
main--sequence companions, for a broad range of subdwarf temperatures
(subject to the colour selections described in Section\,\ref{ss-cuts}).

The population synthesis models of \citet{hanetal03-1} predict that a
significant fraction of subdwarfs will form through a channel involving
stable Roche lobe overflow. These are expected to be $\sim20,000$\,K
subdwarfs with $\sim$F0 or K0-type companions close to the main sequence
\citep[see Figure\,15\,\&\,19 of][]{hanetal03-1}. It is believed that
these have not been found previously because of the ``GK selection
effect'' \citep{hanetal03-1}, where subdwarfs with F, G and K-type
companions were not targeted by the PG survey because they would show
composite spectra (features such as the Ca~II~K line and the G-band).
However, \citet{wadeetal06-1} and \citet{wadeetal09-1} find that only
$\sim3$\,per\,cent of the rejected PG stars show indications of being a
subdwarf with a companion. The majority are (single) metal-poor F stars.
Here we are primarily selecting subdwarfs with F to K-type companions
and therefore we would be sensitive to this peak. 
We do indeed find a significant fraction of subdwarfs with effective
temperatures around $20,000$\,K and some objects below $20,000$\,K.  We
do not see the RLOF systems dominate quite as strongly as they do in
\citet{hanetal03-1}. However, too many cool subdwarfs are found here to
be appropriate for creation solely through the first common envelope
ejection channel (peaking at $\sim30,000$\,K) and thus the RLOF channel
appears to be a significant contributor.

Part of the \citet{liskeretal05-1} SPY survey sample looked at objects
with composite spectra. They do not find a clear contribution from cool
subdwarfs. The SPY survey does, however, suffer from strong pre-selection
biases. The majority of targets were selected from the Hamburg/ESO survey
\citep{friedrichetal00-1} and required not to show evidence of a
companion in the low resolution prism spectroscopy.
The companion types that we are finding in this study also appear to
broadly match the predictions of \citet{hanetal03-1}. The first stable
RLOF channel is very efficient at producing F to K-type companions.
Between F and K-type companions, \citet{hanetal03-1} predict
$\sim$F0-type companions to be the most prevalent (by a factor of
$\sim3$), followed by very few $\sim$G0-type companions, and then two
smaller peaks of approximately equal amplitude at $\sim$\,K0 and
$\sim$M0-type \citep[see Figure\,15 of][]{hanetal03-1}.
Our distribution does not show the feature at F0, but we may not be
sensitive enough to F0-type companions, especially in composite systems
with low temperature subdwarfs. Our colour cuts only select $15,000$\,K
subdwarfs with F0 or later type companions and therefore we may not show
the main peak at $\sim$F0-type (right hand panels of
Figures\,\ref{f-hist1} and \ref{f-hist3}), if it is indeed there.
Equally, most of the K-type companions to subdwarfs predicted by
\citet{hanetal03-1} are evolved and luminous. Therefore they would not be
selected in our colour cuts because the luminosity of the companion would
dominate the subdwarf. If any of these objects are selected, they will be
fitted photometrically as a much hotter companion than K-type, therefore
enhancing the F0-type peak or broadening it. Thus we should not detect
the peak at K0.

In a simulation, we took a theoretical sample of subdwarfs with
companions that matched the distributions from \citet{hanetal03-1} and
used the surface gravities discussed in Section\,\ref{s-fit}. We then
applied the magnitude and colour cuts relevant for the \textit{C2MS} and
\textit{SU} samples. The objects which satisfy these criteria do show a
similar distribution in effective temperature and companion type to that
seen in the real samples, again suggesting that our observed samples are
in broad agreement with the model populations of \citet{hanetal03-1}.

More recently, \citet{clausenetal12-1} present independent population
synthesis calculations of subdwarfs. In their Figure\,13, the
distribution of companion effective temperature is shown using a variety
of input model parameters. Run 6 is the most comparable to the
distribution from \citet{hanetal03-1} in terms of input parameters. In
this run, and the majority of others, \citet{clausenetal12-1} predict a
vast majority of M-type or later companions to the subdwarfs. This does
not agree with our samples, which show a lack of M-type companions and a
significant proportion of K-types. 
This suggests that observational samples such as those presented here
have the ability to directly constrain binary population synthesis
models.

The scale height of subdwarfs is rarely discussed. We used the two
samples here to estimate the scale height at $1\pm0.5$kpc from the peak
in their distance distributions near $2\pm1$kpc (Figure\,\ref{f-histd}).
However, to do so we must assume that the each subdwarf plus companion
system (independent of system parameters) is drawn from the same parent
distance distribution, and we pick each of these with the same frequency.
If the scale height is $\sim1$kpc, it is therefore most consistent with
the Galactic thick disk scale height \citep[e.g.
$0.75\pm0.07$kpc,][]{dejongetal10-1}. If the subdwarf population was
associated with the thin disk, a smaller scale height of $0.3$\,kpc would
be expected \citep{juricetal08-1}, while a rise towards $25$\,kpc would
have indicated a halo population \citep{dejongetal10-1}.
More accurate modelling of individual subdwarfs together with a larger
volume limited sample is required to study the distribution and reliably
quantify the scale height of the subdwarf population. Our methods are
well suited to offer such large samples as ongoing and near-future
surveys cover an increasing part of the sky.

\section{Conclusion}
\label{s-con}

We have developed a method to select hot subdwarfs stars with mid-M to
early-F-type near main--sequence companions using a combination of
ultraviolet, optical and infrared photometry. This selects a
complementary sample to those found from radial velocity surveys, which
typically limit themselves to objects with no obvious evidence for a
companion in the optical range. 
We applied this method to two samples, one selected from a match between
GALEX, CMC and 2MASS (covering a large area), and the other using GALEX,
SDSS and UKIDSS (probing deeper in the $K_s$-band and therefore further
away). We also use the SDSS for fitting in the \textit{C2MS} sample.

A significant number of subdwarfs with F to K-type companions were found
in both samples. The distributions are consistent with the systems being
produced, at least in a significant part, by the very efficient RLOF
channel \citep{hanetal03-1}. However, neither the predictions of
\citet{hanetal03-1} or \citet{clausenetal12-1} match the observed
distribution completely. We find that M-type companions are far less
prevalent than K-type systems. 

It is clear that, at least for a large
fraction of the subdwarf population, prior binary evolution plays an
important role. This group has largely gone unstudied previously. 
With future surveys such as the Southern SkyMapper project and VISTA, the
same procedure as carried out here can be applied to a large field in the
southern sky. This would find many more subdwarfs with early type
companions and allow for a thorough test of our understanding of the
prior binary evolutionary pathways required to form the large subdwarf
populations we see. Similarly, the Wide-field Infrared Survey Explorer
could be an excellent addition to this search, allowing us to probe for
fainter companions and covering the whole sky.

\section*{Acknowledgements}

This work makes use of data products from the Two Micron All Sky Survey,
which is a joint project of the University of Massachusetts and
IPAC/Caltech, funded by NASA and the NSF.
Funding for the Sloan Digital Sky Survey (SDSS) and SDSS-II has been
provided by the Alfred P. Sloan Foundation, the Participating
Institutions, the National Science Foundation, the U.S. Department of
Energy, the National Aeronautics and Space Administration, the Japanese
Monbukagakusho, and the Max Planck Society, and the Higher Education
Funding Council for England. The SDSS Web site is http://www.sdss.org/.
D. Steeghs acknowledges a STFC Advanced Fellowship. BTG and TRM were
supported under an STFC Rolling Grant to Warwick.

\begin{landscape}
\begin{table}
\begin{minipage}{\columnwidth}
\caption{\label{t-full_2m} Example of: Full list of objects from the
\textit{C2M} sample with magnitudes inside the cuts described in
Table\,\ref{t-sel}. Online-only Table.}
\begin{tabular}{llllllllll}
\hline \hline
Name & R.A. & Dec & \fuv & \nuv & \rcmc & $J$ & $H$ & $K_s$ & SIMBAD \\
\hline
\input{full_list_example}
\hline
\end{tabular}
\end{minipage}
\end{table}
\end{landscape}

\begin{landscape}
\begin{table}
\addtolength{\tabcolsep}{-1pt}
\begin{minipage}{\columnwidth}
\caption{\label{t-obj1} Subdwarf and companion effective temperatures,
and distance estimates for the \textit{C2MS} sample of subdwarf plus
companion star systems when fitted using the method described in
Section\,\ref{s-fit}.
A MS\,$\Teff=0$\,K corresponds to the best fit being a single subdwarf,
without the addition of a companion.
The ``\{'' notation is described in Section\,\ref{s-res} and does not
simply represent uncertainties. In all cases, a more realistic error on
the subdwarf temperatures is a few thousand Kelvin (see
Section\,\ref{ss-overlap}).
The ``E(B-V)'' column is the reddening according to the
\citet{schlegeletal98-1} map.
The ``Q'' (Quality) column values correspond to; 1:Good fit, 2:Average
fit, 3:Poor fit, 4:WD/WD+MS/CV and 5:Quasar/Galaxy. Values of three and
above are excluded from the histograms shown in Figure\,\ref{f-hist1},
\ref{f-histd} and \ref{f-hist3}. The classifications in this catagory
between values of 1, 2 and 3 are purely qualitative. The ``SDSS Spec''
column corresponds to the visual classification of the SDSS spectrum.
The ``Known Comp'' highlights objects which were known previously to be
composite subdwarf plus companion systems (1:~\citealt{fergusonetal84-1},
2:~\citealt{kilkennyetal88-1}, 3:~\citealt{allardetal94-2},
4:~\citealt{cuadrado+jeffery01-1}, 5:~\citealt{cuadrado+jeffery02-1},
6:~\citealt{liskeretal05-1}, 7:~\citealt{ostensen06-1},
8:~\citealt{stroeeretal07-1}). The final column shows objects also
included in Table\,\ref{t-ind1}, where ``pWD'' stands for possible white
dwarf, according to the classification of \citet{girvenetal11-1}.
}
\begin{tabular}{llllccccccclclll}
\hline \hline
 & & & & \multicolumn{3}{c}{No Correction} &
\multicolumn{3}{c}{Reddening Corrected} & & & & & & \\
 & & & & sdB \Teff & MS \Teff & d & sdB \Teff & MS \Teff & d & & & &
SDSS & Known & \\
Name & Identifier & R.A. & Dec & (1000\,K) & (1000\,K) & (kpc) &
(1000\,K) & (1000\,K) & (kpc) & E(B-V) & SIMBAD & Q & Spec
& Comp & Table\,\ref{t-ind1} \\
\hline
\input{obj_table}
\hline
\end{tabular}
\end{minipage}
\end{table}
\end{landscape}

\begin{landscape}
\begin{table}
\addtolength{\tabcolsep}{-1pt}
\begin{minipage}{\columnwidth}
\caption{\label{t-obj2} Subdwarf and companion effective temperatures,
and distance estimates for the \textit{SU} sample of subdwarf plus
companion star systems when fitted using the method described in
Section\,\ref{s-fit}, following the same format as Table\,\ref{t-obj1}.
A MS\,$\Teff=0$\,K corresponds to the best fit being a single subdwarf,
without the addition of a companion.
The ``\{'' notation is described in Section\,\ref{s-res} and does not
simply represent uncertainties. In all cases, a more realistic error on
the subdwarf temperatures is a few thousand Kelvin (see
Section\,\ref{ss-overlap}).
The final column shows objects also included in Table\,\ref{t-ind2}.
}
\begin{tabular}{llllccccccclclll}
\hline \hline
 & & & & \multicolumn{3}{c}{No Correction} &
\multicolumn{3}{c}{Reddening Corrected} & & & & & & \\
 & & & & sdB \Teff & MS \Teff & d & sdB \Teff & MS \Teff & d & & & &
SDSS & Known & \\
Name & Identifier & R.A. & Dec & (1000\,K) & (1000\,K) & (kpc) &
(1000\,K) & (1000\,K) & (kpc) & E(B-V) & SIMBAD & Q & Spec &
Comp & Table\,\ref{t-ind2} \\
\hline
\input{ukidss_obj_table}
\hline
\end{tabular}
\end{minipage}
\end{table}
\end{landscape}

\end{document}

%% file: full_list_example.tex
0004$+$2301 & 00:04:06.09 & $+$23:01:50.3 & $13.62\pm0.01$ & $14.33\pm0.01$ & $15.09$ & $14.58$ & $14.42$ & $14.42$ &  \\
0010$+$4313 & 00:10:00.55 & $+$43:13:18.9 & $16.44\pm0.03$ & $16.27\pm0.02$ & $15.14$ & $14.71$ & $14.66$ & $14.54$ &  \\
0016$+$3157 & 00:16:31.06 & $+$31:57:40.8 & $14.92\pm0.01$ & $15.31\pm0.01$ & $15.57$ & $15.08$ & $14.80$ & $14.65$ &  \\
0018$+$0101 & 00:18:43.50 & $+$01:01:25.5 & $13.43\pm0.01$ & $14.23\pm0.01$ & $15.11$ & $15.05$ & $14.88$ & $14.71$ & sdB \\
0031$-$2535 & 00:31:03.29 & $-$25:35:39.5 & $15.46\pm0.01$ & $15.56\pm0.01$ & $15.38\pm0.05$ & $14.78$ & $14.54$ & $14.52$ &  \\
0032$+$3714 & 00:32:31.93 & $+$37:14:54.3 & $15.49\pm0.01$ & $15.52\pm0.00$ & $15.34$ & $14.53$ & $14.37$ & $14.27$ &  \\
0040$-$0021 & 00:40:22.88 & $-$00:21:28.8 & $15.44\pm0.00$ & $15.28\pm0.00$ & $15.03\pm0.09$ & $14.90$ & $14.85$ & $14.70$ & WD \\
0041$+$3726 & 00:41:40.77 & $+$37:26:38.9 & $16.09\pm0.01$ & $15.96\pm0.00$ & $14.78$ & $14.20$ & $14.06$ & $13.98$ &  \\
0046$+$4550 & 00:46:59.60 & $+$45:50:49.1 & $16.49\pm0.03$ & $16.60\pm0.02$ & $15.82\pm0.07$ & $14.95$ & $14.73$ & $14.67$ &  \\
0048$+$3856 & 00:48:57.39 & $+$38:56:28.0 & $16.93\pm0.01$ & $16.75\pm0.01$ & $15.33$ & $14.81$ & $14.75$ & $14.48$ &  \\
0050$+$4251 & 00:50:29.44 & $+$42:51:53.8 & $13.18\pm0.00$ & $13.79\pm0.00$ & $13.23$ & $12.50$ & $12.28$ & $12.24$ &  \\
0051$+$0921 & 00:51:26.89 & $+$09:21:32.6 & $13.73\pm0.01$ & $14.17\pm0.01$ & $14.35\pm0.06$ & $13.71$ & $13.50$ & $13.44$ & Var* \\
0053$+$2229 & 00:53:16.89 & $+$22:29:39.3 & $15.27\pm0.01$ & $15.56\pm0.01$ & $15.42\pm0.01$ & $14.83$ & $14.65$ & $14.43$ &  \\
0054$+$1508 & 00:54:11.12 & $+$15:08:19.5 & $16.47\pm0.01$ & $16.51\pm0.00$ & $15.29\pm0.04$ & $14.45$ & $14.29$ & $14.20$ &  \\
0057$+$3538 & 00:57:20.35 & $+$35:38:59.2 & $14.90\pm0.02$ & $15.02\pm0.01$ & $14.76\pm0.07$ & $14.07$ & $13.87$ & $13.88$ &  \\
0103$+$1332 & 01:03:41.71 & $+$13:32:48.9 & $13.37\pm0.01$ & $13.74\pm0.01$ & $13.20\pm0.03$ & $12.51$ & $12.31$ & $12.36$ &  \\
0107$+$3940 & 01:07:12.57 & $+$39:40:24.6 & $14.44\pm0.02$ & $14.48\pm0.01$ & $13.12\pm0.05$ & $12.30$ & $12.11$ & $12.09$ &  \\
0109$+$4203 & 01:09:16.13 & $+$42:03:04.8 & $13.60\pm0.01$ & $13.72\pm0.01$ & $13.41\pm0.04$ & $12.86$ & $12.69$ & $12.68$ &  \\
0115$+$1922 & 01:15:25.92 & $+$19:22:49.6 & $12.52\pm0.01$ & $12.85\pm0.00$ & $13.18\pm0.03$ & $12.66$ & $12.58$ & $12.58$ &  \\
0115$-$2406 & 01:15:47.49 & $-$24:06:50.9 & $15.12\pm0.02$ & $15.25\pm0.01$ & $14.65\pm0.01$ & $14.16$ & $14.00$ & $13.97$ & WD \\
0116$+$1317 & 01:16:44.63 & $+$13:17:42.9 & $14.92\pm0.01$ & $15.09\pm0.01$ & $14.22$ & $13.63$ & $13.51$ & $13.42$ &  \\
0121$+$4558 & 01:21:29.49 & $+$45:58:52.2 & $13.95\pm0.01$ & $14.41\pm0.01$ & $14.66$ & $13.86$ & $13.55$ & $13.47$ &  \\
0122$+$2150 & 01:22:06.25 & $+$21:50:18.1 & $15.68\pm0.02$ & $15.76\pm0.01$ & $14.60\pm0.03$ & $14.12$ & $14.03$ & $13.98$ &  \\
0129$+$3202 & 01:29:52.69 & $+$32:02:10.2 & $12.65\pm0.00$ & $13.07\pm0.00$ & $14.53$ & $14.42$ & $14.29$ & $14.25$ & Comp \\
0138$+$2430 & 01:38:08.67 & $+$24:30:13.8 & $15.05\pm0.01$ & $15.15\pm0.00$ & $15.25$ & $14.69$ & $14.46$ & $14.30$ &  \\
0138$+$0339 & 01:38:26.97 & $+$03:39:37.6 & $12.17\pm0.00$ & $12.18\pm0.00$ & $13.40\pm0.01$ & $12.67$ & $12.25$ & $12.19$ &  \\
0141$+$0614 & 01:41:39.91 & $+$06:14:37.3 & $16.59\pm0.04$ & $16.26\pm0.02$ & $15.11$ & $14.91$ & $14.84$ & $14.63$ & Nova \\
0143$+$3234 & 01:43:26.27 & $+$32:34:39.5 & $13.93\pm0.01$ & $14.17\pm0.01$ & $15.47\pm0.07$ & $15.42$ & $15.42$ & $15.14$ &  \\
0147$+$3032 & 01:47:10.65 & $+$30:32:15.0 & $14.38\pm0.01$ & $14.28\pm0.01$ & $14.79$ & $14.71$ & $14.66$ & $14.77$ &  \\
0147$-$2156 & 01:47:21.84 & $-$21:56:51.7 & $16.40\pm0.02$ & $15.65\pm0.01$ & $15.28\pm0.01$ & $14.92$ & $14.45$ & $14.34$ & DA \\
0149$-$2741 & 01:49:30.81 & $-$27:41:59.6 & $16.69\pm0.01$ & $16.38\pm0.01$ & $15.01\pm0.04$ & $15.10$ & $14.55$ & $14.05$ & Galaxy \\
0151$+$4631 & 01:51:27.57 & $+$46:31:22.0 & $14.19\pm0.01$ & $14.69\pm0.01$ & $14.13$ & $13.50$ & $13.31$ & $13.30$ &  \\
0152$-$1913 & 01:52:30.93 & $-$19:13:02.9 & $11.75\pm0.00$ & $13.04\pm0.00$ & $14.22$ & $14.02$ & $13.89$ & $13.96$ &  \\
0204$+$2729 & 02:04:47.13 & $+$27:29:03.6 & $12.65\pm0.01$ & $13.26\pm0.00$ & $14.02$ & $13.51$ & $13.28$ & $13.27$ &  \\
0208$+$4712 & 02:08:01.24 & $+$47:12:59.5 & $15.10\pm0.01$ & $15.24\pm0.01$ & $14.40$ & $13.67$ & $13.46$ & $13.46$ &  \\
0209$-$1955 & 02:09:24.50 & $-$19:55:16.3 & $14.73\pm0.01$ & $14.92\pm0.01$ & $14.33$ & $13.74$ & $13.65$ & $13.51$ &  \\
0210$+$0830 & 02:10:21.88 & $+$08:30:59.0 & $13.41\pm0.01$ & $13.76\pm0.01$ & $13.49$ & $12.83$ & $12.68$ & $12.65$ &  \\
0211$+$2851 & 02:11:55.12 & $+$28:51:05.3 & $12.38\pm0.01$ & $12.41\pm0.00$ & $11.55\pm0.02$ & $10.91$ & $10.79$ & $10.72$ &  \\
0217$+$0906 & 02:17:52.30 & $+$09:06:02.7 & $14.32\pm0.01$ & $14.87\pm0.01$ & $14.78\pm0.04$ & $14.03$ & $13.78$ & $13.88$ & Comp \\
0218$+$1831 & 02:18:15.64 & $+$18:31:37.7 & $11.65\pm0.01$ & $12.93\pm0.01$ & $13.62$ & $13.68$ & $13.71$ & $13.76$ &  \\
0219$+$0150 & 02:19:02.46 & $+$01:50:57.1 & $14.81\pm0.01$ & $14.56\pm0.01$ & $14.20\pm0.04$ & $14.04$ & $13.91$ & $13.84$ &  \\
0220$+$0635 & 02:20:48.95 & $+$06:35:13.0 & $14.74\pm0.01$ & $15.03\pm0.01$ & $14.49$ & $13.76$ & $13.55$ & $13.40$ &  \\
0221$-$0713 & 02:21:57.84 & $-$07:13:11.8 & $14.08\pm0.01$ & $14.36\pm0.01$ & $14.51$ & $13.88$ & $13.73$ & $13.71$ &  \\
0224$+$2340 & 02:24:45.41 & $+$23:40:47.4 & $15.70\pm0.03$ & $15.84\pm0.02$ & $14.45\pm0.02$ & $13.58$ & $13.38$ & $13.37$ &  \\
0230$+$4209 & 02:30:31.41 & $+$42:09:30.9 & $15.11\pm0.02$ & $15.08\pm0.01$ & $14.54\pm0.07$ & $13.93$ & $13.74$ & $13.71$ &  \\
0234$+$2534 & 02:34:15.15 & $+$25:34:45.2 & $14.84\pm0.00$ & $15.04\pm0.00$ & $13.79\pm0.04$ & $12.94$ & $12.74$ & $12.71$ &  \\
0241$+$4117 & 02:41:24.63 & $+$41:17:49.3 & $14.34\pm0.01$ & $14.27\pm0.01$ & $13.25$ & $12.72$ & $12.63$ & $12.62$ &  \\
0245$-$1242 & 02:45:53.34 & $-$12:42:21.2 & $13.26\pm0.01$ & $14.00\pm0.01$ & $15.14$ & $14.34$ & $13.90$ & $13.59$ &  \\

%% file: obj_table.tex
0018$+$0101 & HE\,0016$+$0044 & 00:18:43.50 & $+$01:01:25.5 & $40\{_{39}^{40}$ & $5.50\{_{5.25}^{5.75}$ & $1.5\{_{1.4}^{1.5}$ & $40\{_{39}^{40}$ & $5.50\{_{5.25}^{5.75}$ & $1.4\{_{1.4}^{1.5}$ & $0.029$ & sdB & 2 &  &  & SD \\
0040$-$0021 & PG\,0037$-$006 & 00:40:22.88 & $-$00:21:28.8 & $14\{_{13}^{15}$ & $12.50\{_{12.25}^{12.75}$ & $6.8\{_{6.6}^{7.7}$ & $14\{_{13}^{15}$ & $12.75\{_{12.50}^{13.00}$ & $6.9\{_{6.7}^{7.8}$ & $0.020$ & WD & 4 & WD &  &  \\
0054$+$1508 &  & 00:54:11.12 & $+$15:08:19.5 & $21\{_{20}^{22}$ & $7.25\{_{7.00}^{7.50}$ & $3.2\{_{3.2}^{4.8}$ & $21\{_{20}^{22}$ & $7.00\{_{6.75}^{7.25}$ & $2.9\{_{2.9}^{4.2}$ & $0.059$ &  & 2 &  &  &  \\
0138$+$2430 & PG\,0135$+$242 & 01:38:08.67 & $+$24:30:13.8 & $17\{_{16}^{18}$ & $5.50\{_{5.25}^{5.75}$ & $1.9\{_{1.9}^{3.0}$ & $21\{_{20}^{22}$ & $4.50\{_{4.25}^{4.75}$ & $1.0\{_{1.0}^{1.7}$ & $0.126$ &  & 1 &  &  &  \\
0141$+$0614 & HS\,0139$+$0559 & 01:41:39.91 & $+$06:14:37.3 & $12\{_{11}^{13}$ & $7.25\{_{7.00}^{7.50}$ & $4.8\{_{4.3}^{5.4}$ & $12\{_{11}^{13}$ & $6.75\{_{6.50}^{7.00}$ & $4.5\{_{3.9}^{5.0}$ & $0.048$ & NL & 1 &  &  & CV \\
0316$+$0042 & PG\,0313$+$005 & 03:16:20.12 & $+$00:42:22.3 & $28\{_{27}^{29}$ & $6.25\{_{6.00}^{6.50}$ & $2.2\{_{1.4}^{2.2}$ & $27\{_{26}^{28}$ & $6.00\{_{5.75}^{6.25}$ & $1.9\{_{1.8}^{2.1}$ & $0.087$ & WD & 1 & SD &  &  \\
0643$+$3744 &  & 06:43:03.41 & $+$37:44:14.7 & $22\{_{21}^{23}$ & $6.00\{_{5.75}^{6.25}$ & $1.9\{_{1.7}^{2.1}$ & $27\{_{26}^{28}$ & $5.75\{_{5.50}^{6.00}$ & $1.6\{_{1.5}^{1.8}$ & $0.140$ &  & 1 &  &  &  \\
0710$+$2938 &  & 07:10:29.29 & $+$29:38:52.3 & $21\{_{20}^{22}$ & $6.75\{_{6.50}^{7.00}$ & $1.5\{_{1.5}^{2.1}$ & $25\{_{24}^{26}$ & $6.75\{_{6.50}^{7.00}$ & $1.5\{_{1.3}^{1.6}$ & $0.074$ &  & 1 &  &  &  \\
0735$+$2012 &  & 07:35:46.24 & $+$20:12:35.6 & $21\{_{20}^{22}$ & $5.50\{_{5.25}^{5.75}$ & $2.1\{_{2.1}^{3.4}$ & $23\{_{21}^{24}$ & $5.25\{_{5.00}^{5.50}$ & $1.9\{_{1.7}^{2.1}$ & $0.041$ &  & 2 &  &  &  \\
0737$+$2642 &  & 07:37:12.24 & $+$26:42:25.3 & $25\{_{24}^{26}$ & $5.50\{_{5.25}^{5.75}$ & $1.6\{_{1.5}^{1.8}$ & $27\{_{26}^{28}$ & $5.25\{_{5.00}^{5.50}$ & $1.5\{_{1.4}^{1.6}$ & $0.039$ & WD & 1 & SD &  &  \\
0754$+$1822 &  & 07:54:04.24 & $+$18:22:40.4 & $22\{_{21}^{23}$ & $7.25\{_{7.00}^{7.50}$ & $3.4\{_{3.1}^{3.7}$ & $25\{_{24}^{26}$ & $7.50\{_{7.25}^{7.75}$ & $3.5\{_{3.3}^{3.8}$ & $0.045$ &  & 1 &  &  &  \\
0755$+$2128 &  & 07:55:49.51 & $+$21:28:18.0 & $17\{_{16}^{18}$ & $7.25\{_{7.00}^{7.50}$ & $2.3\{_{2.3}^{3.5}$ & $20\{_{19}^{21}$ & $7.75\{_{7.50}^{8.00}$ & $2.5\{_{1.7}^{2.5}$ & $0.065$ &  & 1 &  &  &  \\
0804$+$2250 &  & 08:04:20.93 & $+$22:50:18.0 & $37\{_{36}^{40}$ & $5.75\{_{5.50}^{6.00}$ & $1.5\{_{1.5}^{1.7}$ & $31\{_{30}^{32}$ & $4.50\{_{4.25}^{4.75}$ & $1.0\{_{0.9}^{1.0}$ & $0.048$ &  & 3 &  &  &  \\
0805$-$0741 &  & 08:05:16.32 & $-$07:41:50.6 & $29\{_{28}^{30}$ & $7.00\{_{6.75}^{7.25}$ & $2.4\{_{2.2}^{2.8}$ & $28\{_{27}^{29}$ & $7.00\{_{6.75}^{7.25}$ & $2.1\{_{1.7}^{2.2}$ & $0.117$ &  & 1 &  &  &  \\
0812$+$1911 &  & 08:12:56.86 & $+$19:11:57.9 & $14\{_{13}^{15}$ & $7.25\{_{7.00}^{7.50}$ & $4.8\{_{4.8}^{6.9}$ & $15\{_{14}^{16}$ & $7.00\{_{6.75}^{7.25}$ & $4.4\{_{4.0}^{4.8}$ & $0.035$ & CV & 4 & CV &  & CV \\
0814$+$2019 &  & 08:14:06.84 & $+$20:19:01.0 & $21\{_{20}^{22}$ & $5.50\{_{5.25}^{5.75}$ & $2.0\{_{2.0}^{3.2}$ & $23\{_{21}^{24}$ & $5.50\{_{5.25}^{5.75}$ & $2.1\{_{1.8}^{2.2}$ & $0.042$ &  & 1 & SD &  &  \\
0815$+$4740 & PG\,0812$+$478 & 08:15:48.88 & $+$47:40:40.4 & $40\{_{39}^{40}$ & $5.00\{_{4.75}^{5.25}$ & $1.4\{_{1.3}^{1.4}$ & $40\{_{39}^{40}$ & $4.75\{_{4.50}^{5.00}$ & $1.2\{_{1.2}^{1.3}$ & $0.067$ & WD & 2 &  &  &  \\
0818$-$0701 &  & 08:18:06.86 & $-$07:01:23.9 & $22\{_{21}^{23}$ & $7.75\{_{7.50}^{8.00}$ & $3.4\{_{3.1}^{3.6}$ & $14\{_{13}^{15}$ & $5.75\{_{5.50}^{6.00}$ & $2.4\{_{2.4}^{3.8}$ & $0.097$ &  & 1 &  &  &  \\
0820$+$1739 &  & 08:20:03.34 & $+$17:39:14.0 & $20\{_{19}^{21}$ & $6.75\{_{6.50}^{7.00}$ & $2.9\{_{1.8}^{2.9}$ & $20\{_{19}^{21}$ & $6.50\{_{6.25}^{6.75}$ & $2.6\{_{1.6}^{2.6}$ & $0.033$ &  & 1 & SD &  &  \\
0824$+$3028 & PG\,0821$+$306 & 08:24:34.03 & $+$30:28:54.6 & $21\{_{20}^{22}$ & $5.25\{_{5.00}^{5.50}$ & $1.4\{_{1.4}^{2.4}$ & $21\{_{20}^{22}$ & $5.00\{_{4.75}^{5.25}$ & $1.3\{_{1.3}^{2.2}$ & $0.044$ &  & 1 & SD &  &  \\
0825$+$2006 &  & 08:25:07.22 & $+$20:06:36.5 & $24\{_{23}^{26}$ & $6.00\{_{5.75}^{6.25}$ & $1.7\{_{1.5}^{2.0}$ & $27\{_{26}^{28}$ & $6.00\{_{5.75}^{6.25}$ & $1.7\{_{1.6}^{1.9}$ & $0.037$ &  & 1 &  &  &  \\
0825$+$1202 &  & 08:25:44.73 & $+$12:02:45.2 & $22\{_{21}^{23}$ & $8.25\{_{8.00}^{8.50}$ & $2.5\{_{2.3}^{2.7}$ & $26\{_{25}^{27}$ & $8.75\{_{8.50}^{9.00}$ & $2.7\{_{2.5}^{2.8}$ & $0.044$ &  & 1 &  &  &  \\
0825$+$1307 &  & 08:25:56.86 & $+$13:07:54.3 & $24\{_{23}^{25}$ & $5.00\{_{4.75}^{5.25}$ & $1.0\{_{1.0}^{1.1}$ & $27\{_{26}^{28}$ & $5.00\{_{4.75}^{5.25}$ & $1.1\{_{1.0}^{1.1}$ & $0.034$ &  & 2 &  &  &  \\
0829$+$2246 &  & 08:29:02.64 & $+$22:46:37.6 & $26\{_{24}^{27}$ & $6.00\{_{5.75}^{6.25}$ & $2.7\{_{2.4}^{3.0}$ & $28\{_{26}^{29}$ & $6.00\{_{5.75}^{6.25}$ & $2.7\{_{1.8}^{2.7}$ & $0.036$ &  & 1 & SD &  &  \\
0833$-$0006 &  & 08:33:37.88 & $-$00:06:21.4 & $29\{_{28}^{30}$ & $7.25\{_{7.00}^{7.50}$ & $3.1\{_{2.9}^{3.6}$ & $29\{_{28}^{30}$ & $7.00\{_{6.75}^{7.25}$ & $2.5\{_{2.3}^{3.0}$ & $0.041$ &  & 2 &  &  &  \\
0844$+$3102 & PG\,0841$+$312 & 08:44:08.18 & $+$31:02:09.3 & $22\{_{21}^{23}$ & $4.75\{_{4.50}^{5.00}$ & $1.0\{_{0.9}^{1.1}$ & $22\{_{21}^{23}$ & $4.75\{_{4.50}^{5.00}$ & $1.0\{_{0.9}^{1.1}$ & $0.049$ &  & 1 &  &  &  \\
0849$+$1337 &  & 08:49:51.40 & $+$13:37:00.4 & $21\{_{20}^{22}$ & $6.75\{_{6.50}^{7.00}$ & $2.4\{_{2.4}^{3.6}$ & $21\{_{20}^{22}$ & $6.50\{_{6.25}^{6.75}$ & $2.0\{_{2.0}^{3.1}$ & $0.040$ &  & 2 &  &  &  \\
0907$+$2739 &  & 09:07:34.26 & $+$27:39:03.4 & $21\{_{20}^{22}$ & $5.50\{_{5.25}^{5.75}$ & $2.5\{_{2.5}^{4.0}$ & $21\{_{20}^{22}$ & $5.50\{_{5.25}^{5.75}$ & $2.3\{_{2.3}^{3.8}$ & $0.026$ & WD & 3 &  &  &  \\
0923$+$0652 &  & 09:23:58.62 & $+$06:52:18.3 & $29\{_{28}^{30}$ & $6.75\{_{6.50}^{7.00}$ & $2.1\{_{2.0}^{2.9}$ & $29\{_{28}^{30}$ & $6.25\{_{6.00}^{6.50}$ & $1.5\{_{1.4}^{2.2}$ & $0.054$ &  & 1 &  &  & pWD \\
0924$+$2035 & PG\,0921$+$208 & 09:24:05.20 & $+$20:35:46.8 & $19\{_{18}^{20}$ & $4.25\{_{3.50}^{4.50}$ & $1.6\{_{1.5}^{1.8}$ & $19\{_{18}^{20}$ & $3.00\{_{3.00}^{0.00}$ & $1.6\{_{1.4}^{1.7}$ & $0.041$ &  & 3 &  &  &  \\
0929$+$0603 &  & 09:29:20.48 & $+$06:03:47.1 & $21\{_{20}^{30}$ & $5.75\{_{5.50}^{6.00}$ & $1.6\{_{1.5}^{2.5}$ & $29\{_{28}^{30}$ & $5.50\{_{5.25}^{5.75}$ & $1.2\{_{1.2}^{2.0}$ & $0.052$ &  & 2 &  &  &  \\
0935$+$1621 & PG\,0932$+$166 & 09:35:41.37 & $+$16:21:11.0 & $30\{_{29}^{31}$ & $5.00\{_{4.75}^{5.25}$ & $1.0\{_{0.9}^{1.0}$ & $29\{_{28}^{30}$ & $4.75\{_{4.50}^{5.00}$ & $0.9\{_{0.9}^{1.4}$ & $0.033$ &  & 1 &  &  &  \\
0937$+$0813 & PG\,0935$+$084 & 09:37:40.95 & $+$08:13:20.5 & $23\{_{22}^{24}$ & $6.00\{_{5.75}^{6.25}$ & $2.0\{_{1.8}^{2.3}$ & $23\{_{22}^{25}$ & $5.75\{_{5.50}^{6.00}$ & $1.8\{_{1.6}^{2.0}$ & $0.042$ & sdB & 1 & SD &  &  \\
0941$+$0657 & PG\,0939$+$072 & 09:41:59.35 & $+$06:57:17.2 & $21\{_{20}^{22}$ & $6.25\{_{6.00}^{6.50}$ & $1.7\{_{1.6}^{2.5}$ & $21\{_{20}^{22}$ & $6.00\{_{5.75}^{6.25}$ & $1.4\{_{1.4}^{2.1}$ & $0.040$ & WD & 1 &  &  &  \\
0958$+$2236 &  & 09:58:15.97 & $+$22:36:04.2 & $33\{_{32}^{34}$ & $6.75\{_{6.50}^{7.00}$ & $1.8\{_{1.7}^{2.0}$ & $34\{_{33}^{35}$ & $6.75\{_{6.50}^{7.00}$ & $1.8\{_{1.7}^{2.0}$ & $0.033$ &  & 1 &  &  &  \\
1003$+$3716 & PG\,1000$+$375 & 10:03:19.69 & $+$37:16:35.1 & $30\{_{29}^{31}$ & $5.50\{_{5.25}^{5.75}$ & $1.2\{_{1.1}^{1.3}$ & $29\{_{28}^{31}$ & $5.25\{_{5.00}^{5.50}$ & $1.1\{_{1.1}^{1.7}$ & $0.016$ & WD & 1 & SD &  &  \\
1005$+$4317 &  & 10:05:05.07 & $+$43:17:36.5 & $29\{_{28}^{30}$ & $6.75\{_{6.50}^{7.00}$ & $1.9\{_{1.8}^{2.5}$ & $29\{_{28}^{30}$ & $6.75\{_{6.50}^{7.00}$ & $1.9\{_{1.7}^{2.4}$ & $0.012$ &  & 1 &  &  &  \\
1015$-$0308 & SW\,Sex & 10:15:09.39 & $-$03:08:32.3 & $18\{_{17}^{19}$ & $7.00\{_{6.75}^{7.25}$ & $2.6\{_{2.3}^{2.9}$ & $20\{_{19}^{21}$ & $7.00\{_{6.75}^{7.25}$ & $2.6\{_{1.7}^{2.6}$ & $0.033$ & NL & 1 &  &  & CV \\
\hline
\end{tabular}
\end{minipage}
\end{table}
\end{landscape}

\begin{landscape}
\begin{table}
\addtolength{\tabcolsep}{-1pt}
\begin{minipage}{\columnwidth}
\contcaption{}
\begin{tabular}{llllccccccclclll}
\hline\hline
 & & & & \multicolumn{3}{c}{No Correction} &
\multicolumn{3}{c}{Reddening Corrected} & & & & & & \\
 & & & & sdB \Teff & MS \Teff & d & sdB \Teff & MS \Teff & d & & & &
SDSS & Known & \\
Name & Identifier & R.A. & Dec & (1000\,K) & (1000\,K) & (kpc) &
(1000\,K) & (1000\,K) & (kpc) & E(B-V) & SIMBAD & Q & Spec
& Comp & Table\,\ref{t-ind1} \\
\hline
1018$+$0721 &  & 10:18:01.55 & $+$07:21:24.4 & $29\{_{28}^{30}$ & $6.00\{_{5.75}^{6.25}$ & $1.9\{_{1.9}^{3.0}$ & $30\{_{29}^{32}$ & $5.75\{_{5.50}^{6.00}$ & $1.6\{_{1.5}^{1.8}$ & $0.027$ &  & 3 & SD &  &  \\
1018$+$0953 &  & 10:18:33.15 & $+$09:53:36.0 & $28\{_{27}^{29}$ & $5.75\{_{5.00}^{6.00}$ & $1.6\{_{1.0}^{1.6}$ & $29\{_{28}^{30}$ & $4.50\{_{4.25}^{4.75}$ & $0.8\{_{0.8}^{1.3}$ & $0.037$ & WD & 1 &  &  &  \\
1027$+$2409 & PG\,1025$+$244 & 10:27:51.19 & $+$24:09:17.0 & $24\{_{23}^{27}$ & $6.00\{_{5.75}^{6.50}$ & $1.9\{_{1.7}^{2.4}$ & $26\{_{25}^{28}$ & $6.25\{_{6.00}^{6.50}$ & $2.1\{_{1.9}^{2.4}$ & $0.017$ &  & 1 & SD &  &  \\
1049$+$1842 & PG\,1046$+$189 & 10:49:33.53 & $+$18:42:41.5 & $20\{_{19}^{21}$ & $5.75\{_{5.50}^{6.00}$ & $2.1\{_{1.3}^{2.1}$ & $20\{_{19}^{21}$ & $5.50\{_{4.75}^{5.75}$ & $2.0\{_{1.2}^{2.0}$ & $0.033$ &  & 1 &  &  &  \\
1100$-$2113 & EC\,10583$-$2057 & 11:00:46.69 & $-$21:13:12.3 & $30\{_{22}^{31}$ & $6.25\{_{6.00}^{6.50}$ & $2.1\{_{2.0}^{2.4}$ & $35\{_{32}^{36}$ & $5.75\{_{5.50}^{6.00}$ & $1.4\{_{1.2}^{1.5}$ & $0.053$ &  & 1 &  &  &  \\
1102$+$2616 &  & 11:02:11.09 & $+$26:16:46.3 & $22\{_{21}^{23}$ & $6.25\{_{6.00}^{6.50}$ & $2.0\{_{1.8}^{2.2}$ & $21\{_{20}^{22}$ & $6.00\{_{5.75}^{6.25}$ & $1.7\{_{1.7}^{2.6}$ & $0.019$ &  & 1 & SD &  &  \\
1113$+$0413 & PG\,1110$+$045 & 11:13:17.31 & $+$04:13:14.7 & $29\{_{28}^{30}$ & $4.75\{_{4.50}^{5.00}$ & $0.9\{_{0.9}^{1.4}$ & $29\{_{28}^{30}$ & $4.50\{_{4.25}^{4.75}$ & $0.8\{_{0.8}^{1.3}$ & $0.051$ &  & 1 &  & 2,7 &  \\
1131$+$0932 & PG\,1128$+$098 & 11:31:14.37 & $+$09:32:20.4 & $38\{_{37}^{39}$ & $5.75\{_{5.50}^{6.00}$ & $1.2\{_{1.1}^{1.2}$ & $40\{_{39}^{40}$ & $5.75\{_{5.50}^{6.00}$ & $1.1\{_{1.1}^{1.2}$ & $0.039$ &  & 2 & SD &  &  \\
1149$+$2231 & PG\,1146$+$228 & 11:49:00.50 & $+$22:31:05.9 & $23\{_{22}^{25}$ & $5.25\{_{5.00}^{5.50}$ & $1.4\{_{1.3}^{1.5}$ & $24\{_{23}^{25}$ & $5.25\{_{5.00}^{5.50}$ & $1.4\{_{1.3}^{1.5}$ & $0.022$ &  & 1 & SD &  &  \\
1203$+$0909 & PG\,1200$+$094 & 12:03:19.46 & $+$09:09:51.6 & $27\{_{25}^{28}$ & $5.75\{_{5.50}^{6.00}$ & $1.5\{_{1.3}^{1.6}$ & $28\{_{26}^{29}$ & $5.75\{_{5.50}^{6.00}$ & $1.5\{_{1.0}^{1.5}$ & $0.020$ &  & 1 &  &  &  \\
1212$+$4240 & PG\,1210$+$429 & 12:12:38.56 & $+$42:40:02.1 & $23\{_{22}^{24}$ & $5.75\{_{5.50}^{6.00}$ & $1.5\{_{1.4}^{1.7}$ & $24\{_{23}^{26}$ & $5.75\{_{5.50}^{6.00}$ & $1.5\{_{1.4}^{1.7}$ & $0.015$ &  & 1 & SD & 1,2,7 & SD \\
1233$+$0834 &  & 12:33:09.68 & $+$08:34:34.1 & $30\{_{29}^{31}$ & $6.00\{_{5.75}^{6.25}$ & $1.9\{_{1.7}^{2.1}$ & $29\{_{28}^{30}$ & $5.75\{_{5.50}^{6.00}$ & $1.6\{_{1.6}^{2.5}$ & $0.019$ &  & 2 & SD &  &  \\
1316$+$4359 & PG\,1314$+$442 & 13:16:33.00 & $+$43:59:04.9 & $28\{_{26}^{29}$ & $5.25\{_{5.00}^{5.50}$ & $1.7\{_{1.1}^{1.7}$ & $28\{_{27}^{29}$ & $5.00\{_{4.75}^{5.25}$ & $1.6\{_{1.0}^{1.6}$ & $0.021$ &  & 1 &  &  &  \\
1325$+$1212 & PG\,1323$+$125 & 13:25:57.21 & $+$12:12:20.6 & $26\{_{25}^{27}$ & $5.75\{_{5.50}^{6.00}$ & $2.1\{_{1.9}^{2.3}$ & $27\{_{26}^{28}$ & $5.75\{_{5.50}^{6.00}$ & $2.1\{_{1.9}^{2.3}$ & $0.034$ &  & 1 &  &  &  \\
1326$+$0357 & PG\,1323$+$042 & 13:26:19.95 & $+$03:57:54.3 & $24\{_{23}^{25}$ & $5.00\{_{4.75}^{5.25}$ & $1.5\{_{1.4}^{1.7}$ & $26\{_{25}^{27}$ & $5.00\{_{4.75}^{5.25}$ & $1.6\{_{1.5}^{1.7}$ & $0.025$ & sdO & 2 & SD &  &  \\
1402$+$3215 &  & 14:02:32.86 & $+$32:15:21.5 & $22\{_{21}^{23}$ & $6.25\{_{6.00}^{6.50}$ & $1.9\{_{1.7}^{2.1}$ & $22\{_{21}^{23}$ & $6.00\{_{5.75}^{6.25}$ & $1.7\{_{1.5}^{1.9}$ & $0.015$ &  & 1 & SD &  &  \\
1404$+$2450 & PG\,1402$+$251 & 14:04:29.98 & $+$24:50:20.6 & $27\{_{26}^{28}$ & $6.25\{_{6.00}^{6.50}$ & $1.8\{_{1.6}^{2.0}$ & $27\{_{26}^{28}$ & $6.00\{_{5.75}^{6.25}$ & $1.6\{_{1.5}^{1.7}$ & $0.017$ &  & 1 &  &  &  \\
1407$+$3103 &  & 14:07:47.63 & $+$31:03:18.3 & $20\{_{19}^{21}$ & $5.25\{_{5.00}^{5.50}$ & $1.5\{_{0.9}^{1.5}$ & $20\{_{19}^{21}$ & $4.75\{_{4.50}^{5.00}$ & $1.4\{_{0.8}^{1.4}$ & $0.011$ &  & 1 &  &  &  \\
1421$+$0753 & KN\,Boo & 14:21:38.21 & $+$07:53:20.9 & $27\{_{26}^{28}$ & $5.25\{_{5.00}^{5.50}$ & $1.6\{_{1.5}^{1.7}$ & $29\{_{28}^{30}$ & $4.50\{_{4.25}^{4.75}$ & $0.9\{_{0.9}^{1.5}$ & $0.028$ & sdB & 1 & SD &  &  \\
1502$-$0245 & PG\,1459$-$026 & 15:02:12.13 & $-$02:45:56.7 & $24\{_{22}^{25}$ & $6.25\{_{6.00}^{6.50}$ & $1.8\{_{1.5}^{1.9}$ & $24\{_{23}^{26}$ & $6.00\{_{5.75}^{6.25}$ & $1.5\{_{1.4}^{1.7}$ & $0.124$ &  & 1 & SD &  &  \\
1517$+$0310 & PG\,1514$+$034 & 15:17:14.30 & $+$03:10:27.6 & $40\{_{39}^{40}$ & $6.00\{_{5.75}^{6.25}$ & $1.1\{_{1.0}^{1.1}$ & $39\{_{38}^{40}$ & $5.75\{_{5.50}^{6.00}$ & $0.9\{_{0.9}^{1.0}$ & $0.039$ & WD & 1 &  & 1,7 & SD \\
1518$+$2019 & PG\,1516$+$205 & 15:18:38.81 & $+$20:19:47.0 & $24\{_{23}^{26}$ & $6.00\{_{5.75}^{6.25}$ & $1.5\{_{1.4}^{1.7}$ & $26\{_{25}^{27}$ & $6.00\{_{5.75}^{6.25}$ & $1.5\{_{1.4}^{1.7}$ & $0.051$ &  & 1 &  &  &  \\
1524$+$0134 &  & 15:24:03.04 & $+$01:34:21.3 & $27\{_{26}^{28}$ & $6.25\{_{6.00}^{6.50}$ & $1.3\{_{1.2}^{1.4}$ & $28\{_{27}^{29}$ & $6.00\{_{5.75}^{6.25}$ & $1.1\{_{0.7}^{1.1}$ & $0.065$ &  & 1 &  &  &  \\
1528$+$1300 &  & 15:28:33.90 & $+$13:00:57.2 & $40\{_{39}^{40}$ & $6.00\{_{5.75}^{6.25}$ & $1.3\{_{1.3}^{1.3}$ & $40\{_{39}^{40}$ & $5.75\{_{5.50}^{6.00}$ & $1.1\{_{1.1}^{1.1}$ & $0.040$ &  & 3 &  &  &  \\
1530$+$1204 &  & 15:30:05.00 & $+$12:04:02.0 & $11\{_{11}^{12}$ & $8.00\{_{7.50}^{8.50}$ & $4.8\{_{4.5}^{5.5}$ & $14\{_{13}^{21}$ & $9.25\{_{8.75}^{9.75}$ & $4.8\{_{3.7}^{6.1}$ & $0.038$ &  & 2 &  &  &  \\
1542$+$0056 &  & 15:42:18.31 & $+$00:56:12.6 & $29\{_{28}^{30}$ & $6.50\{_{6.25}^{6.75}$ & $1.5\{_{1.4}^{2.1}$ & $21\{_{20}^{22}$ & $6.25\{_{6.00}^{6.50}$ & $1.3\{_{1.3}^{2.0}$ & $0.098$ &  & 1 &  &  &  \\
1602$+$0725 & PG\,1559$+$076 & 16:02:08.96 & $+$07:25:10.8 & $32\{_{30}^{35}$ & $5.25\{_{5.00}^{5.50}$ & $0.9\{_{0.8}^{1.1}$ & $37\{_{36}^{38}$ & $5.25\{_{5.00}^{5.50}$ & $1.0\{_{0.9}^{1.0}$ & $0.047$ &  & 1 &  &  &  \\
1603$+$0954 &  & 16:03:24.51 & $+$09:54:42.9 & $29\{_{28}^{30}$ & $5.75\{_{5.50}^{6.00}$ & $2.0\{_{2.0}^{3.2}$ & $32\{_{29}^{33}$ & $5.75\{_{5.50}^{6.00}$ & $2.0\{_{1.7}^{2.1}$ & $0.059$ &  & 3 & SD &  &  \\
1610$+$3450 &  & 16:10:40.72 & $+$34:50:44.1 & $29\{_{28}^{30}$ & $7.50\{_{7.25}^{7.75}$ & $2.0\{_{2.0}^{2.5}$ & $31\{_{30}^{32}$ & $7.50\{_{7.25}^{7.75}$ & $2.0\{_{1.9}^{2.1}$ & $0.018$ &  & 1 &  &  &  \\
1618$+$2141 &  & 16:18:06.46 & $+$21:41:25.0 & $22\{_{21}^{23}$ & $6.00\{_{5.75}^{6.25}$ & $1.6\{_{1.4}^{1.7}$ & $24\{_{21}^{25}$ & $6.00\{_{5.50}^{6.25}$ & $1.5\{_{1.2}^{1.7}$ & $0.069$ &  & 1 &  &  &  \\
1619$+$1453 & PG\,1617$+$150 & 16:19:49.35 & $+$14:53:09.5 & $29\{_{28}^{30}$ & $6.00\{_{5.75}^{6.25}$ & $1.3\{_{1.3}^{1.9}$ & $29\{_{28}^{30}$ & $5.75\{_{5.50}^{6.00}$ & $1.1\{_{1.1}^{1.7}$ & $0.051$ &  & 1 &  &  &  \\
1629$+$0026 &  & 16:29:06.76 & $+$00:26:19.9 & $29\{_{28}^{30}$ & $7.25\{_{7.00}^{7.50}$ & $2.5\{_{2.4}^{3.0}$ & $32\{_{22}^{33}$ & $7.25\{_{7.00}^{7.50}$ & $2.4\{_{2.2}^{2.6}$ & $0.095$ &  & 1 &  &  &  \\
1640$+$3842 & PN\,G061.9$+$41.3 & 16:40:18.20 & $+$38:42:20.6 & $15\{_{14}^{16}$ & $3.00\{_{3.00}^{4.25}$ & $1.8\{_{1.6}^{2.0}$ & $16\{_{15}^{17}$ & $3.00\{_{3.00}^{3.25}$ & $1.8\{_{1.1}^{1.8}$ & $0.012$ & PN & 3 &  &  &  \\
1709$+$4054 & PG\,1708$+$409 & 17:09:59.23 & $+$40:54:49.5 & $26\{_{25}^{28}$ & $5.50\{_{5.25}^{5.75}$ & $1.7\{_{1.6}^{1.9}$ & $28\{_{27}^{29}$ & $5.50\{_{5.25}^{5.75}$ & $1.8\{_{1.1}^{1.8}$ & $0.029$ & WD & 1 &  &  & SD \\
1710$+$2238 &  & 17:10:36.45 & $+$22:38:07.4 & $31\{_{30}^{32}$ & $7.25\{_{7.00}^{7.50}$ & $1.4\{_{1.3}^{1.5}$ & $34\{_{33}^{35}$ & $7.50\{_{7.25}^{7.75}$ & $1.4\{_{1.4}^{1.5}$ & $0.058$ &  & 1 &  &  &  \\
1734$+$3213 &  & 17:34:49.30 & $+$32:13:43.5 & $23\{_{22}^{24}$ & $6.00\{_{5.75}^{6.25}$ & $1.7\{_{1.6}^{1.9}$ & $25\{_{24}^{26}$ & $6.00\{_{5.75}^{6.25}$ & $1.7\{_{1.5}^{1.9}$ & $0.054$ &  & 1 &  &  &  \\
1822$+$4320 &  & 18:22:42.87 & $+$43:20:37.4 & $31\{_{30}^{32}$ & $3.00\{_{3.00}^{3.25}$ & $0.5\{_{0.5}^{0.6}$ & $40\{_{39}^{40}$ & $0.00\{_{25.00}^{0.00}$ & $0.6\{_{0.6}^{10.7}$ & $0.050$ &  & 1 &  &  &  \\
1834$+$4237 &  & 18:34:14.46 & $+$42:37:27.3 & $22\{_{21}^{23}$ & $6.25\{_{6.00}^{6.50}$ & $2.7\{_{2.4}^{3.0}$ & $23\{_{22}^{25}$ & $6.00\{_{5.75}^{6.25}$ & $2.3\{_{2.1}^{2.6}$ & $0.055$ &  & 3 &  &  &  \\
2020$+$0704 &  & 20:20:27.24 & $+$07:04:14.5 & $23\{_{22}^{24}$ & $5.75\{_{5.50}^{6.00}$ & $1.1\{_{1.0}^{1.2}$ & $38\{_{37}^{39}$ & $4.50\{_{4.25}^{4.75}$ & $0.6\{_{0.5}^{0.6}$ & $0.143$ &  & 1 &  &  &  \\
2023$+$1230 &  & 20:23:14.11 & $+$12:30:56.9 & $21\{_{20}^{22}$ & $6.00\{_{5.75}^{6.25}$ & $1.7\{_{1.7}^{2.6}$ & $35\{_{34}^{36}$ & $4.75\{_{4.50}^{5.00}$ & $0.8\{_{0.7}^{0.8}$ & $0.123$ &  & 2 &  &  &  \\
2047$-$0542 &  & 20:47:42.36 & $-$05:42:32.0 & $22\{_{21}^{23}$ & $6.00\{_{5.75}^{6.25}$ & $1.6\{_{1.5}^{1.8}$ & $24\{_{23}^{25}$ & $5.75\{_{5.50}^{6.00}$ & $1.4\{_{1.3}^{1.5}$ & $0.050$ &  & 1 &  &  &  \\
2052$-$0457 &  & 20:52:26.19 & $-$04:57:46.0 & $29\{_{28}^{30}$ & $6.00\{_{5.75}^{6.25}$ & $1.2\{_{1.1}^{1.7}$ & $37\{_{36}^{38}$ & $6.25\{_{6.00}^{6.50}$ & $1.3\{_{1.2}^{1.5}$ & $0.104$ &  & 1 &  &  &  \\
2056$+$0425 &  & 20:56:19.33 & $+$04:25:23.6 & $22\{_{21}^{23}$ & $6.50\{_{6.25}^{6.75}$ & $3.0\{_{2.7}^{3.4}$ & $21\{_{20}^{22}$ & $6.00\{_{5.75}^{6.25}$ & $2.2\{_{2.1}^{3.3}$ & $0.092$ &  & 1 &  &  &  \\
2105$+$1635 &  & 21:05:15.38 & $+$16:35:18.3 & $21\{_{20}^{22}$ & $6.00\{_{5.75}^{6.25}$ & $1.8\{_{1.8}^{2.7}$ & $30\{_{29}^{31}$ & $4.75\{_{4.50}^{5.00}$ & $0.8\{_{0.8}^{0.9}$ & $0.075$ &  & 3 &  &  &  \\
2117$-$0015 &  & 21:17:15.90 & $-$00:15:47.7 & $13\{_{12}^{14}$ & $6.75\{_{6.50}^{7.00}$ & $3.5\{_{2.3}^{3.5}$ & $14\{_{13}^{22}$ & $12.75\{_{12.50}^{13.00}$ & $5.4\{_{4.9}^{6.0}$ & $0.055$ &  & 1 &  &  & pWD \\
2117$-$0006 &  & 21:17:42.22 & $-$00:06:19.9 & $21\{_{20}^{22}$ & $6.25\{_{6.00}^{6.50}$ & $2.0\{_{1.9}^{2.9}$ & $22\{_{21}^{23}$ & $6.00\{_{5.75}^{6.25}$ & $1.6\{_{1.5}^{1.8}$ & $0.074$ &  & 1 & SD &  & pWD \\
\hline
\end{tabular}
\end{minipage}
\end{table}
\end{landscape}

\begin{landscape}
\begin{table}
\addtolength{\tabcolsep}{-1pt}
\begin{minipage}{\columnwidth}
\contcaption{}
\begin{tabular}{llllccccccclclll}
\hline\hline
 & & & & \multicolumn{3}{c}{No Correction} &
\multicolumn{3}{c}{Reddening Corrected} & & & & & & \\
 & & & & sdB \Teff & MS \Teff & d & sdB \Teff & MS \Teff & d & & & &
SDSS & Known & \\
Name & Identifier & R.A. & Dec & (1000\,K) & (1000\,K) & (kpc) &
(1000\,K) & (1000\,K) & (kpc) & E(B-V) & SIMBAD & Q & Spec
& Comp & Table\,\ref{t-ind1} \\
\hline
2129$+$0045 &  & 21:29:06.05 & $+$00:45:09.6 & $25\{_{24}^{26}$ & $5.50\{_{5.25}^{5.75}$ & $2.0\{_{1.9}^{2.2}$ & $29\{_{28}^{30}$ & $4.75\{_{4.50}^{5.00}$ & $1.2\{_{1.2}^{1.9}$ & $0.044$ & WD & 1 & SD &  &  \\
2129$+$1039 &  & 21:29:29.11 & $+$10:39:09.9 & $21\{_{20}^{23}$ & $5.50\{_{5.25}^{5.75}$ & $1.5\{_{1.5}^{2.4}$ & $25\{_{24}^{26}$ & $5.50\{_{5.25}^{5.75}$ & $1.5\{_{1.4}^{1.6}$ & $0.064$ &  & 1 &  &  &  \\
2135$+$2026 &  & 21:35:51.03 & $+$20:26:45.3 & $21\{_{20}^{22}$ & $5.75\{_{5.50}^{6.00}$ & $1.8\{_{1.8}^{2.8}$ & $22\{_{21}^{23}$ & $5.00\{_{4.75}^{5.25}$ & $1.4\{_{1.3}^{1.5}$ & $0.124$ &  & 1 &  &  &  \\
2138$+$0442 & PG\,2135$+$045 & 21:38:00.77 & $+$04:42:11.5 & $25\{_{24}^{26}$ & $5.00\{_{4.75}^{5.25}$ & $1.2\{_{1.1}^{1.3}$ & $27\{_{26}^{28}$ & $4.00\{_{3.75}^{4.25}$ & $1.0\{_{1.0}^{1.1}$ & $0.056$ &  & 1 &  & 3,4,5 & SD \\
2143$+$1244 &  & 21:43:54.65 & $+$12:44:58.3 & $30\{_{29}^{31}$ & $6.25\{_{6.00}^{6.50}$ & $2.9\{_{2.6}^{3.3}$ & $29\{_{28}^{30}$ & $5.75\{_{5.50}^{6.00}$ & $2.1\{_{2.0}^{3.1}$ & $0.095$ & CV & 4 & CV &  & CV \\
2147$-$0837 &  & 21:47:08.07 & $-$08:37:47.5 & $29\{_{28}^{30}$ & $6.50\{_{6.25}^{6.75}$ & $2.3\{_{2.1}^{3.1}$ & $29\{_{21}^{30}$ & $6.25\{_{6.00}^{6.50}$ & $1.9\{_{1.8}^{2.2}$ & $0.044$ &  & 1 & SD &  &  \\
2223$+$3850 &  & 22:23:26.78 & $+$38:50:16.7 & $32\{_{31}^{33}$ & $7.75\{_{7.50}^{8.00}$ & $2.9\{_{2.6}^{3.1}$ & $28\{_{27}^{29}$ & $8.00\{_{7.75}^{8.25}$ & $2.8\{_{2.2}^{2.9}$ & $0.108$ &  & 1 &  &  &  \\
2346$+$3657 &  & 23:46:21.39 & $+$36:57:27.6 & $29\{_{28}^{30}$ & $7.25\{_{7.00}^{7.50}$ & $2.5\{_{2.4}^{3.3}$ & $22\{_{21}^{23}$ & $6.75\{_{6.50}^{7.00}$ & $2.0\{_{1.8}^{2.2}$ & $0.142$ &  & 2 &  &  &  \\
2346$+$0344 &  & 23:46:55.71 & $+$03:44:29.4 & $30\{_{29}^{31}$ & $6.75\{_{6.50}^{7.00}$ & $1.7\{_{1.5}^{1.8}$ & $29\{_{28}^{30}$ & $6.50\{_{6.25}^{6.75}$ & $1.5\{_{1.4}^{2.2}$ & $0.054$ &  & 1 &  &  &  \\

%% file: ukidss_obj_table.tex
0018$+$0101 & HE\,0016$+$0044 & 00:18:43.52 & $+$01:01:23.6 & $24\{_{23}^{25}$ & $4.25\{_{4.00}^{4.50}$ & $1.2\{_{1.2}^{1.3}$ & $28\{_{27}^{29}$ & $4.75\{_{4.50}^{5.00}$ & $1.5\{_{1.0}^{1.6}$ & $0.029$ & sdB & 1 &  &  & SD \\
0032$+$0739 & PB\,6015 & 00:32:21.87 & $+$07:39:34.4 & $21\{_{20}^{22}$ & $5.25\{_{5.00}^{5.50}$ & $4.4\{_{4.4}^{6.9}$ & $21\{_{20}^{22}$ & $5.00\{_{4.75}^{5.25}$ & $4.1\{_{4.1}^{6.5}$ & $0.040$ & Comp & 3 &  &  & WDMS \\
0051$+$0059 &  & 00:51:49.65 & $+$00:59:50.7 & $27\{_{26}^{28}$ & $5.00\{_{4.75}^{5.25}$ & $4.5\{_{4.2}^{4.8}$ & $28\{_{27}^{29}$ & $5.00\{_{4.75}^{5.25}$ & $4.6\{_{2.8}^{4.6}$ & $0.028$ & WD & 5 &  &  &  \\
0054$+$1508 &  & 00:54:11.11 & $+$15:08:19.3 & $29\{_{28}^{30}$ & $7.25\{_{7.00}^{7.50}$ & $2.8\{_{2.7}^{3.3}$ & $29\{_{28}^{30}$ & $7.25\{_{7.00}^{7.50}$ & $2.7\{_{2.6}^{3.2}$ & $0.059$ &  & 3 &  &  &  \\
0141$+$0614 & HS\,0139$+$0559 & 01:41:39.92 & $+$06:14:37.6 & $14\{_{13}^{15}$ & $7.25\{_{7.00}^{7.50}$ & $3.7\{_{3.7}^{4.9}$ & $14\{_{13}^{15}$ & $7.00\{_{6.75}^{7.25}$ & $3.6\{_{3.5}^{4.7}$ & $0.048$ & NL & 2 &  &  & CV \\
0205$+$0712 & PB\,6645 & 02:05:59.87 & $+$07:12:39.9 & $25\{_{23}^{26}$ & $5.75\{_{5.50}^{6.00}$ & $7.3\{_{6.5}^{8.1}$ & $26\{_{24}^{28}$ & $5.75\{_{5.50}^{6.00}$ & $7.3\{_{6.5}^{8.3}$ & $0.056$ &  & 1 &  &  &  \\
0300$-$0023 & WD\,0257$-$005 & 03:00:24.57 & $-$00:23:42.1 & $38\{_{34}^{39}$ & $5.00\{_{4.75}^{5.25}$ & $3.2\{_{2.8}^{3.4}$ & $39\{_{38}^{40}$ & $4.50\{_{4.25}^{4.75}$ & $2.6\{_{2.6}^{2.7}$ & $0.120$ & Comp & 1 &  &  & WDMS \\
0316$+$0042 & PG\,0313$+$005 & 03:16:20.13 & $+$00:42:22.8 & $26\{_{25}^{27}$ & $6.00\{_{5.75}^{6.25}$ & $2.0\{_{1.8}^{2.2}$ & $27\{_{26}^{28}$ & $6.00\{_{5.75}^{6.25}$ & $2.0\{_{1.8}^{2.2}$ & $0.087$ & WD & 1 & SD &  &  \\
0737$+$2642 &  & 07:37:12.27 & $+$26:42:24.7 & $25\{_{24}^{26}$ & $5.50\{_{5.25}^{5.75}$ & $1.6\{_{1.5}^{1.8}$ & $26\{_{25}^{27}$ & $5.00\{_{4.75}^{5.25}$ & $1.4\{_{1.3}^{1.5}$ & $0.039$ & WD & 1 & SD &  &  \\
0744$+$2103 &  & 07:44:41.80 & $+$21:03:52.6 & $40\{_{39}^{40}$ & $4.75\{_{4.50}^{5.00}$ & $2.8\{_{2.7}^{3.0}$ & $39\{_{38}^{40}$ & $4.00\{_{3.75}^{4.25}$ & $2.3\{_{2.2}^{2.3}$ & $0.053$ &  & 1 &  &  &  \\
0752$+$2535 &  & 07:52:39.82 & $+$25:35:50.0 & $17\{_{16}^{18}$ & $4.25\{_{4.00}^{4.50}$ & $4.8\{_{4.8}^{7.8}$ & $11\{_{11}^{15}$ & $18.00\{_{17.00}^{19.00}$ & $36.9\{_{32.2}^{41.9}$ & $0.061$ & WD & 2 &  &  &  \\
0755$+$2128 &  & 07:55:49.49 & $+$21:28:18.5 & $21\{_{20}^{22}$ & $7.00\{_{6.75}^{7.25}$ & $2.0\{_{1.9}^{2.4}$ & $17\{_{16}^{18}$ & $6.50\{_{6.25}^{6.75}$ & $1.7\{_{1.7}^{2.4}$ & $0.065$ &  & 1 &  &  &  \\
0758$+$2818 &  & 07:58:13.60 & $+$28:18:16.0 & $23\{_{22}^{24}$ & $6.50\{_{6.25}^{6.75}$ & $3.6\{_{3.2}^{3.9}$ & $18\{_{17}^{19}$ & $6.25\{_{6.00}^{6.50}$ & $3.5\{_{3.1}^{3.9}$ & $0.039$ &  & 1 & SD &  &  \\
0809$+$1924 &  & 08:09:21.89 & $+$19:24:00.1 & $40\{_{39}^{40}$ & $0.00\{_{25.00}^{0.00}$ & $2.3\{_{2.2}^{36.6}$ & $40\{_{39}^{40}$ & $0.00\{_{25.00}^{0.00}$ & $2.0\{_{1.9}^{31.4}$ & $0.038$ &  & 3 & WD &  &  \\
0813$+$2813 &  & 08:13:52.02 & $+$28:13:17.3 & $20\{_{19}^{21}$ & $6.25\{_{6.00}^{6.50}$ & $6.4\{_{4.6}^{6.7}$ & $20\{_{19}^{21}$ & $6.00\{_{5.75}^{6.25}$ & $5.8\{_{4.1}^{6.0}$ & $0.032$ & CV & 4 & CV &  & CV \\
0814$+$2019 &  & 08:14:06.80 & $+$20:19:01.7 & $20\{_{19}^{21}$ & $6.25\{_{6.00}^{6.50}$ & $3.5\{_{2.7}^{3.7}$ & $20\{_{19}^{21}$ & $5.75\{_{5.50}^{6.00}$ & $2.9\{_{2.1}^{3.1}$ & $0.042$ &  & 1 & SD &  &  \\
0814$+$2811 &  & 08:14:53.92 & $+$28:11:22.5 & $22\{_{21}^{23}$ & $6.00\{_{5.75}^{6.25}$ & $3.7\{_{3.3}^{4.0}$ & $22\{_{21}^{23}$ & $5.75\{_{5.50}^{6.00}$ & $3.2\{_{2.9}^{3.5}$ & $0.030$ &  & 1 & SD &  & pWD \\
0829$+$2246 &  & 08:29:02.62 & $+$22:46:36.8 & $21\{_{20}^{22}$ & $5.25\{_{5.00}^{5.50}$ & $1.9\{_{1.8}^{2.6}$ & $22\{_{21}^{23}$ & $5.25\{_{5.00}^{5.50}$ & $1.9\{_{1.7}^{2.0}$ & $0.036$ &  & 1 & SD &  &  \\
0833$-$0006 &  & 08:33:37.87 & $-$00:06:21.4 & $29\{_{28}^{30}$ & $6.75\{_{6.50}^{7.00}$ & $2.4\{_{2.1}^{2.7}$ & $29\{_{28}^{30}$ & $6.75\{_{6.50}^{7.00}$ & $2.3\{_{2.1}^{2.6}$ & $0.041$ &  & 2 &  &  &  \\
0843$-$0048 &  & 08:43:51.07 & $-$00:48:24.6 & $24\{_{23}^{27}$ & $6.50\{_{6.25}^{7.00}$ & $3.0\{_{2.7}^{3.6}$ & $19\{_{18}^{20}$ & $6.00\{_{5.75}^{6.25}$ & $2.7\{_{2.4}^{3.0}$ & $0.033$ &  & 1 &  &  &  \\
0846$+$0142 &  & 08:46:28.66 & $+$01:42:17.0 & $24\{_{23}^{25}$ & $4.50\{_{4.25}^{4.75}$ & $4.6\{_{4.3}^{4.9}$ & $28\{_{27}^{29}$ & $4.00\{_{3.75}^{4.25}$ & $4.5\{_{2.7}^{4.5}$ & $0.042$ &  & 1 &  &  &  \\
0854$+$0853 & PN\,A66\,31 & 08:54:13.16 & $+$08:53:52.9 & $40\{_{39}^{40}$ & $3.00\{_{3.00}^{3.25}$ & $1.2\{_{1.2}^{1.2}$ & $40\{_{39}^{40}$ & $0.00\{_{25.00}^{0.00}$ & $1.1\{_{1.1}^{17.6}$ & $0.065$ & PN & 3 &  &  & pWD \\
0856$+$0518 &  & 08:56:33.17 & $+$05:18:39.6 & $29\{_{28}^{30}$ & $3.25\{_{3.00}^{3.50}$ & $2.6\{_{2.6}^{4.3}$ & $40\{_{38}^{40}$ & $3.50\{_{3.25}^{3.75}$ & $3.1\{_{3.0}^{3.1}$ & $0.050$ &  & 1 &  &  &  \\
0859$+$0759 &  & 08:59:26.08 & $+$07:59:13.3 & $22\{_{21}^{23}$ & $5.75\{_{5.50}^{6.00}$ & $2.9\{_{2.6}^{3.2}$ & $23\{_{22}^{24}$ & $5.50\{_{5.25}^{5.75}$ & $2.5\{_{2.3}^{2.8}$ & $0.080$ &  & 1 &  &  &  \\
0902$+$0734 &  & 09:02:25.06 & $+$07:34:04.0 & $21\{_{20}^{22}$ & $6.50\{_{6.25}^{6.75}$ & $3.5\{_{3.3}^{4.7}$ & $22\{_{21}^{23}$ & $6.25\{_{6.00}^{6.50}$ & $3.1\{_{2.8}^{3.5}$ & $0.072$ &  & 1 & SD &  &  \\
0906$+$0251 &  & 09:06:40.00 & $+$02:51:46.4 & $39\{_{38}^{40}$ & $6.75\{_{6.50}^{7.00}$ & $5.7\{_{5.1}^{6.1}$ & $28\{_{27}^{29}$ & $6.50\{_{6.25}^{6.75}$ & $5.2\{_{4.0}^{5.6}$ & $0.034$ &  & 1 & SD &  &  \\
0906$+$0437 &  & 09:06:00.86 & $+$04:37:45.1 & $21\{_{20}^{22}$ & $5.75\{_{5.50}^{6.00}$ & $2.4\{_{2.3}^{3.4}$ & $27\{_{25}^{28}$ & $6.25\{_{6.00}^{6.50}$ & $3.0\{_{2.7}^{3.3}$ & $0.037$ &  & 1 & SD &  &  \\
0920$+$1057 &  & 09:20:48.04 & $+$10:57:34.5 & $34\{_{33}^{35}$ & $4.75\{_{4.50}^{5.00}$ & $3.3\{_{3.1}^{3.5}$ & $40\{_{39}^{40}$ & $4.75\{_{4.50}^{5.00}$ & $3.3\{_{3.2}^{3.4}$ & $0.036$ & Comp & 1 &  &  & WDMS \\
0920$+$3356 & BK\,Lyn & 09:20:11.21 & $+$33:56:42.4 & $20\{_{18}^{21}$ & $6.75\{_{6.25}^{7.00}$ & $2.3\{_{1.7}^{2.3}$ & $20\{_{19}^{21}$ & $6.50\{_{6.25}^{6.75}$ & $2.1\{_{1.6}^{2.2}$ & $0.017$ & NL & 3 & WD &  & CV \\
0925$-$0140 &  & 09:25:35.00 & $-$01:40:46.8 & $17\{_{16}^{18}$ & $5.50\{_{5.25}^{5.75}$ & $9.4\{_{9.4}^{14.9}$ & $20\{_{19}^{21}$ & $5.50\{_{5.25}^{5.75}$ & $9.3\{_{5.8}^{9.3}$ & $0.031$ &  & 2 &  &  &  \\
0929$+$0603 &  & 09:29:20.43 & $+$06:03:46.2 & $29\{_{28}^{30}$ & $6.00\{_{5.75}^{6.25}$ & $1.6\{_{1.4}^{2.0}$ & $29\{_{28}^{30}$ & $5.50\{_{5.25}^{5.75}$ & $1.2\{_{1.1}^{1.6}$ & $0.052$ &  & 1 & SD &  &  \\
0937$+$0813 & PG\,0935$+$084 & 09:37:40.93 & $+$08:13:20.9 & $21\{_{20}^{22}$ & $5.75\{_{5.50}^{6.00}$ & $1.7\{_{1.6}^{2.3}$ & $23\{_{22}^{24}$ & $5.75\{_{5.50}^{6.00}$ & $1.7\{_{1.5}^{1.9}$ & $0.042$ & sdB & 1 & SD &  &  \\
0939$+$3038 &  & 09:39:14.38 & $+$30:38:17.3 & $39\{_{23}^{40}$ & $6.25\{_{5.75}^{6.50}$ & $5.0\{_{4.1}^{5.6}$ & $37\{_{26}^{38}$ & $6.00\{_{5.75}^{6.25}$ & $4.4\{_{4.0}^{5.2}$ & $0.017$ &  & 1 & SD &  &  \\
0941$+$0657 & PG\,0939$+$072 & 09:41:59.32 & $+$06:57:17.2 & $29\{_{28}^{30}$ & $6.50\{_{6.25}^{6.75}$ & $1.7\{_{1.5}^{2.0}$ & $30\{_{29}^{31}$ & $6.25\{_{6.00}^{6.50}$ & $1.5\{_{1.3}^{1.6}$ & $0.040$ &  & 2 &  &  &  \\
0951$+$0347 &  & 09:51:01.29 & $+$03:47:57.0 & $23\{_{22}^{24}$ & $4.00\{_{3.75}^{4.25}$ & $1.9\{_{1.7}^{2.0}$ & $28\{_{27}^{29}$ & $4.00\{_{3.75}^{4.25}$ & $2.0\{_{1.2}^{2.0}$ & $0.039$ & WD & 1 & SD &  &  \\
0959$+$0330 & PG\,0957$+$037 & 09:59:52.01 & $+$03:30:32.8 & $31\{_{30}^{32}$ & $3.00\{_{3.00}^{3.25}$ & $1.1\{_{1.1}^{1.2}$ & $40\{_{39}^{40}$ & $3.00\{_{3.00}^{3.25}$ & $1.3\{_{1.2}^{1.3}$ & $0.025$ &  & 2 &  &  & pWD \\
1006$+$0032 & PG\,1004$+$008 & 10:06:45.75 & $+$00:32:04.5 & $26\{_{25}^{27}$ & $5.00\{_{4.75}^{5.25}$ & $3.3\{_{3.1}^{3.6}$ & $29\{_{28}^{30}$ & $4.25\{_{4.00}^{4.50}$ & $1.9\{_{1.9}^{3.1}$ & $0.036$ &  & 1 &  &  & pWD \\
1011$-$0212 &  & 10:11:36.23 & $-$02:12:14.6 & $25\{_{24}^{26}$ & $4.75\{_{4.50}^{5.00}$ & $4.0\{_{3.7}^{4.4}$ & $28\{_{27}^{29}$ & $4.75\{_{4.50}^{5.00}$ & $4.2\{_{2.6}^{4.2}$ & $0.040$ &  & 1 &  &  &  \\
1012$+$0044 &  & 10:12:18.95 & $+$00:44:13.4 & $26\{_{25}^{27}$ & $4.75\{_{4.50}^{5.00}$ & $5.0\{_{4.7}^{5.4}$ & $29\{_{28}^{30}$ & $3.75\{_{3.50}^{4.00}$ & $2.9\{_{2.9}^{4.8}$ & $0.033$ & WD & 1 & SD &  &  \\
1015$-$0308 & SW\,Sex & 10:15:09.38 & $-$03:08:32.8 & $21\{_{20}^{22}$ & $6.75\{_{6.50}^{7.00}$ & $2.0\{_{1.9}^{2.5}$ & $23\{_{22}^{24}$ & $6.50\{_{6.25}^{6.75}$ & $1.8\{_{1.6}^{2.0}$ & $0.033$ & NL & 2 &  &  & CV \\
1016$+$0443 &  & 10:16:42.94 & $+$04:43:17.7 & $29\{_{28}^{30}$ & $4.50\{_{4.25}^{4.75}$ & $4.8\{_{4.8}^{7.9}$ & $29\{_{28}^{30}$ & $4.00\{_{3.75}^{4.25}$ & $4.3\{_{4.3}^{7.1}$ & $0.024$ & Comp & 4 & WD &  &  \\
1018$+$0953 &  & 10:18:33.11 & $+$09:53:36.1 & $35\{_{34}^{36}$ & $5.50\{_{5.25}^{5.75}$ & $1.3\{_{1.2}^{1.5}$ & $38\{_{37}^{39}$ & $5.25\{_{5.00}^{5.50}$ & $1.2\{_{1.1}^{1.3}$ & $0.037$ & WD & 1 & WD &  &  \\
1034$+$0327 & HS\,1031$+$0343 & 10:34:30.16 & $+$03:27:36.4 & $18\{_{17}^{19}$ & $4.25\{_{4.00}^{4.50}$ & $4.7\{_{4.3}^{5.0}$ & $20\{_{19}^{40}$ & $4.50\{_{4.25}^{19.00}$ & $5.0\{_{3.8}^{37.9}$ & $0.038$ & WD & 3 &  &  &  \\
\hline
\end{tabular}
\end{minipage}
\end{table}
\end{landscape}

\begin{landscape}
\begin{table}
\addtolength{\tabcolsep}{-1pt}
\begin{minipage}{\columnwidth}
\contcaption{}
\begin{tabular}{llllccccccclclll}
\hline\hline
 & & & & sdB \Teff & MS \Teff & d & sdB \Teff & MS \Teff & d & & & &
SDSS & Known & \\
Name & Identifier & R.A. & Dec & (1000\,K) & (1000\,K) & (kpc) &
(1000\,K) & (1000\,K) & (kpc) & E(B-V) & SIMBAD & Q & Spec &
Comp & Table\,\ref{t-ind2} \\
\hline
1040$+$0217 &  & 10:40:32.74 & $+$02:17:29.7 & $30\{_{29}^{33}$ & $4.25\{_{4.00}^{4.50}$ & $2.4\{_{2.3}^{2.8}$ & $39\{_{38}^{40}$ & $4.50\{_{4.25}^{4.75}$ & $3.0\{_{2.8}^{3.1}$ & $0.035$ &  & 1 &  &  &  \\
1055$+$0930 &  & 10:55:25.88 & $+$09:30:56.3 & $28\{_{27}^{29}$ & $6.50\{_{6.25}^{6.75}$ & $7.0\{_{4.8}^{7.1}$ & $27\{_{26}^{28}$ & $6.00\{_{5.75}^{6.25}$ & $5.5\{_{5.0}^{5.9}$ & $0.030$ &  & 1 & SD &  &  \\
1057$-$0230 &  & 10:57:59.29 & $-$02:30:02.1 & $27\{_{25}^{28}$ & $5.50\{_{5.25}^{5.75}$ & $5.6\{_{5.0}^{6.1}$ & $27\{_{26}^{28}$ & $5.25\{_{5.00}^{5.50}$ & $5.1\{_{4.7}^{5.4}$ & $0.042$ &  & 2 &  &  &  \\
1100$+$0346 &  & 11:00:53.55 & $+$03:46:22.8 & $34\{_{33}^{36}$ & $3.75\{_{3.50}^{4.25}$ & $2.7\{_{2.6}^{2.9}$ & $40\{_{39}^{40}$ & $3.75\{_{3.50}^{4.00}$ & $2.9\{_{2.9}^{3.0}$ & $0.044$ & WD & 1 & SD &  & pWD \\
1113$+$0413 & PG\,1110$+$045 & 11:13:17.32 & $+$04:13:14.5 & $30\{_{29}^{31}$ & $4.75\{_{4.50}^{5.00}$ & $0.9\{_{0.8}^{1.0}$ & $34\{_{32}^{35}$ & $5.00\{_{4.75}^{5.25}$ & $1.0\{_{0.9}^{1.1}$ & $0.051$ &  & 1 &  & 2,7 &  \\
1116$+$0755 &  & 11:16:16.37 & $+$07:55:32.5 & $28\{_{27}^{29}$ & $5.00\{_{4.75}^{5.25}$ & $2.3\{_{1.4}^{2.3}$ & $29\{_{28}^{30}$ & $3.75\{_{3.50}^{4.00}$ & $1.2\{_{1.2}^{1.9}$ & $0.042$ &  & 1 &  &  & pWD \\
1131$+$0932 & PG\,1128$+$098 & 11:31:14.32 & $+$09:32:19.0 & $40\{_{39}^{40}$ & $5.75\{_{5.50}^{6.00}$ & $1.2\{_{1.1}^{1.4}$ & $39\{_{38}^{40}$ & $5.50\{_{5.25}^{5.75}$ & $1.1\{_{1.0}^{1.2}$ & $0.039$ &  & 1 &  &  &  \\
1134$+$0153 &  & 11:34:18.00 & $+$01:53:22.1 & $38\{_{21}^{39}$ & $6.50\{_{5.75}^{6.75}$ & $7.0\{_{5.1}^{7.8}$ & $24\{_{23}^{26}$ & $6.00\{_{5.75}^{6.25}$ & $5.8\{_{5.2}^{6.5}$ & $0.031$ & WD & 1 & SD &  &  \\
1135$+$0731 &  & 11:35:36.86 & $+$07:31:28.3 & $29\{_{28}^{30}$ & $6.25\{_{6.00}^{6.50}$ & $6.4\{_{5.7}^{8.2}$ & $29\{_{28}^{30}$ & $6.00\{_{5.75}^{6.25}$ & $5.5\{_{4.9}^{7.3}$ & $0.041$ &  & 1 & SD &  & pWD \\
1203$+$0909 & PG\,1200$+$094 & 12:03:19.38 & $+$09:09:51.6 & $27\{_{26}^{28}$ & $5.75\{_{5.50}^{6.00}$ & $1.5\{_{1.3}^{1.6}$ & $28\{_{27}^{29}$ & $5.75\{_{5.50}^{6.00}$ & $1.5\{_{1.1}^{1.6}$ & $0.020$ &  & 1 &  &  &  \\
1215$+$1351 &  & 12:15:23.73 & $+$13:51:02.3 & $21\{_{20}^{22}$ & $4.50\{_{4.25}^{4.75}$ & $3.1\{_{3.1}^{5.1}$ & $21\{_{20}^{22}$ & $4.00\{_{3.75}^{4.25}$ & $2.9\{_{2.9}^{4.8}$ & $0.032$ &  & 2 & SD &  & pWD \\
1228$+$1040 & WD\,1226$+$110 & 12:28:59.93 & $+$10:40:33.0 & $21\{_{20}^{22}$ & $3.00\{_{3.00}^{3.25}$ & $2.2\{_{2.2}^{3.8}$ & $23\{_{22}^{24}$ & $3.00\{_{3.00}^{3.25}$ & $2.3\{_{2.2}^{2.4}$ & $0.028$ & WD & 4 & WD &  & pWD \\
1233$+$0834 &  & 12:33:09.62 & $+$08:34:34.5 & $30\{_{29}^{31}$ & $6.00\{_{5.75}^{6.25}$ & $1.9\{_{1.7}^{2.2}$ & $30\{_{29}^{31}$ & $5.75\{_{5.50}^{6.00}$ & $1.7\{_{1.5}^{1.9}$ & $0.019$ &  & 1 & SD &  &  \\
1235$+$1029 &  & 12:35:13.03 & $+$10:29:59.5 & $23\{_{22}^{24}$ & $5.75\{_{5.50}^{6.00}$ & $3.0\{_{2.7}^{3.4}$ & $24\{_{23}^{26}$ & $5.75\{_{5.50}^{6.00}$ & $3.0\{_{2.7}^{3.4}$ & $0.026$ &  & 1 & SD &  &  \\
1237$-$0151 &  & 12:37:04.70 & $-$01:51:23.0 & $23\{_{22}^{25}$ & $4.75\{_{4.50}^{5.00}$ & $3.9\{_{3.6}^{4.3}$ & $25\{_{24}^{26}$ & $4.75\{_{4.50}^{5.00}$ & $3.9\{_{3.6}^{4.2}$ & $0.028$ &  & 1 &  &  & pWD \\
1247$-$0039 & PG\,1244$-$004 & 12:47:06.79 & $-$00:39:25.8 & $33\{_{32}^{34}$ & $0.00\{_{25.00}^{0.00}$ & $1.8\{_{1.8}^{43.0}$ & $39\{_{37}^{40}$ & $0.00\{_{25.00}^{0.00}$ & $2.1\{_{2.0}^{41.0}$ & $0.033$ & WD & 3 & SD &  &  \\
1300$+$0045 & PG\,1257$+$010 & 13:00:25.52 & $+$00:45:30.1 & $29\{_{28}^{30}$ & $4.75\{_{4.50}^{5.00}$ & $1.4\{_{1.3}^{2.0}$ & $29\{_{28}^{30}$ & $4.50\{_{4.25}^{4.75}$ & $1.2\{_{1.2}^{1.9}$ & $0.026$ &  & 1 & SD &  & pWD \\
1300$+$0057 & HE\,1258$+$0113 & 13:00:59.21 & $+$00:57:11.8 & $30\{_{29}^{31}$ & $3.50\{_{3.25}^{3.75}$ & $1.7\{_{1.6}^{1.8}$ & $32\{_{31}^{33}$ & $3.50\{_{3.25}^{3.75}$ & $1.8\{_{1.7}^{1.9}$ & $0.025$ & WD & 1 & SD & 8 & SD \\
1312$+$2245 &  & 13:12:42.62 & $+$22:45:04.2 & $26\{_{25}^{27}$ & $3.00\{_{3.00}^{0.00}$ & $3.7\{_{3.5}^{3.8}$ & $29\{_{28}^{30}$ & $0.00\{_{25.00}^{0.00}$ & $2.1\{_{2.1}^{59.1}$ & $0.013$ &  & 3 &  &  &  \\
1315$+$0245 &  & 13:15:12.39 & $+$02:45:31.7 & $33\{_{32}^{34}$ & $3.25\{_{3.00}^{3.50}$ & $0.9\{_{0.8}^{1.0}$ & $36\{_{35}^{37}$ & $3.25\{_{3.00}^{3.50}$ & $0.9\{_{0.9}^{1.0}$ & $0.023$ &  & 1 &  &  & pWD \\
1316$+$0348 & PG\,1314$+$041 & 13:16:38.48 & $+$03:48:18.5 & $29\{_{28}^{30}$ & $4.75\{_{4.50}^{5.00}$ & $1.5\{_{1.4}^{2.1}$ & $29\{_{28}^{30}$ & $4.50\{_{4.25}^{4.75}$ & $1.3\{_{1.2}^{1.9}$ & $0.030$ &  & 1 & SD &  &  \\
1316$+$0739 &  & 13:16:33.59 & $+$07:39:41.3 & $29\{_{28}^{30}$ & $6.75\{_{6.50}^{7.00}$ & $6.0\{_{5.4}^{7.1}$ & $29\{_{28}^{30}$ & $6.50\{_{6.25}^{6.75}$ & $5.2\{_{4.7}^{6.4}$ & $0.026$ &  & 1 & SD &  &  \\
1319$-$0141 &  & 13:19:32.20 & $-$01:41:31.2 & $30\{_{29}^{31}$ & $6.25\{_{6.00}^{6.50}$ & $7.8\{_{6.8}^{8.8}$ & $29\{_{28}^{32}$ & $6.00\{_{5.75}^{6.25}$ & $6.6\{_{6.1}^{8.9}$ & $0.024$ &  & 1 & SD &  &  \\
1323$+$2615 &  & 13:23:57.28 & $+$26:15:02.5 & $20\{_{19}^{21}$ & $5.00\{_{4.75}^{5.25}$ & $5.8\{_{3.6}^{5.8}$ & $21\{_{20}^{22}$ & $4.25\{_{4.00}^{4.50}$ & $3.3\{_{3.3}^{5.4}$ & $0.018$ &  & 1 &  &  & pWD \\
1325$+$1212 & PG\,1323$+$125 & 13:25:57.24 & $+$12:12:21.3 & $26\{_{25}^{28}$ & $5.75\{_{5.50}^{6.00}$ & $2.1\{_{1.9}^{2.4}$ & $27\{_{26}^{28}$ & $5.75\{_{5.50}^{6.00}$ & $2.1\{_{1.9}^{2.3}$ & $0.034$ &  & 1 &  &  &  \\
1326$+$0357 & PG\,1323$+$042 & 13:26:19.95 & $+$03:57:54.4 & $22\{_{21}^{23}$ & $4.75\{_{4.50}^{5.00}$ & $1.4\{_{1.2}^{1.5}$ & $25\{_{24}^{26}$ & $5.00\{_{4.75}^{5.25}$ & $1.5\{_{1.4}^{1.6}$ & $0.025$ & sdO & 1 & SD &  &  \\
1328$+$3108 &  & 13:28:56.72 & $+$31:08:46.0 & $40\{_{39}^{40}$ & $4.75\{_{4.50}^{5.00}$ & $4.2\{_{3.9}^{4.4}$ & $38\{_{37}^{39}$ & $4.50\{_{4.25}^{4.75}$ & $3.8\{_{3.5}^{4.0}$ & $0.011$ &  & 4 & Galaxy &  &  \\
1336$+$1126 & PG\,1334$+$117 & 13:36:53.99 & $+$11:26:05.4 & $29\{_{28}^{30}$ & $4.25\{_{4.00}^{4.50}$ & $1.6\{_{1.6}^{2.5}$ & $30\{_{29}^{31}$ & $4.25\{_{4.00}^{4.50}$ & $1.6\{_{1.6}^{1.8}$ & $0.031$ &  & 1 &  &  &  \\
1341$+$0317 &  & 13:41:22.97 & $+$03:17:51.6 & $28\{_{24}^{29}$ & $6.50\{_{6.00}^{6.75}$ & $8.9\{_{6.0}^{9.3}$ & $27\{_{25}^{28}$ & $6.25\{_{6.00}^{6.50}$ & $7.8\{_{6.9}^{8.7}$ & $0.024$ & WD & 1 & SD &  &  \\
1351$+$0234 &  & 13:51:40.69 & $+$02:34:29.2 & $29\{_{28}^{30}$ & $4.25\{_{4.00}^{4.50}$ & $2.5\{_{2.5}^{4.1}$ & $30\{_{29}^{31}$ & $3.75\{_{3.50}^{4.00}$ & $2.4\{_{2.3}^{2.5}$ & $0.027$ & WD & 1 & SD &  &  \\
1352$+$0910 &  & 13:52:28.14 & $+$09:10:39.1 & $29\{_{28}^{30}$ & $4.00\{_{3.75}^{4.25}$ & $4.0\{_{4.0}^{6.5}$ & $29\{_{28}^{30}$ & $4.00\{_{3.75}^{4.25}$ & $3.9\{_{3.9}^{6.3}$ & $0.028$ & DA+M & 4 & CV &  & WDMS \\
1402$+$0725 & PG\,1359$+$077 & 14:02:03.86 & $+$07:25:39.1 & $26\{_{25}^{27}$ & $5.00\{_{4.75}^{5.25}$ & $2.3\{_{2.1}^{2.4}$ & $27\{_{26}^{28}$ & $5.00\{_{4.75}^{5.25}$ & $2.3\{_{2.1}^{2.4}$ & $0.024$ &  & 1 & SD &  &  \\
1402$+$3215 &  & 14:02:32.83 & $+$32:15:22.2 & $22\{_{21}^{23}$ & $6.25\{_{6.00}^{6.50}$ & $1.9\{_{1.7}^{2.1}$ & $23\{_{22}^{24}$ & $6.25\{_{6.00}^{6.50}$ & $1.9\{_{1.7}^{2.1}$ & $0.015$ &  & 1 & SD &  &  \\
1421$+$0753 & KN\,Boo & 14:21:38.17 & $+$07:53:19.6 & $27\{_{26}^{28}$ & $5.25\{_{5.00}^{5.50}$ & $1.6\{_{1.5}^{1.7}$ & $27\{_{26}^{28}$ & $5.00\{_{4.75}^{5.25}$ & $1.5\{_{1.3}^{1.6}$ & $0.028$ & sdB & 1 & SD &  &  \\
1422$+$0920 &  & 14:22:11.11 & $+$09:20:43.6 & $26\{_{25}^{27}$ & $4.75\{_{4.50}^{5.00}$ & $3.5\{_{3.2}^{3.7}$ & $28\{_{27}^{29}$ & $4.75\{_{4.50}^{5.00}$ & $3.5\{_{2.3}^{3.6}$ & $0.024$ &  & 1 & SD &  & pWD \\
1425$+$0302 &  & 14:25:26.81 & $+$03:02:00.8 & $17\{_{16}^{18}$ & $6.25\{_{6.00}^{6.50}$ & $7.0\{_{7.0}^{10.4}$ & $17\{_{16}^{18}$ & $6.00\{_{5.75}^{6.25}$ & $6.2\{_{6.2}^{9.4}$ & $0.035$ &  & 5 &  &  &  \\
1429$+$0643 &  & 14:29:47.00 & $+$06:43:35.0 & $22\{_{21}^{23}$ & $6.50\{_{6.25}^{6.75}$ & $9.3\{_{8.3}^{10.3}$ & $21\{_{20}^{22}$ & $6.25\{_{6.00}^{6.50}$ & $7.9\{_{7.5}^{11.3}$ & $0.025$ & HII & 4 & Galaxy &  &  \\
1440$+$1223 &  & 14:40:10.10 & $+$12:23:34.3 & $21\{_{20}^{22}$ & $5.00\{_{4.75}^{5.25}$ & $5.3\{_{5.2}^{8.3}$ & $21\{_{20}^{22}$ & $5.00\{_{4.75}^{5.25}$ & $5.2\{_{5.1}^{8.1}$ & $0.029$ &  & 3 &  &  &  \\
1442$+$0910 &  & 14:42:10.30 & $+$09:10:07.6 & $26\{_{24}^{27}$ & $5.00\{_{4.75}^{5.25}$ & $6.6\{_{5.9}^{7.1}$ & $26\{_{25}^{27}$ & $4.75\{_{4.50}^{5.00}$ & $6.2\{_{5.8}^{6.6}$ & $0.023$ &  & 1 &  &  & pWD \\
1443$+$0931 &  & 14:43:07.70 & $+$09:31:34.0 & $28\{_{27}^{29}$ & $4.50\{_{4.25}^{4.75}$ & $4.5\{_{2.7}^{4.5}$ & $30\{_{29}^{31}$ & $3.25\{_{3.00}^{3.50}$ & $2.6\{_{2.4}^{2.7}$ & $0.031$ &  & 1 & SD &  & pWD \\
1445$+$0002 & V594\,Vir & 14:45:14.93 & $+$00:02:48.9 & $25\{_{24}^{26}$ & $5.50\{_{5.25}^{5.75}$ & $4.6\{_{4.3}^{5.0}$ & $27\{_{26}^{28}$ & $5.50\{_{5.25}^{5.75}$ & $4.6\{_{4.3}^{5.0}$ & $0.041$ & Var* & 1 & SD &  &  \\
1456$+$0330 &  & 14:56:01.20 & $+$03:30:28.8 & $18\{_{17}^{19}$ & $4.50\{_{4.25}^{4.75}$ & $5.6\{_{5.3}^{6.0}$ & $19\{_{18}^{20}$ & $4.50\{_{4.25}^{4.75}$ & $5.6\{_{5.3}^{6.1}$ & $0.041$ &  & 3 &  &  &  \\
1500$+$0642 &  & 15:00:11.77 & $+$06:42:11.5 & $27\{_{26}^{28}$ & $4.25\{_{3.75}^{4.50}$ & $3.9\{_{3.7}^{4.2}$ & $34\{_{33}^{35}$ & $3.25\{_{3.00}^{3.50}$ & $2.5\{_{2.4}^{2.6}$ & $0.034$ &  & 1 & SD &  & pWD \\
1501$+$0537 &  & 15:01:15.02 & $+$05:37:39.4 & $28\{_{27}^{29}$ & $6.50\{_{6.25}^{6.75}$ & $5.1\{_{3.9}^{5.4}$ & $28\{_{27}^{29}$ & $6.25\{_{6.00}^{6.50}$ & $4.4\{_{3.3}^{4.8}$ & $0.038$ &  & 1 & SD &  &  \\
1502$-$0245 & PG\,1459$-$026 & 15:02:12.12 & $-$02:45:57.8 & $30\{_{29}^{31}$ & $6.00\{_{5.75}^{6.25}$ & $1.4\{_{1.3}^{1.6}$ & $22\{_{21}^{23}$ & $5.75\{_{5.50}^{6.00}$ & $1.3\{_{1.2}^{1.5}$ & $0.124$ &  & 1 & SD &  &  \\
1507$+$0724 &  & 15:07:37.71 & $+$07:24:16.5 & $27\{_{26}^{28}$ & $4.50\{_{4.25}^{4.75}$ & $4.1\{_{3.8}^{4.4}$ & $28\{_{27}^{29}$ & $4.00\{_{3.75}^{4.25}$ & $3.8\{_{2.3}^{3.8}$ & $0.030$ &  & 1 &  &  & pWD \\
1509$-$0143 &  & 15:09:02.07 & $-$01:43:54.4 & $29\{_{28}^{30}$ & $6.50\{_{6.25}^{6.75}$ & $3.7\{_{3.4}^{4.8}$ & $30\{_{29}^{31}$ & $6.50\{_{6.25}^{6.75}$ & $3.6\{_{3.2}^{4.0}$ & $0.072$ &  & 1 & SD &  &  \\
\hline
\end{tabular}
\end{minipage}
\end{table}
\end{landscape}

\begin{landscape}
\begin{table}
\addtolength{\tabcolsep}{-1pt}
\begin{minipage}{\columnwidth}
\contcaption{}
\begin{tabular}{llllccccccclclll}
\hline\hline
 & & & & sdB \Teff & MS \Teff & d & sdB \Teff & MS \Teff & d & & & &
SDSS & Known & \\
Name & Identifier & R.A. & Dec & (1000\,K) & (1000\,K) & (kpc) &
(1000\,K) & (1000\,K) & (kpc) & E(B-V) & SIMBAD & Q & Spec &
Comp & Table\,\ref{t-ind2} \\
\hline
1510$+$0409 &  & 15:10:42.06 & $+$04:09:55.6 & $26\{_{25}^{27}$ & $4.00\{_{3.75}^{4.25}$ & $3.4\{_{3.2}^{3.6}$ & $30\{_{29}^{31}$ & $3.00\{_{3.00}^{3.25}$ & $2.0\{_{1.9}^{2.1}$ & $0.039$ & WD & 1 & SD &  & pWD \\
1516$+$0926 &  & 15:16:46.27 & $+$09:26:31.7 & $28\{_{27}^{29}$ & $5.25\{_{5.00}^{5.50}$ & $4.4\{_{3.0}^{4.5}$ & $32\{_{31}^{33}$ & $4.25\{_{4.00}^{4.50}$ & $2.4\{_{2.3}^{2.6}$ & $0.039$ &  & 1 &  &  &  \\
1517$+$0310 & PG\,1514$+$034 & 15:17:14.27 & $+$03:10:28.0 & $40\{_{39}^{40}$ & $6.00\{_{5.75}^{6.25}$ & $1.1\{_{1.0}^{1.2}$ & $28\{_{27}^{29}$ & $5.75\{_{5.50}^{6.00}$ & $1.0\{_{0.8}^{1.1}$ & $0.039$ & WD & 1 &  & 1,7 & SD \\
1518$+$0410 & PG\,1515$+$044 & 15:18:08.48 & $+$04:10:43.8 & $26\{_{25}^{27}$ & $5.50\{_{5.25}^{5.75}$ & $1.8\{_{1.7}^{2.0}$ & $28\{_{27}^{29}$ & $5.25\{_{5.00}^{5.50}$ & $1.7\{_{1.3}^{1.8}$ & $0.047$ & sdO & 1 & SD & 1,7 & SD \\
1520$-$0009 &  & 15:20:20.40 & $-$00:09:48.3 & $17\{_{16}^{18}$ & $5.75\{_{5.50}^{6.50}$ & $4.2\{_{4.2}^{6.8}$ & $18\{_{17}^{19}$ & $5.50\{_{5.25}^{5.75}$ & $3.8\{_{3.5}^{4.2}$ & $0.062$ &  & 1 &  &  &  \\
1520$+$0713 &  & 15:20:00.81 & $+$07:13:48.8 & $24\{_{23}^{25}$ & $5.00\{_{4.75}^{5.25}$ & $2.4\{_{2.2}^{2.5}$ & $27\{_{26}^{28}$ & $4.75\{_{4.50}^{5.00}$ & $2.2\{_{2.1}^{2.4}$ & $0.037$ &  & 1 &  &  &  \\
1522$+$0803 &  & 15:22:12.20 & $+$08:03:40.9 & $21\{_{14}^{29}$ & $22.00\{_{21.00}^{23.00}$ & $78.3\{_{69.8}^{86.9}$ & $21\{_{17}^{33}$ & $24.00\{_{23.00}^{25.00}$ & $80.0\{_{73.1}^{86.5}$ & $0.034$ &  & 4 & CV &  &  \\
1524$+$1020 &  & 15:24:28.45 & $+$10:20:51.6 & $25\{_{23}^{27}$ & $6.25\{_{6.00}^{6.50}$ & $8.0\{_{7.0}^{9.1}$ & $28\{_{26}^{29}$ & $6.25\{_{6.00}^{6.50}$ & $7.9\{_{5.8}^{8.5}$ & $0.034$ &  & 1 & SD &  &  \\
1525$+$0958 &  & 15:25:34.15 & $+$09:58:51.0 & $29\{_{28}^{30}$ & $3.25\{_{3.00}^{4.25}$ & $2.8\{_{2.8}^{4.8}$ & $33\{_{32}^{34}$ & $3.25\{_{3.00}^{3.50}$ & $3.0\{_{2.9}^{3.1}$ & $0.036$ &  & 1 & SD &  & pWD \\
1527$+$1016 &  & 15:27:07.20 & $+$10:16:12.5 & $23\{_{22}^{24}$ & $5.50\{_{5.25}^{5.75}$ & $2.5\{_{2.3}^{2.7}$ & $24\{_{23}^{25}$ & $5.25\{_{5.00}^{5.50}$ & $2.2\{_{2.1}^{2.4}$ & $0.039$ &  & 1 & SD &  &  \\
1536$+$0218 &  & 15:36:13.08 & $+$02:18:09.4 & $24\{_{22}^{28}$ & $5.75\{_{5.50}^{6.25}$ & $3.9\{_{3.5}^{5.0}$ & $25\{_{24}^{28}$ & $5.75\{_{5.50}^{6.00}$ & $3.9\{_{3.5}^{4.5}$ & $0.059$ &  & 1 &  &  &  \\
1538$+$0644 & HS\,1536$+$0944 & 15:38:18.87 & $+$06:44:38.7 & $14\{_{13}^{15}$ & $7.25\{_{7.00}^{7.50}$ & $6.5\{_{6.4}^{8.6}$ & $14\{_{13}^{15}$ & $7.00\{_{6.75}^{7.25}$ & $6.0\{_{5.9}^{8.2}$ & $0.052$ &  & 2 & SD & 6 & pWD \\
1538$+$0934 &  & 15:38:42.85 & $+$09:34:42.3 & $23\{_{22}^{24}$ & $5.00\{_{4.75}^{5.25}$ & $1.8\{_{1.7}^{2.0}$ & $39\{_{38}^{40}$ & $5.25\{_{5.00}^{5.50}$ & $1.8\{_{1.7}^{2.0}$ & $0.038$ &  & 1 &  &  & SD \\
1539$+$0933 &  & 15:39:24.44 & $+$09:33:28.3 & $19\{_{18}^{20}$ & $5.75\{_{5.50}^{6.25}$ & $1.8\{_{1.7}^{2.2}$ & $20\{_{19}^{21}$ & $5.50\{_{5.25}^{5.75}$ & $1.7\{_{1.3}^{1.8}$ & $0.035$ &  & 1 &  &  &  \\
1540$+$0005 & PG\,1538$+$002 & 15:40:50.58 & $+$00:05:17.8 & $26\{_{25}^{27}$ & $4.25\{_{4.00}^{4.50}$ & $1.8\{_{1.7}^{2.0}$ & $30\{_{29}^{31}$ & $3.25\{_{3.00}^{3.50}$ & $1.1\{_{1.0}^{1.2}$ & $0.087$ &  & 1 &  &  &  \\
1542$+$0056 &  & 15:42:18.25 & $+$00:56:11.8 & $29\{_{28}^{30}$ & $6.50\{_{6.25}^{6.75}$ & $1.5\{_{1.3}^{1.7}$ & $21\{_{20}^{22}$ & $6.25\{_{6.00}^{6.50}$ & $1.3\{_{1.2}^{1.6}$ & $0.098$ &  & 1 &  &  &  \\
1542$+$0155 &  & 15:42:10.89 & $+$01:55:57.2 & $21\{_{20}^{23}$ & $5.75\{_{5.50}^{6.00}$ & $2.2\{_{2.1}^{3.0}$ & $25\{_{24}^{26}$ & $6.00\{_{5.75}^{6.25}$ & $2.5\{_{2.2}^{2.8}$ & $0.069$ &  & 1 & SD &  &  \\
1543$+$0012 & WD\,1541$+$003 & 15:43:38.69 & $+$00:12:02.1 & $21\{_{20}^{22}$ & $4.75\{_{4.50}^{5.00}$ & $2.9\{_{2.8}^{4.5}$ & $25\{_{24}^{26}$ & $5.00\{_{4.75}^{5.25}$ & $3.2\{_{3.0}^{3.4}$ & $0.086$ & WD & 1 & SD &  & pWD \\
1545$+$0132 &  & 15:45:45.57 & $+$01:32:29.3 & $29\{_{28}^{30}$ & $6.00\{_{5.75}^{6.25}$ & $3.0\{_{2.7}^{3.8}$ & $29\{_{28}^{30}$ & $5.75\{_{5.50}^{6.00}$ & $2.6\{_{2.3}^{3.4}$ & $0.093$ &  & 1 & SD &  &  \\
1546$+$0625 &  & 15:46:41.89 & $+$06:25:39.3 & $29\{_{28}^{30}$ & $4.75\{_{4.50}^{5.00}$ & $3.5\{_{3.4}^{5.5}$ & $30\{_{29}^{31}$ & $4.50\{_{4.25}^{4.75}$ & $3.1\{_{2.9}^{3.4}$ & $0.050$ & WD & 4 & WD &  &  \\
1548$+$0334 &  & 15:48:52.88 & $+$03:34:29.4 & $23\{_{22}^{24}$ & $4.75\{_{4.50}^{5.00}$ & $3.6\{_{3.3}^{3.8}$ & $25\{_{24}^{26}$ & $4.75\{_{4.50}^{5.00}$ & $3.6\{_{3.4}^{3.9}$ & $0.107$ &  & 1 &  &  &  \\
1550$-$0104 &  & 15:50:21.35 & $-$01:04:53.5 & $24\{_{23}^{25}$ & $5.25\{_{5.00}^{5.50}$ & $4.0\{_{3.8}^{4.4}$ & $28\{_{27}^{29}$ & $5.25\{_{5.00}^{5.50}$ & $4.1\{_{2.7}^{4.2}$ & $0.120$ &  & 1 &  &  &  \\
1551$+$0029 & PG\,1549$+$006 & 15:51:44.88 & $+$00:29:48.8 & $23\{_{22}^{24}$ & $3.00\{_{3.00}^{3.25}$ & $1.3\{_{1.3}^{1.4}$ & $27\{_{26}^{28}$ & $3.00\{_{3.00}^{3.25}$ & $1.4\{_{1.4}^{1.5}$ & $0.074$ &  & 3 &  &  &  \\
1554$+$0616 &  & 15:54:32.27 & $+$06:16:17.8 & $29\{_{28}^{30}$ & $4.75\{_{4.50}^{5.00}$ & $3.8\{_{3.6}^{5.8}$ & $29\{_{28}^{30}$ & $4.50\{_{4.25}^{4.75}$ & $3.3\{_{3.1}^{5.1}$ & $0.041$ &  & 3 &  &  & pWD \\
1619$+$2407 &  & 16:19:42.83 & $+$24:07:15.7 & $24\{_{23}^{25}$ & $6.50\{_{6.25}^{6.75}$ & $4.3\{_{3.8}^{4.8}$ & $27\{_{26}^{28}$ & $6.75\{_{6.50}^{7.00}$ & $4.7\{_{4.3}^{5.1}$ & $0.067$ &  & 1 & SD &  & pWD \\
1635$+$2952 &  & 16:35:18.31 & $+$29:52:03.3 & $25\{_{24}^{27}$ & $5.25\{_{5.00}^{5.50}$ & $3.4\{_{3.1}^{3.7}$ & $27\{_{26}^{28}$ & $5.25\{_{5.00}^{5.50}$ & $3.4\{_{3.2}^{3.7}$ & $0.021$ &  & 1 & SD &  &  \\
1644$+$3123 &  & 16:44:44.95 & $+$31:23:45.3 & $26\{_{25}^{27}$ & $5.75\{_{5.50}^{6.00}$ & $3.1\{_{2.8}^{3.4}$ & $27\{_{26}^{28}$ & $5.50\{_{5.25}^{5.75}$ & $2.8\{_{2.6}^{3.0}$ & $0.028$ & WD & 1 & SD &  &  \\
1650$+$3127 & PG\,1648$+$315 & 16:50:22.05 & $+$31:27:49.7 & $29\{_{28}^{30}$ & $5.25\{_{5.00}^{5.50}$ & $1.6\{_{1.5}^{2.1}$ & $29\{_{28}^{30}$ & $5.00\{_{4.75}^{5.25}$ & $1.5\{_{1.4}^{1.9}$ & $0.030$ & WD & 1 & SD &  &  \\
2045$+$0024 &  & 20:45:37.81 & $+$00:24:40.5 & $29\{_{28}^{30}$ & $6.00\{_{5.75}^{6.25}$ & $5.7\{_{5.2}^{7.9}$ & $29\{_{28}^{30}$ & $5.50\{_{5.25}^{5.75}$ & $4.0\{_{3.9}^{6.0}$ & $0.096$ & WD & 2 & SD &  &  \\
2046$-$0006 &  & 20:46:43.28 & $-$00:06:30.2 & $17\{_{16}^{18}$ & $6.00\{_{5.75}^{6.25}$ & $7.4\{_{7.2}^{11.2}$ & $18\{_{17}^{19}$ & $5.75\{_{5.50}^{6.00}$ & $6.5\{_{5.9}^{7.1}$ & $0.079$ &  & 2 &  &  &  \\
2049$-$0001 &  & 20:49:22.58 & $-$00:01:34.7 & $18\{_{17}^{19}$ & $5.25\{_{5.00}^{5.50}$ & $6.0\{_{5.6}^{6.4}$ & $20\{_{19}^{21}$ & $5.00\{_{4.75}^{5.25}$ & $5.8\{_{3.6}^{5.8}$ & $0.093$ &  & 1 &  &  & pWD \\
2050$+$0057 &  & 20:50:51.37 & $+$00:57:12.2 & $22\{_{21}^{23}$ & $6.00\{_{5.75}^{6.25}$ & $6.8\{_{6.0}^{7.6}$ & $26\{_{24}^{28}$ & $6.25\{_{6.00}^{6.50}$ & $7.4\{_{6.5}^{8.4}$ & $0.107$ &  & 1 & SD &  &  \\
2051$+$0112 &  & 20:51:01.72 & $+$01:12:59.7 & $23\{_{21}^{24}$ & $6.00\{_{5.75}^{6.25}$ & $5.9\{_{5.1}^{6.5}$ & $24\{_{23}^{26}$ & $5.75\{_{5.50}^{6.00}$ & $5.1\{_{4.7}^{5.8}$ & $0.111$ &  & 1 &  &  &  \\
2052$-$0050 &  & 20:52:54.69 & $-$00:50:31.8 & $23\{_{22}^{24}$ & $5.75\{_{5.50}^{6.00}$ & $6.2\{_{5.7}^{6.8}$ & $24\{_{23}^{25}$ & $5.50\{_{5.25}^{5.75}$ & $5.6\{_{5.2}^{6.0}$ & $0.093$ & WD & 1 & SD &  &  \\
2057$+$0108 &  & 20:57:58.45 & $+$01:08:17.7 & $22\{_{21}^{23}$ & $5.75\{_{5.50}^{6.00}$ & $3.6\{_{3.3}^{3.9}$ & $23\{_{21}^{24}$ & $5.50\{_{5.25}^{5.75}$ & $3.3\{_{2.9}^{3.5}$ & $0.083$ & WD & 1 & SD &  &  \\
2059$+$0105 &  & 20:59:54.78 & $+$01:05:57.0 & $23\{_{22}^{24}$ & $6.00\{_{5.75}^{6.25}$ & $3.5\{_{3.1}^{3.9}$ & $25\{_{24}^{26}$ & $6.00\{_{5.75}^{6.25}$ & $3.5\{_{3.1}^{3.9}$ & $0.075$ &  & 1 & SD &  &  \\
2117$-$0006 &  & 21:17:42.22 & $-$00:06:19.9 & $30\{_{29}^{31}$ & $6.50\{_{6.25}^{6.75}$ & $2.1\{_{1.8}^{2.3}$ & $24\{_{23}^{25}$ & $6.25\{_{6.00}^{6.50}$ & $1.9\{_{1.7}^{2.1}$ & $0.074$ &  & 1 & SD &  & pWD \\
2120$+$0037 &  & 21:20:14.38 & $+$00:37:56.4 & $22\{_{21}^{23}$ & $3.00\{_{3.00}^{3.25}$ & $2.9\{_{2.7}^{3.1}$ & $28\{_{27}^{29}$ & $3.00\{_{3.00}^{3.25}$ & $3.1\{_{1.9}^{3.1}$ & $0.086$ &  & 3 &  &  &  \\
2147$-$0112 & FBS\,2145$-$014 & 21:47:43.59 & $-$01:12:02.9 & $25\{_{24}^{26}$ & $3.25\{_{3.00}^{3.50}$ & $1.6\{_{1.5}^{1.7}$ & $28\{_{27}^{29}$ & $0.00\{_{25.00}^{0.00}$ & $1.5\{_{0.9}^{22.7}$ & $0.047$ &  & 2 &  &  & pWD \\
2236$+$0640 & PG\,2234$+$064 & 22:36:41.97 & $+$06:40:17.5 & $28\{_{11}^{40}$ & $19.00\{_{18.00}^{20.00}$ & $26.0\{_{23.4}^{29.2}$ & $27\{_{26}^{28}$ & $0.00\{_{25.00}^{0.00}$ & $2.1\{_{2.0}^{34.8}$ & $0.132$ &  & 3 &  &  &  \\
2244$+$0106 & PB\,5146 & 22:44:51.81 & $+$01:06:31.0 & $24\{_{22}^{26}$ & $5.75\{_{5.50}^{6.00}$ & $6.9\{_{6.1}^{7.9}$ & $26\{_{24}^{28}$ & $5.75\{_{5.50}^{6.00}$ & $6.9\{_{6.1}^{7.8}$ & $0.079$ & sdB & 1 & SD &  &  \\
2245$+$0611 &  & 22:45:11.78 & $+$06:11:43.7 & $30\{_{29}^{31}$ & $6.00\{_{5.75}^{6.25}$ & $6.4\{_{5.6}^{7.3}$ & $39\{_{38}^{40}$ & $5.75\{_{5.50}^{6.00}$ & $5.3\{_{4.8}^{5.8}$ & $0.095$ &  & 1 &  &  &  \\
2333$+$1522 &  & 23:33:25.92 & $+$15:22:22.2 & $17\{_{16}^{18}$ & $6.75\{_{6.50}^{7.00}$ & $12.6\{_{12.3}^{17.8}$ & $17\{_{16}^{18}$ & $6.50\{_{6.25}^{6.75}$ & $10.9\{_{10.7}^{15.9}$ & $0.068$ & CV & 4 & CV &  & CV \\
2346$+$0344 &  & 23:46:55.70 & $+$03:44:28.5 & $29\{_{28}^{30}$ & $6.75\{_{6.50}^{7.00}$ & $1.6\{_{1.5}^{1.9}$ & $29\{_{28}^{30}$ & $6.50\{_{6.25}^{6.75}$ & $1.5\{_{1.3}^{1.7}$ & $0.054$ &  & 1 &  &  &  \\